# Is DNS Ready for Prime Time?: The State of ANSYS Fluent DNS in 2025

Reid Prichard[1], Wayne Strasser*[2], and Keith Walters[3]

[1]*BWXT Advanced Technologies LLC*

[2]*Liberty University*

[3]*University of Arkansas*

**Corresponding author*

# Contents

# Abstract

Can an industrial computational fluid dynamics (CFD) tool like ANSYS Fluent perform Direct Numerical Simulation (DNS) reliably? We answer by providing a combination of a feasibility report and a best-practice guide that will be useful for researchers and practicing engineers as pressure mounts to transition DNS from academia to industry. DNS has historically been the realm of efficient but somewhat restrictive spectral solvers. DNS has been validated on Finite Volume Method (FVM) solvers, but many of these works have been on structured grids with high-order stencils. This paper explores a canonical channel flow at $\mathrm{Re}_\tau = 180$ using ANSYS Fluent, a commercial FVM code, with lower order methods. Additionally, for the first time, this paper provides direct comparisons of the diffusion and dispersion of Fluent's momentum discretization schemes on six easily reproducible grids: two grid spacings each of grids composed of triangular prisms, hexahedrons, and polygonal prisms. We also demonstrate the substantial increase in solution speed offered by GPUs. While velocity profiles are predicted with moderate error, wall-normal profiles of turbulent kinetic energy budgets exhibited significant error, and large amounts of spectral pileup were visible. Therefore, it can be concluded that, for a canonical flow configuration, ANSYS Fluent is only partially “ready”. While it cannot be said that all FVM DNS is problematic for all possible flows, we prove, through a simple flow scenario, that healthy skepticism should be applied when the term “DNS” is used in conjunction with  ANSYS Fluent and FVM codes which use similar stencils.

## Introduction

The early 1900s brought a confluence of wall-bounded turbulence flows physics revelations, but the first turbulence simulations were not born until the early 1970s at the National Center for Atmospheric Research (NCAR) [1,2]. It is the authors’ understanding that the first turbulent channel validation study was the large eddy simulations (LES) of [3], wherein the sub-grid-scale viscosity constant for the filtered stresses had to be tuned to prevent overdamped large eddies. That work did not resolve the near-wall flows, however, and instead used a logarithmic law of the wall boundary condition. The first author to take the LES approach all the way to the wall was [4] with some pseudospectral method. Very recently, [5] compared Fluent LES to a reference DNS; they showed strong convergence of first- and second-order statistics, the absence of an inertial range, rapid decay of high wavenumber energy density, and energy pile-up close to the grid cutoff wavenumber.

While Reynolds-Averaged Navier Stokes (RANS) turbulence models and LES have improved greatly in recent years, direct numerical simulation (DNS) remains indispensable in certain cases, e.g., studying near-wall physics. As far as the authors are aware, the first validation DNS study of wall-bounded flow was that of [6] using some spectral method. Originally largely confined to academic research, exponentially increasing computing power has made DNS more accessible. In industrial and applied engineering problems, it is generally desired to study complex geometries on arbitrary grids. This rules out the spectral methods typically used for DNS for their high numerical order. When we speak of "spectral methods", we are not necessary emphasizing any particular spectral method. While traditional Fourier or Chebyshev spectral methods are indeed restricted to simple geometries, high-order spectralelement or discontinuous Galerkin spectral element methods (DGSEM) can handle complex geometries effectively.

DNS has been largely validated on Finite Difference Method (FDM) solvers using high-order methods that require a structured grid [7–9]. This restriction removes many of the advantages compared to the spectral method. DNS with the Finite Volume Method (FVM) has been demonstrated with low-order methods on a structured grid. Jin et al. provided good agreement with validation data for velocity profiles (mean and RMS) [10]. Only the streamwise energy spectrum was shown; simulation data matched validation except at high wavenumbers, where energy was an order of magnitude underpredicted. De Azevedo et al. also demonstrated reasonably accurate reproduction of mean and RMS velocity profiles as well as energy spectra [11]. However, in the Turbulent Kinetic Energy (TKE) budget, near-wall dissipation exhibited a maximum error of around 30%, and TKE profiles were not presented.

Literature demonstrating FVM DNS on unstructured grids neglects to show important, sensitive metrics such as the power spectral densities (sometimes for lack of validation data), limiting the scope of their application [12–15]. Furthermore, practitioners in industry are often constrained in their choice of FVM solver; using an academic code or writing their own bespoke code is frequently not an option. As such, it is essential to validate industrial codes for DNS, despite the "black box" nature of their implementation.

Ansys Fluent is one of the most common CFD codes used in industry and many university undergraduate programs. Several papers claim to use Fluent for DNS [9,16–18], but discussion of numerical schemes and validation is not comprehensive. Furthermore, literature appears to be lacking any insight into the numerical dispersiveness and diffusion of Fluent's momentum discretization schemes. Though most of these schemes are well known in literature, Ansys's unique and often undisclosed implementation details merit specific study of Fluent.

## Objectives

In this work, we seek to accomplish three, to our knowledge, not previously reported goals. First, we will quantify the currently undescribed numerical properties of Fluent's discretization schemes via a canonical inviscid 2D study. In addition to comparing all available momentum schemes, we will attempt to shed some light on Fluent's "Bounded

Central Differencing" scheme and compare values of its little-discussed "boundedness" parameter. Second, we will discuss the challenges faced dealing with Fluent's native GPU solver and present our techniques to overcome them. Finally, we will demonstrate DNS of a $\mathrm{Re}_\tau = 180$ (see Equation 1) channel flow, extensively comparing to validation data by Abe et al. [7] and Vreman & Kuerten [8]. To clarify, the purpose of this work is not to provide any new algorithm or a validation of an advanced rigorous flow scenario. Nor do we seek to replace such well-defined and broadly accepted validation test cases from sources like Drag Prediction Workshop and the High-Order CFD Workshop (links provided for both). On the contrary, we intentionally selected a low Reynolds number ($Re_\tau = 180$) and simple geometry to prove a broadly community-relevant point: If an industrial FVM has difficulty resolving this flow scenario, caution should cover any attempt at using industrial FVM for more complicated DNS scenarios.

## Notation

Since turbulence involves discussion of instantaneous, mean, and fluctuating components, we will adopt the following convention: the instantaneous velocity component is represented by $\vec{U}$ or $U_i$, the mean component is represented by $\langle\vec{U}\rangle$ or $\langle U_i\rangle$, and the fluctuating component is represented by $\vec{u}$ or $u_i$. More generally, $\langle\phi\rangle$ represents the mean value of any scalar.

Inner scaling was used to produce dimensionless results as shown in Equations 1-4. Specifically, velocities were normalized by the friction velocity $u_\tau$, obtained via the definition of the friction Reynolds number:

$$\mathrm{Re}_\tau = \frac{u_\tau \delta}{\nu} \tag{1}$$

Lengths were normalized by the viscous length scale:

$$\delta_\nu = \nu / u_\tau \tag{2}$$

And pressures normalized by the wall shear stress:

$$\tau_w = \rho u_\tau^2 \tag{3}$$

Finally, timescales were normalized by the viscous time scale:

$$t^+ = \frac{u_\tau^2}{\nu} \tag{4}$$

# Methodology

Although this paper exhibits results on a regular domain easily representable with a structured grid, the intent was to validate Ansys Fluent 2023R2's GPU-based solver with methods applicable to a arbitrary unstructured grids. This goal rules out many methods that may otherwise be competitive choices, yet reasonably rigorous options remain available. Computations were all performed in a GPU-accelerated, double-precision, segregated, incompressible, pressure-based solver. Fluent refers to direct solution of the discretized Navier-Stokes (Equation 5) and continuity (Equation 6) equations as its “laminar” solver, but we will demonstrate that it is capable of fully resolving a turbulent flow as one would expect. The equations solved for an incompressible, Newtonian fluid are:

$$\rho \frac{D}{Dt}[U_i] = -\frac{\partial p}{\partial x_i} + \mu \frac{\partial^2 U_i}{\partial x_j \partial x_j} + S_i \tag{5}$$

$$\frac{\partial U_i}{\partial x_i} = 0 \tag{6}$$

Where $\rho$ represents density, $p$ is the static pressure, $x_i$ is the position vector, $\mu$ is the molecular viscosity, $S_i$ is the momentum source term, and $U_i$ is the velocity vector, $t$ is time, and $\frac{D}{Dt}$ represents the material derivative operator:

$$\frac{D}{Dt}[\phi] \equiv \frac{\partial \phi_i}{\partial t} + U_j \frac{\partial \phi_i}{\partial x_j} \tag{7}$$

## Computational Domain

We performed validation of DNS on four separate grids (see Table 1), which we will denote S1 (small domain, grid spacing 1), S2 (small domain, grid spacing 2), SP (small domain, unstructured polyhedral grid), and L1 (large domain, grid spacing 1). All grids but SP consisted of structured hexahedra.

*Table 1: Dimensions (in inner and outer units) and cell types of grids used.*

| Grid | Type | | | | Channel half-widths | | |
|---|---|---|---|---|---|---|---|
| | | $x^+$ | $y^+$ | $z^+$ | $L_x$ | $L_y$ | $L_z$ |
| S1 | Structured hexahedra | 2304 | 360 | 1152 | 12.8 | 2 | 6.4 |
| S2 | Structured hexahedra | 2304 | 360 | 1152 | 12.8 | 2 | 6.4 |
| SP | Unstructured polyhedra | 2304 | 360 | 1152 | 12.8 | 2 | 6.4 |
| L1 | Structured hexahedra | 4608 | 360 | 2304 | 25.6 | 2 | 12.8 |

## Structured Grids (S1, S2, L1)

Grid S1 matches the domain size used in Abe et al. [7], with dimensions of $2304\delta_v \times 360\delta_v \times 1152\delta_v$ in the streamwise $(x)$, wall-normal $(y)$, and spanwise $(z)$ directions, respectively. It also matches cell counts along each axis of $256 \times 128 \times 256$. Elements are evenly distributed in the streamwise and spanwise directions, while the wall-normal element distribution matched Abe et al.'s distribution, originally presented by Moin and Kim [19]:

$$y_j = \frac{1}{2}\tanh\left(\xi_j \tanh^{-1} \alpha\right) \tag{8}$$

$$\xi_j = -1 + 2\frac{j-1}{N_2 - 1} \tag{9}$$

Where $y_j$ is the y-coordinate of a mesh node, $j$ is the node number moving upward from the lower wall, and $N_2$ is the node count in the wall-normal direction. As in Abe et al., $\alpha = 0.967$. This sizing function produces narrow elements near the wall that widen moving to the centerline, depicted in Figure 1. Grid S2 was adapted from S1, splitting each cell in half along all three axes. This produces a grid with 8x the cell count of grid S1. Since the sizing function was not recomputed, the resulting wall-normal distribution is a stepwise approximation of the relationship presented in Equations 8 and 9. Grid L1 maintained the same grid spacing as grid S1, but the domain was enlarged by doubling in the streamwise and spanwise direction. The channel height remained the same.

## Unstructured Grid (SP)

While a structured hexahedral grid is generally considered the most appropriate choice for a simple channel flow, our goal is to validate methods applicable to arbitrary geometries. Consequently, we chose to additionally evaluate the performance of a completely unstructured polyhedral grid with no prism layers. One advantage of an unstructured grid is the ability to produce an isotropic grid with small cells near the wall and large cells in the free stream. Hexahedral grids are constrained by geometry and end up being highly stretched in the streamwise and spanwise directions; because of the polyhedral grid's isotropic nature, cell count is higher than might be expected for the grid's wall-normal spacing.

Grid SP's isotropic mesh length scale (calculated as the cube root of the cell volume) was in between that of S1 and S2. However, due to their anisotropy, S1 and S2 had lower wall-normal cell heights throughout the domain despite comparable cell counts as shown by Table 2. Grid SP was used as a stand-in to represent a canonical industrial use case with complex geometries, and Figure 1 shows its grid spatial distribution.

*Table 2: Description of turbulent channel computational grids. Since grid SP had isotropic spacing, the "Dy+" columns represent the mesh length scale in all three axes at those locations. Bolded values draw attention to differences in L1. Grid SP's large near-wall spacing evidences its lack of suitability for DNS.*

| Grid | $\Delta x^+$ | $\Delta y^+$ (centerline) | $\Delta y^+$ (wall) | $\Delta z^+$ | Cells |
|---|---|---|---|---|---|
| S1 | 9 | 10.5 | 0.39 | 4.5 | 9.41e6 |
| S2 | 4.5 | 5.7 | 0.20 | 2.25 | 6.71e7 |
| SP | - | 11.2 | 4.29 | - | 1.89e7 |
| L1 | **9** | 10.5 | 0.39 | **4.5** | 3.36e7 |

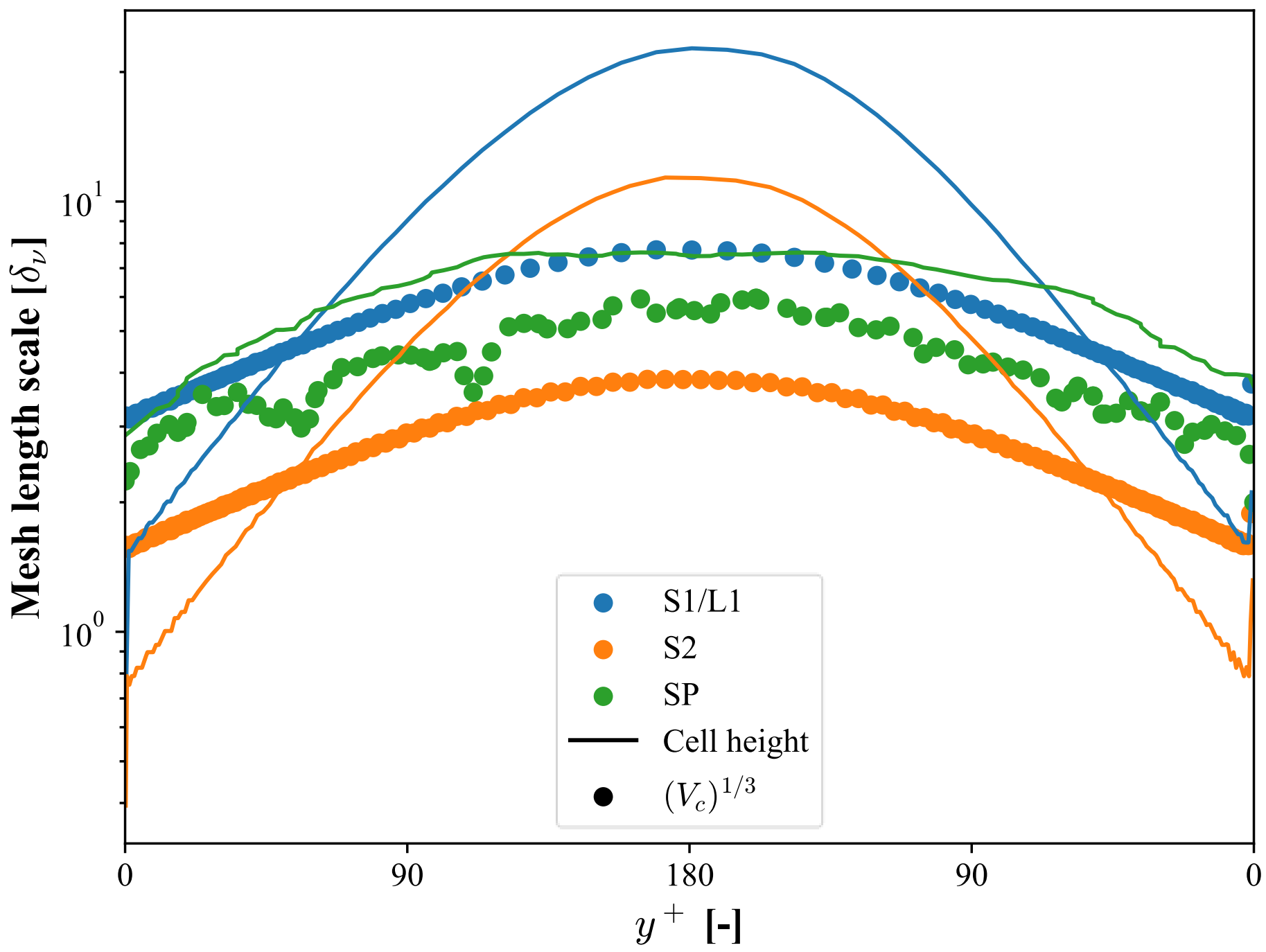

*Figure 1: Mesh length scale (defined as $\sqrt[3]{V_c}$, where $V_c$ is a cell's volume) and wall-normal cell spacing as a function of y+ for all turbulent channel grids. Cell height was measured as twice the difference between cell centroid locations along a wall-normal line. Due to the*

*inherent noisiness of sampling a line along an unstructured grid, the "SP" cell height was smoothed with an 80-point moving average.*

## Kolmogorov Length Scale Profiles

To assess the chosen grid spacing, we will discuss profiles of the computed Kolmogorov length scales in each axis and compare them to the mesh length scale. Ideally, the grid spacing would be less than the Kolmogorov scale for true DNS, but it is generally accepted in practice that 2-4x the Kolmogorov scale is a reasonable threshold [20–22].

Grid S1, grid S1 with CD, and grid L1 produced very similar curves. Grid S1's data can be seen in Figure 2; the data with CD and on L1 can be found in Appendix A. Horizontal dotted lines are at values of 2 and 4. Grid S2 slightly reduced calculated Kolmogorov length scales across the board, especially toward the centerline. Seen in Figure 3, S2 sharpened the near-wall peaks, drawing attention to the edge of the viscous sublayer. In all cases, grid spacing in $x$ and $z$ was sufficient (or nearly so) to resolve the computed Kolmogorov length. However, $\eta_y^+$ was significantly smaller than $\Delta y$, particularly with the coarser grids of S1 and L1. In these cases, the grid spacing at the centerline was 4× the Kolmogorov length in $y$. In grid S2, this ratio dropped to ~2. The near-wall limiting behavior of $\varepsilon_{22}$ is $\sim(y^+)^2$ [23]. Therefore, $\eta_y^+$ should vary as $(y^+)^{-1/2}$, which means it increases as the wall is approached.

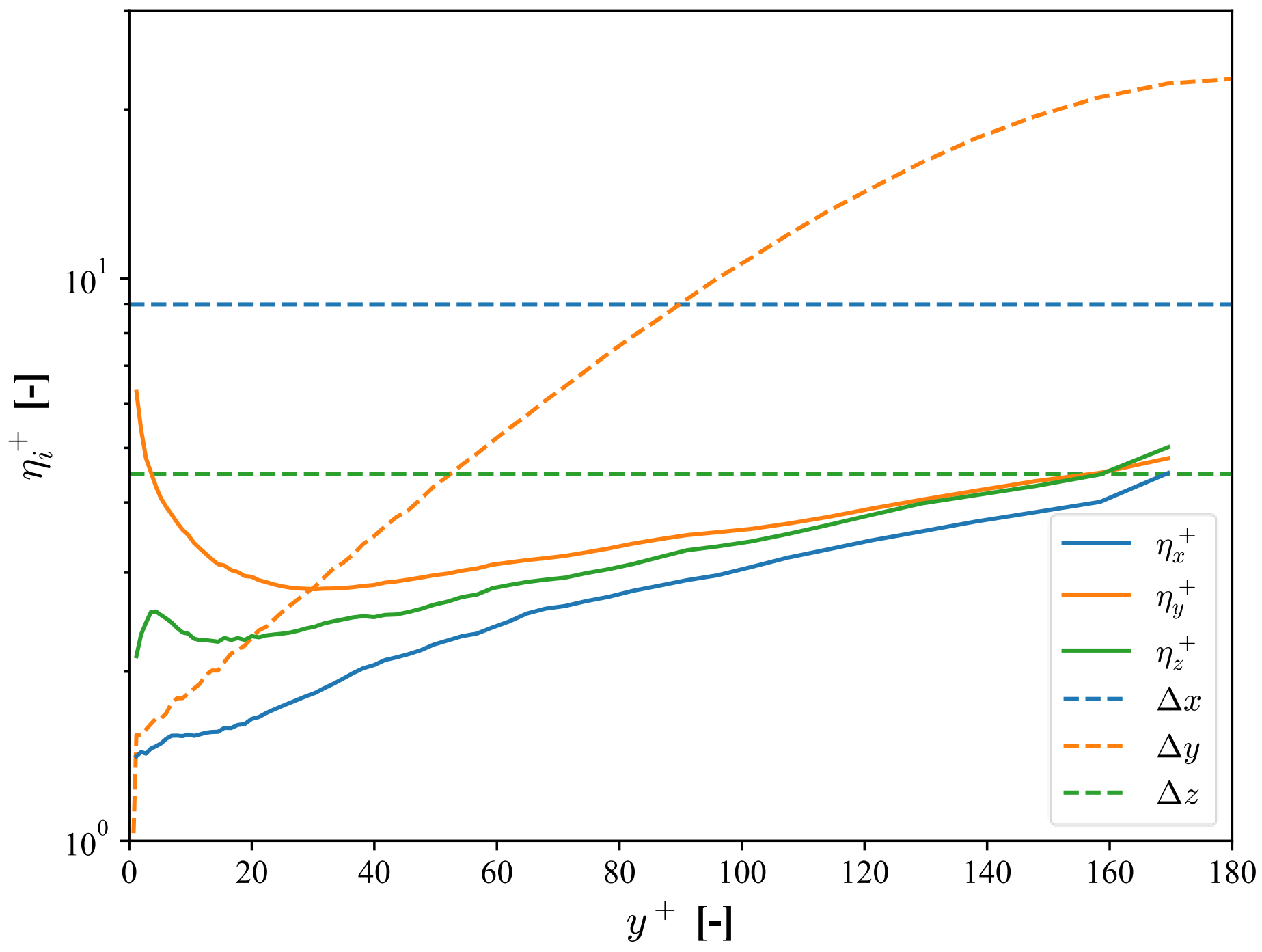

*Figure 2: Turbulent channel Kolmogorov length scales calculated by grid S1*

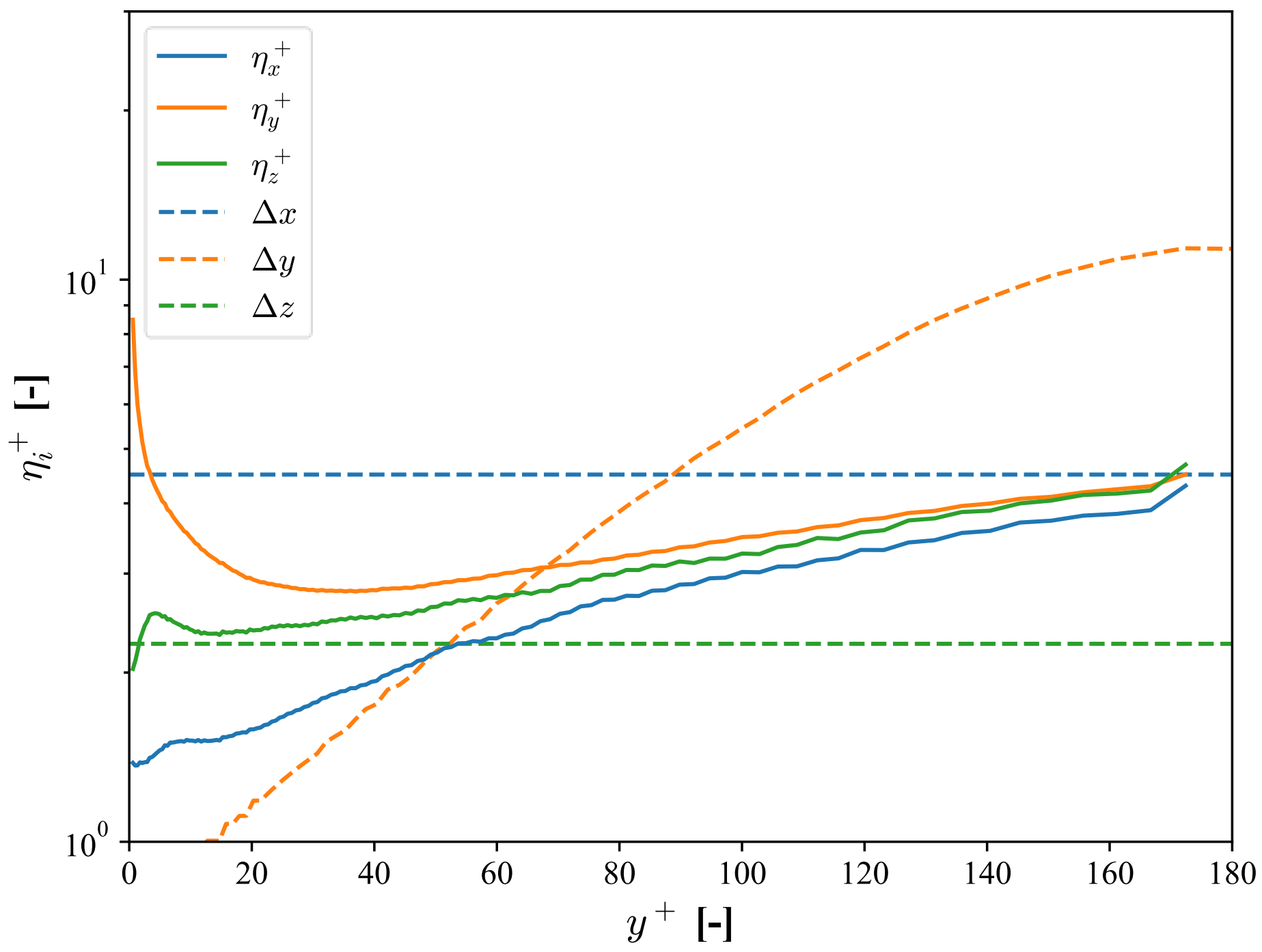

*Figure 3: Turbulent channel Kolmogorov length scales calculated by grid S2.*

### Boundary Conditions

As is customary for a channel flow, the boundary conditions were as follows: The bounding faces in the streamwise and spanwise directions were periodic, and the upper and lower walls were set to no-slip. The flow was driven by a $x$-momentum source term derived for our chosen $\mathrm{Re}_\tau$.

## Discretization and Numerics

### Pressure-Velocity Coupling (PVC)

Fluent's GPU solver does not implement a coupled scheme for transient flows, leaving SIMPLE and SIMPLEC as the only options. With simple geometry and a low Courant number, inter-time step convergence behavior was not a concern, so SIMPLE was used. Doubling time step size (which will later be discussed with the transient scheme) produced minimal changes in results, shown in Appendix A, confirming that SIMPLE achieved sufficient time step level convergence. We performed 20 inner PVC iterations per time step in all cases, which was found to achieve residual convergence. For inner iteration "convergence", a rigorous method was not applied. Generally speaking, residuals flattened to the point of insignificant change with an increasing count of inner iterations. Additionally, we tried cases earlier with 10, 20, and 30 PVC iterations and could not discern any differences other than the obvious run time impact.

## Momentum

A considerable effort was carried out to formally evaluate Fluent's momentum discretization options, noting this information is lacking in the literature. Every equation, except for BCD, can be found in [24]. Assuming every reader does not have convenient access to that manual, we direct to readers to Appendix A for all definitions necessary to understand the findings herein and Appendix B for the detailed numerical evaluations. Additionally, a summary of those findings will be given here: Of the numerical schemes applicable to an arbitrary grid, CD and BCD produce the least false diffusion, making them the best choices for DNS. Fluent has not yet incorporated their CD implementation into the GPU solver, so BCD is the only remaining option. In this study, we opted to use BCD-0.25 in all cases but one; there, BCD-0 (i.e., pure central differencing) was used for comparison. While both of these schemes exhibited visible dispersion, it is expected that the very small mesh length scale should prevent this from causing instability due to the small cell Peclet number.

## Velocity Interpolation

Fluent implements distance-weighted and momentum-weighted forms of Rhie-Chow [25] velocity interpolation to calculate face fluxes. This method adds a third-order damping term to suppress "checkerboarding". Description of the scheme is beyond the scope of this work, but we used the distance-weighted formulation.

## Pressure Interpolation

Fluent offers several pressure interpolation schemes, but their "Second Order" scheme was the only viable option set by the aim to minimize numerical diffusion. This scheme computes face velocities using a second-order central interpolation of the same form as Equation 31 but shown in Equation 10:

$$P_f = \frac{1}{2}(P_0 + P_1) + \frac{1}{2}\left[\overrightarrow{\nabla P}_0 \cdot (\vec{x}_f - \vec{x}_0) + \overrightarrow{\nabla P}_1 \cdot (\vec{x}_f - \vec{x}_1)\right] \tag{10}$$

Where subscripts 0 and 1 refer to the cells on either side of the face.

## Gradient Reconstruction

Least-Squares cell-based (LS) gradient reconstruction is the only available option in the GPU solver. This scheme is known to be first-order accurate on an arbitrary grid and second-order accurate on a structured grid [26]. For more detail of its implementation, see the appendix.

## Transient Terms

Fluent's GPU solver provides the first- and second-order implicit transient formulations. We chose to use the latter, which computes time derivatives as Equation 11:

$$\frac{\partial \phi}{\partial t} = \frac{3\phi^{n+1} - 4\phi^{n} + \phi^{n-1}}{2\Delta t} \tag{11}$$

We used a time step of $0.08t^+$ in all results, producing Courant numbers listed in Table 3: Courant numbers produced by $\Delta t = 0.08t^+$ on grids tested. As previously discussed, the time step size was chosen after comparison to time steps of $0.16t^+$ and $0.32t^+$.

*Table 3: Courant numbers produced by $\Delta t = 0.08t^{+}$ on grids tested.*

| Grid | Maximum Courant number | Volume average Courant number |
|---|---|---|
| S1/L1 | 0.40 | 0.16 |
| SP | 5.9 | 0.41 |
| S2 | 0.20 | 0.08 |

## Summary of Numerics Used

For the reader's convenience, Table 4 lists the schemes that were used in this paper's direct numerical simulation of a turbulent channel, which will follow.

*Table 4: Summary of numerical schemes used for direct numerical simulation in the latter half of this work.*

| | |
|---|---|
| Pressure-velocity coupling | SIMPLE |
| Momentum discretization | Bounded central differencing (BCD) |
| Velocity interpolation | Rhie-Chow distance-based |
| Pressure interpolation | Second-order |
| Gradient reconstruction | Least-squares cell-based |
| Transient | Second-order implicit |

## Data Analysis

Because of the outsized impact of export frequency on solver speed, we exported data only every 60 time steps, or $4.8t^+$. Data was then gathered for $10^4\ t^+$; we found this combination of parameters sufficiently minimized both low-frequency and high-frequency error.

### Initialization and Convergence

An initial perturbation was required to build turbulent fluctuations via imposing a non-Gaussian fluctuating component on the velocity field in the form of the Langevin equation. Once a turbulent velocity profile was observed, this forcing term was removed and the system allowed to converge to a statistically stationary state. Convergence was assessed by plotting the Power Spectral Density (PSD) of the velocity field at several consecutive time divisions. The one-dimensional turbulent energy PSD is referred to as $E_{ij}$. In the current study, it is computed over both streamwise and spanwise directions and is defined as:

$$E_{ij}(x_2, \kappa) = \frac{1}{\pi}\int_{-\infty}^{+\infty} R_{ij}(e_k r) e^{-i\kappa r}\, dr \tag{12a}$$

$$R_{ij}(x_2, e_k r) = \langle u_i(x_1, x_2, x_3, t) u_j(x_1 + e_1 r, x_2, x_3 + e_3 r, t)\rangle \tag{12b}$$

Where $\kappa$ represents the wavenumber, $R_{ij}$ represents the two-point velocity correlation for fluctuating velocity components $u_i$ and $u_j$, $e_k$ is the basis vector for the direction along which the spectrum is being computed, $r$ is the separation distance in the two-point correlation, and the subscript $k$ denotes the direction, where $k$ = 1 for streamwise and $k$ = 3 for spanwise correlation, respectively. Angle brackets denote infinite ensemble averaging over $x$, $z$, and $t$. In practice, the PSD is computed via a Discrete Fourier Transform (DFT) along axes of interest.

For example, we might plot and compare 30,000-40,000, 40,000-50,000, and 50,000-60,000 viscous timescales of flow time. An example of this is shown in Figure 4 for mesh S1, and repeated indices do not imply summation. Darker colors indicate later 10,000 viscous units-sized chunks of time. Once there was no directional trend in our sampled data, we considered the solution statistically stationary. We found that it took on the order of 30,000$t^+$ (~30 flow seconds) to reach a stationary state after turbulence was initiated. Then, after the development of a stationary state, time-averaged (TA) was carried out over another 30,000$t^+$ worth of data. With an outer scaling ($t_{out}$) which scales with large eddy turnover time and having time units of $h/u_{tau}$~$0.6s$, the TA accumulated for $50t_{out}$. We found that the PSDs took significantly longer to converge than did wall-normal velocity variances and covariances. Figure 4 indicates that our accumulation time was sufficient. However, as will be shown later, $50t_{out}$ may have been too little of time. Perhaps a finer grid, with more energetic motions, requires more. Other PSDs will be discussed herein.

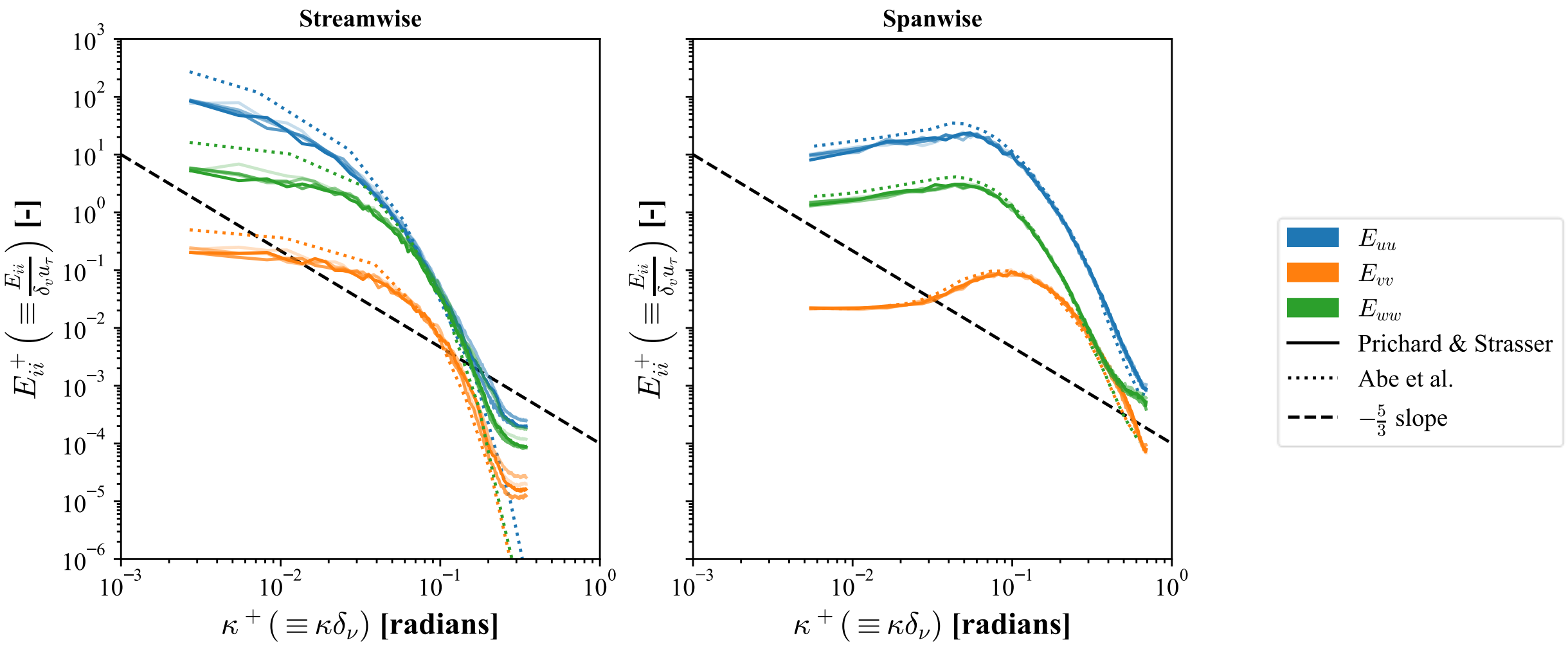

*Figure 4: An energy spectrum plot illustrating statistical variation typical of a stationary solution for mesh S1. Repeated lines of each color represent consecutive “chunks” of flow time with darker for later times; their agreement shows that integration time is sufficient and that flow is statistically stationary.*

## Other Measures of Interest

Additional validation measures considered are profiles of mean streamwise velocity and all three components of RMS fluctuating velocities. These will be compared to DNS data by Abe et al [7]. Computing these was trivial, so it will not be discussed.

Velocity PSDs were computed along the streamwise and spanwise directions at two wall distances: $y^+ = 5.4$ and $y^+ = 30$. The former will again be validated against Abe et al., while the latter will be compared to DNS data from Vreman & Kuerten [8]. Computing energy spectra was performed via the FFTW3 library [27] in a custom C++ code. Because grid SP was unstructured, linear interpolation was used to sample regularly-spaced values at the two wall distances: the streamwise sampling lines consist of 1000 interpolated points, and the spanwise lines consist of 512 sampling points. In all other grids, cell-centered values were used.

Another validation measure is the Turbulence Kinetic Energy (TKE) budget:

$$\frac{\bar{D}\langle u_i u_i\rangle}{\bar{D}t} = \mathcal{P} + \Pi - \varepsilon + T + D \tag{13}$$

where $\frac{\bar{D}}{\bar{D}t}$ represents the mean material derivative, $\mathcal{P}$ is the production scalar, $\varepsilon$ is the dissipation scalar, $\Pi$ is the velocity-pressure-gradient scalar, $T$ is the turbulent diffusion scalar, and $D$ is the molecular diffusion scalar. Though they contract to scalars for the TKE budget, these terms are generally defined in their full tensorial form in Equations 14-19,

$$\frac{\bar{D}}{\bar{D}t}[\phi] \equiv \frac{\partial \phi}{\partial t} + \langle U_i\rangle \frac{\partial \langle \phi\rangle}{\partial x_i} \tag{14}$$

$$\mathcal{P}_{ij} \equiv -\left(\langle u_i u_k\rangle \frac{\partial \langle U_j\rangle}{\partial x_k} + \langle u_j u_k\rangle \frac{\partial \langle U_i\rangle}{\partial x_k}\right) \tag{15}$$

$$\varepsilon_{ij} \equiv 2\nu \left\langle \frac{\partial u_i}{\partial x_k} \frac{\partial u_j}{\partial x_k} \right\rangle \tag{16}$$

$$\Pi_{ij} \equiv -\frac{1}{\rho}\left\langle u_i \frac{\partial p'}{\partial x_j} + u_j \frac{\partial p'}{\partial x_i} \right\rangle \tag{17}$$

$$T_{ij} \equiv -\frac{\partial}{\partial x_k}\langle u_i u_j u_k\rangle \tag{18}$$

$$D_{ij} = \nu \frac{\partial^2}{\partial x^2}\langle u_i u_j\rangle \tag{19}$$

where $p'$ is the fluctuating component of the pressure field. It is useful to decompose the velocity-pressure-gradient tensor into the pressure strain tensor, $\mathcal{R}_{ij}$, and the pressure diffusion tensor, $\frac{\partial T_{kij}^{(p)}}{\partial x_k}$:

$$\mathcal{R}_{ij} \equiv \left\langle \frac{p'}{\rho} \left( \frac{\partial u_i}{\partial x_j} + \frac{\partial u_j}{\partial x_i} \right) \right\rangle \tag{20}$$

$$\frac{\partial T_{kij}^{(p)}}{\partial x_k} \equiv \frac{\partial}{\partial x_i} \langle u_j p' \rangle + \frac{\partial}{\partial x_j} \langle u_i p' \rangle \tag{21}$$

Because the TKE budget involves high-order terms, e.g. the gradient of a triple velocity correlation in the turbulent diffusion tensor, it was necessary to perform these computations post facto. To enable gradient reconstruction, we exported transient solution data along a 3x3 grid of cells extending from the wall to the channel center. Green-Gauss node-based gradient reconstruction was then used to compute gradients along the center of the 3x3 column of cells and is given in Equation 22,

$$(\nabla \phi_c)_i = \frac{1}{V_c} \sum_{\text{cell faces}} \phi_f \left( A_f \right)_i \tag{22}$$

where $\phi_c$ is the cell-centered value of scalar $\phi$, $\phi_f$ is the scalar's value at one of the cell's faces,$V_c$ is the cell volume, and $A_f$ is a face's outward-facing area vector. The face value, $\phi_f$, is calculated as the mean of interpolated vertex values $\phi_v$, themselves calculated as the mean of neighboring cell values:

$$\phi_f = \frac{1}{N} \sum_{N \text{ vertices}} \phi_v \,, \qquad \phi_v = \frac{1}{M} \sum_{M \text{ cells}} \phi_c \tag{23}$$

Green-Gauss node-based was chosen over Least Squares cell-based due to its simplicity of implementation and comparable numeric order [26]. No special treatment was given to the nine wall-adjacent cells, which assumes that the wall-normal gradient was zero. Therefore, there is the potential for error at the nearest points to the wall. Due to the complexities introduced with an unstructured polyhedral grid, these analyses were not performed for grid SP. It is categorically more challenging to write our own CFD code to extract the necessary data from an unstructured mesh, so we opted to relegate that to future interesting research.

Lastly, we will examine the computed Kolmogorov length scale:

$$\eta \equiv \left(\frac{\nu^3}{\varepsilon}\right)^{1/4} \tag{24}$$

Since we compute the full dissipation tensor, we can compute the Kolmogorov scale along each axis, as follows:

$$\eta_x = \left(\frac{\nu^3}{\varepsilon_{11}}\right)^{1/4}, \qquad \eta_y = \left(\frac{\nu^3}{\varepsilon_{22}}\right)^{1/4}, \qquad \eta_z = \left(\frac{\nu^3}{\varepsilon_{33}}\right)^{1/4} \qquad (25-27)$$

# Results

We will present four types of results: velocity profiles, energy spectra compared to data from Abe et al., energy spectra compared with data from Vreman & Kuerten, and finally TKE budgets compared with data from Abe at al.. We begin with the velocity profiles produced by the three grids compared to validation data in Figure 5 and Figure 6 which shows the percent error compared to the validation data. Following those is Table 5, showing that all models exhibit RMS error of around 3%, while maximum error is up to 7.8%.

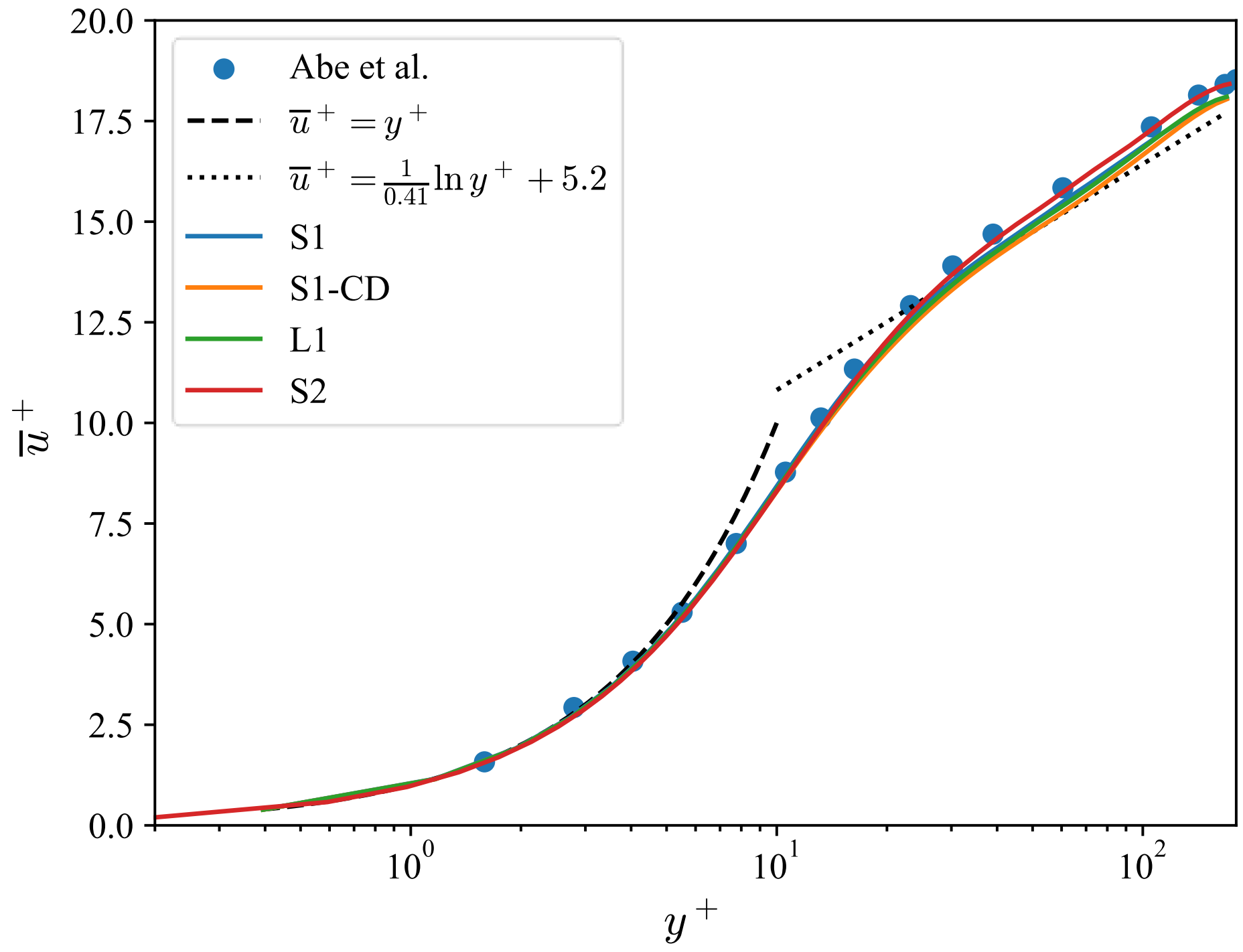

*Figure 5: Mean streamwise turbulent channel velocity profiles in comparison with the data from Abe et al..*

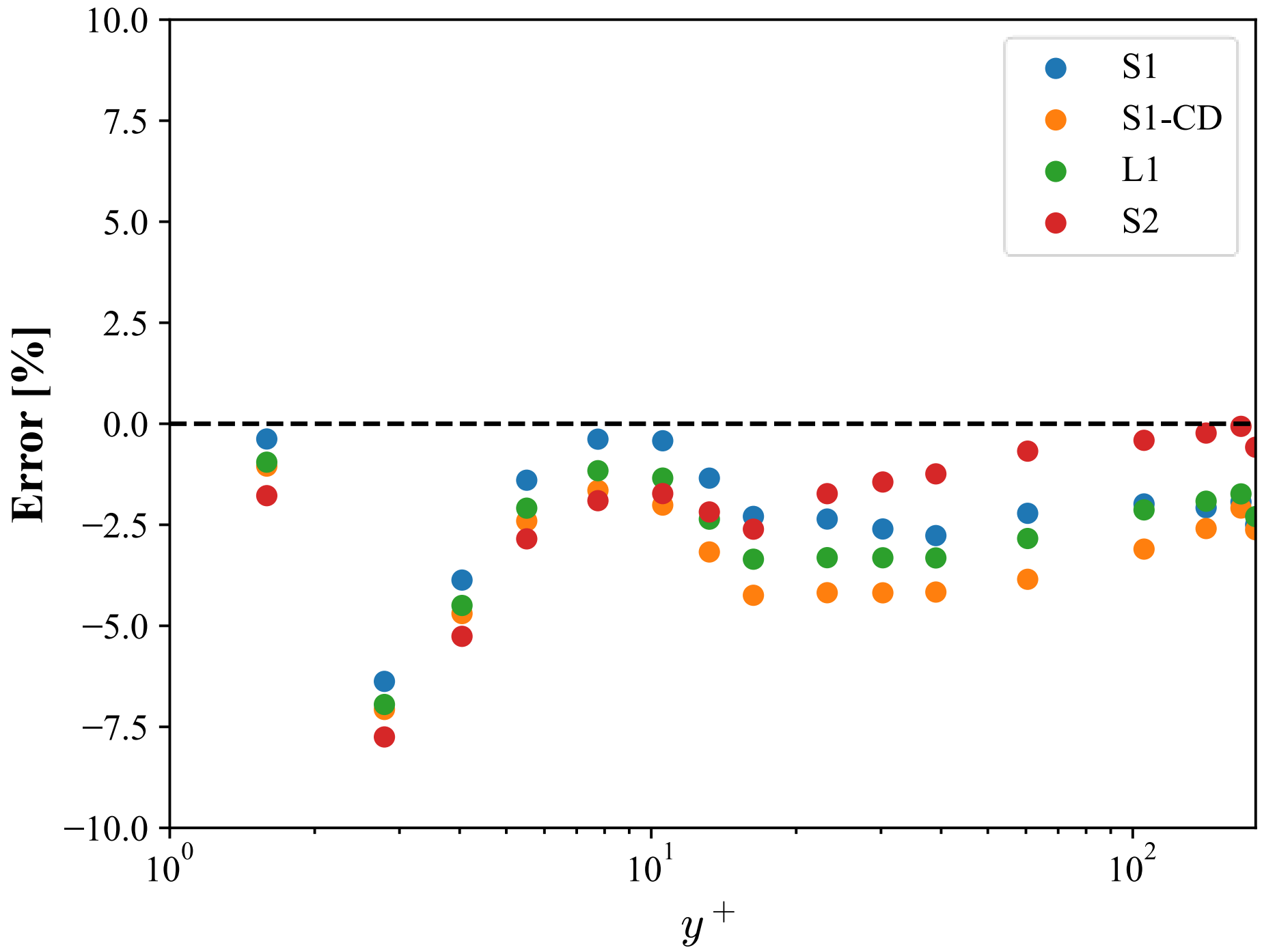

*Figure 6: Error in velocity profile compared to Abe et al.*

*Table 5: Error statistics for mean velocity profiles in comparison with the data from Abe et al..*

| Model | RMS Error | Max Error |
|---|---|---|
| S1-BCD | 2.6% | 6.4% |
| S1-CD | 3.6% | 7.1% |
| L1-BCD | 3.1% | 7.0% |
| S2-BCD | 2.8% | 7.8% |

Notably, the largest errors occur within the viscous sublayer, and the finest structured mesh (S2) exhibited the largest error. This indicates that at TA duration of $t_{out} = 50$ could have been improved upon. The similarity between the profiles produced by grids S1 and L1 suggest that the smaller domain is sufficient, though such a coarse metric is not strong evidence for sufficiency [28]. Velocity RMS fluctuation profiles are more similar between the models and match validation data well (Figure 7). While S2 produces marginally higher peaks of $u'$ and $v'$ (possibly related to TA durationTable 5: Error statistics for mean velocity profiles in comparison with the data from Abe et al..), the differences are near the range of variability seen in the other models.

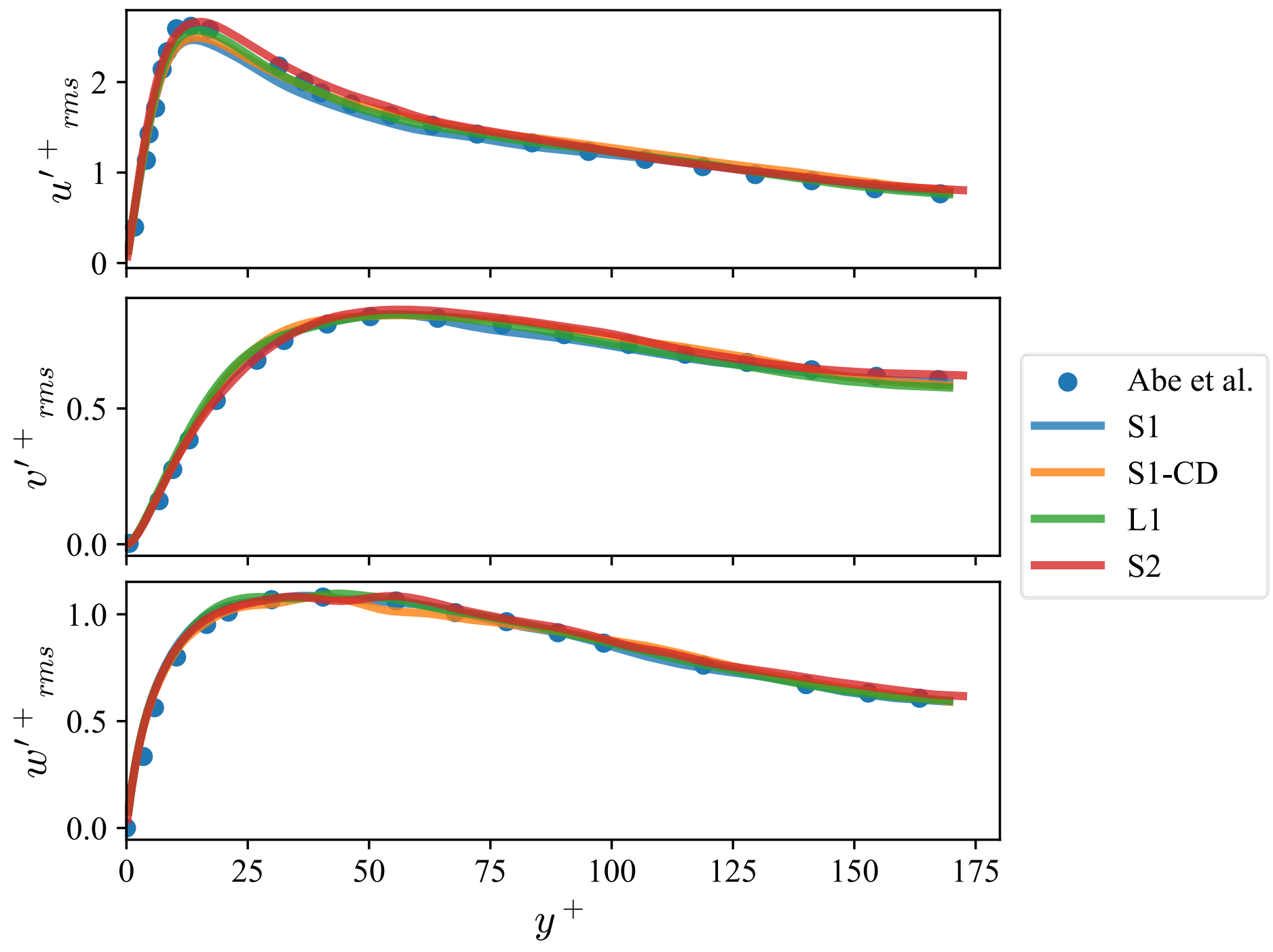

Figure 7: Turbulent channel RMS velocity fluctuation *in comparison with the data from Abe et al.*.

Energy Spectra at $y^+ = 5.4$ are now compared with Abe et al.. Figure 8 shows that coarser structured grid S1 has notable deficiencies in the energy spectra, particularly in the streamwise direction. Along that axis, all three components are overpredicted at wavenumbers greater than 0.1 radians, with pile-up starting at around 0.2 radians. A significant issue is the underprediction of energy at less than 0.1 radians. This difference is nearly an order of magnitude at the smallest wavenumbers resolved. However, Abe et al.'s data exhibit a noticeable slope at the lowest wavenumbers, suggesting their domain was not large enough to resolve the largest turbulent motions along that axis. Since our domain has the same proportions as theirs, it is unclear why our data do not show the same issue.

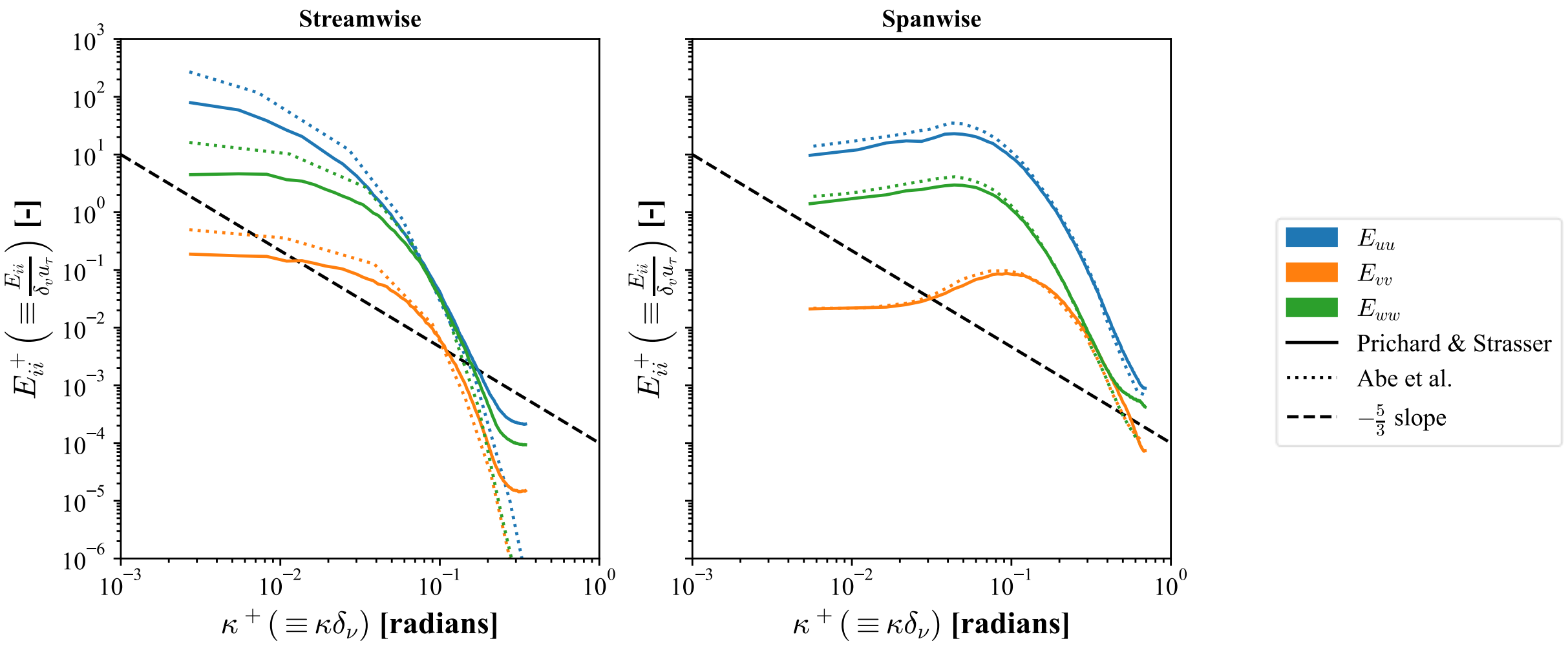

*Figure 8: Turbulent channel energy spectra of grid S1 at $y^+ = 5.4$ in comparison with the data from Abe et al..*

Our spanwise profiles show much closer agreement to Abe et al., albeit still with a slight underprediction of low-wavenumber energy and over-prediction of high-wavenumber energy. However, there is significant error in *ww* at the highest wavenumbers. In contrast to the pile-up seen in the streamwise direction, this error has a somewhat sigmoidal shape. While pile-up was found in the aforementioned LES study [5], direct pile-up comparisons between this study and that would require full similitude between mesh shape and size distribution, time-step size, and discretization schemes.

Moving to pure central differencing (boundedness = 0) did not substantially affect the energy spectra at this wall distance, as is depicted in Figure 9. While it is difficult to draw conclusions from such minor differences, it appears that CD produced slightly higher energy content. Likewise, doubling the domain size (with BCD again) along the streamwise and spanwise directions in grid L1 similarly had little effect (in comparison with Figure 8) as is shown in Figure 10.

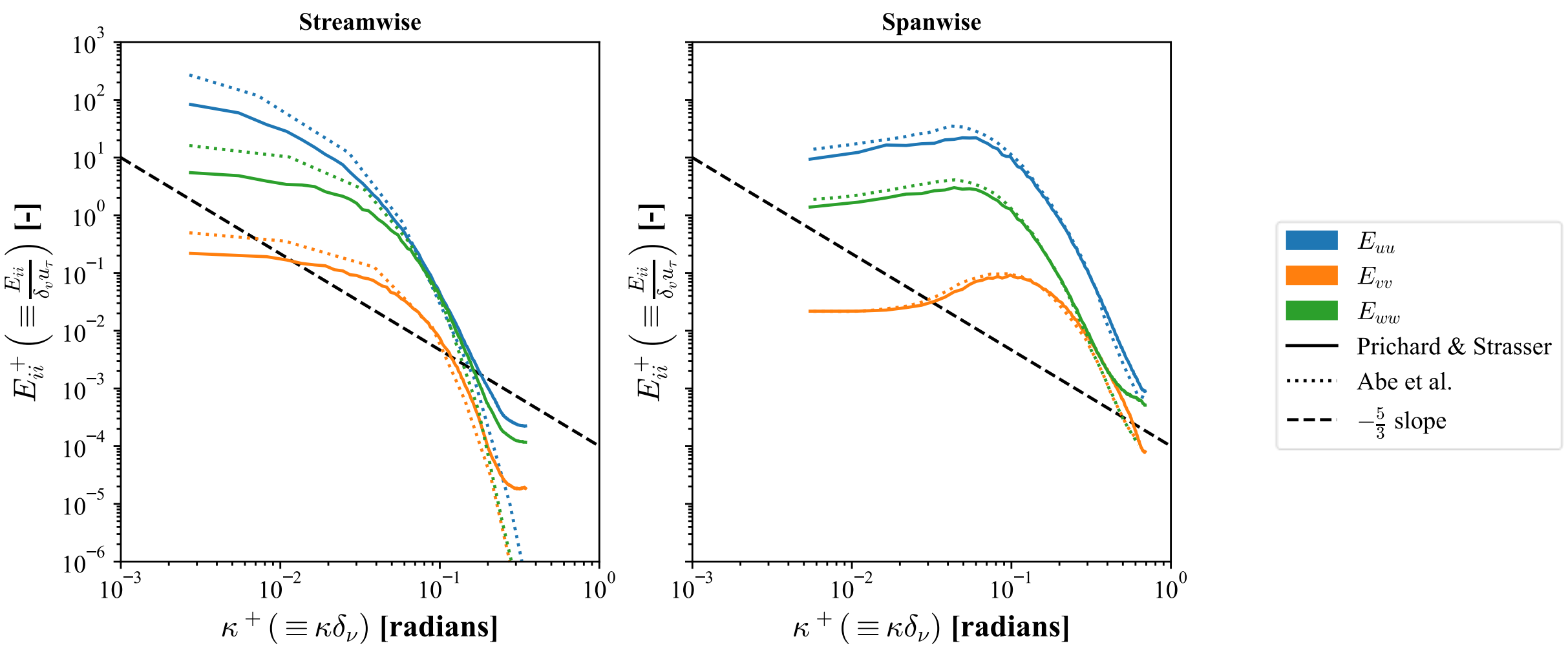

*Figure 9: Turbulent channel energy spectra of grid S1 using pure CD in comparison with the data from Abe et al..*

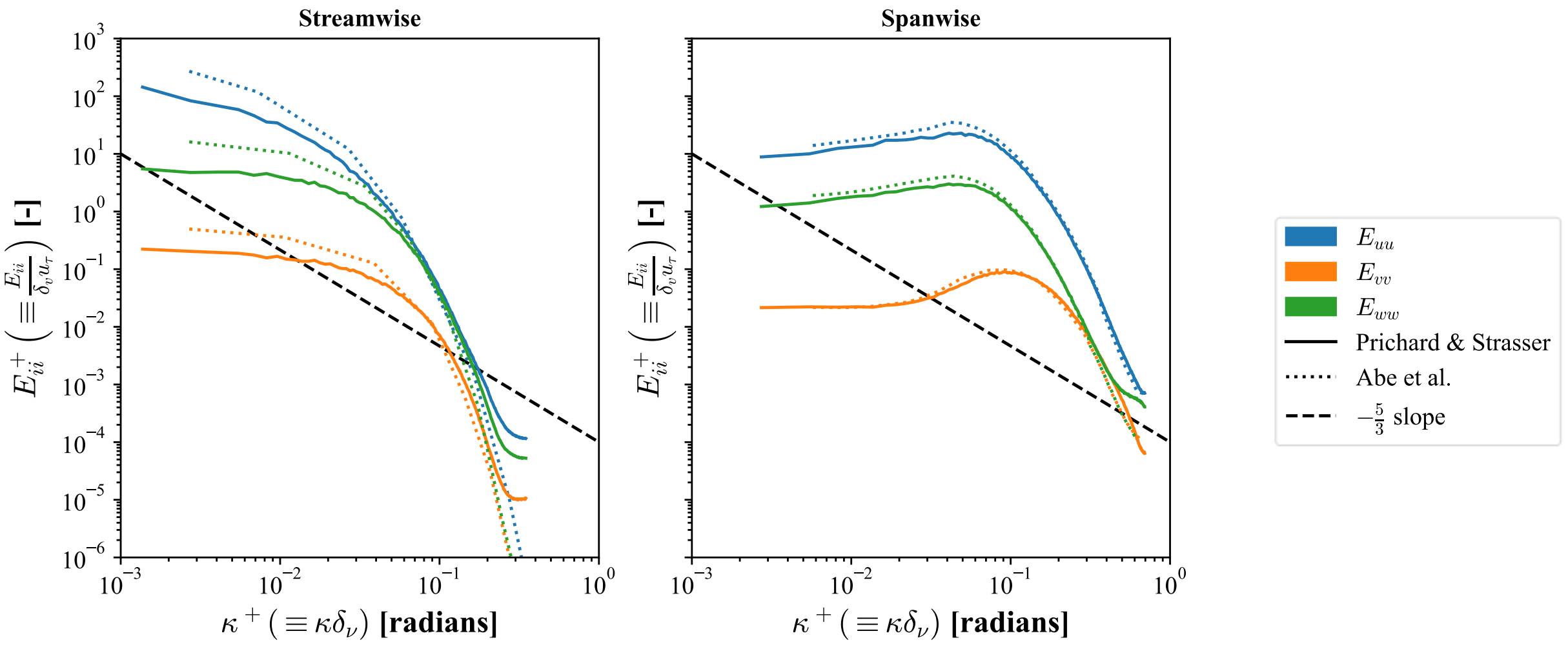

*Figure 10: Turbulent channel energy spectra of grid L1 at $y^+ = 5.4$ in comparison with the data from Abe et al..*

As with other results presented thus far, a more noticeable difference was seen by grid S2 (shown in Figure 11). In the streamwise direction, low-wavenumber energy content matches Abe et al., and pile-up has been extended to higher wavenumbers. However, high-wavenumber energy content is substantially higher than shown in Abe at wavenumbers over $5 \times 10^{-2}$.

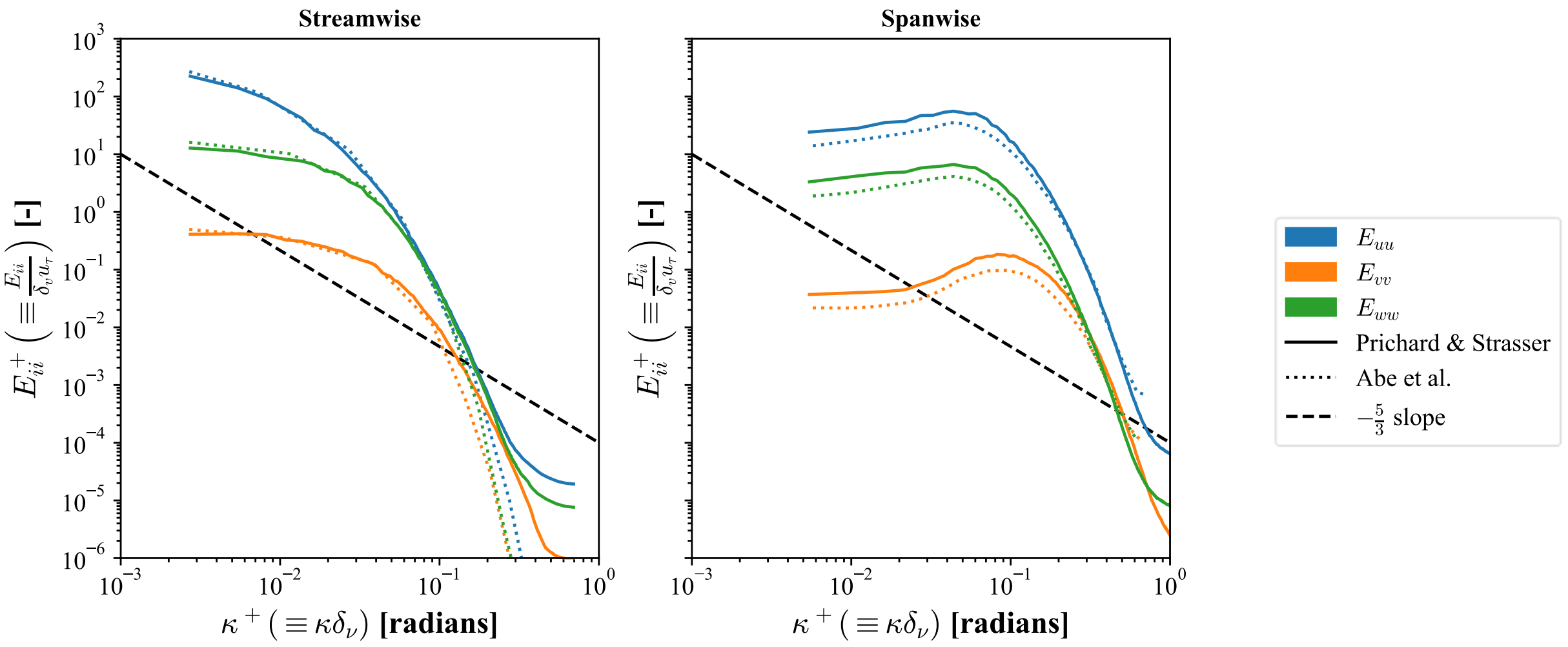

*Figure 11: Turbulent channel energy spectra of grid S2 at $y^+ = 5.4$ in comparison with the data from Abe et al..*

Energy content is similarly elevated in the spanwise direction. While grids S1 and L1 slightly underproduced energy at $\kappa^+ < 0.2$, grid S2 overproduces spanwise energy at all but the greatest wavenumbers ($\kappa^+ > 0.4$). This shift comes with a mild steepening of the energy cascade. Lastly, traditional pile-up (in contrast with the white noise-like pile-up previously observed) is seen in all three components as opposed to the deviation that was previously seen in just the *ww* component.

Finally, grid SP is examined in Figure 12. Low-wavenumber energy content is quite accurate – arguably nearly as close as grid S2, which is reasonable given a matching domain size. However, the high-wavenumber content deviates substantially from reference data. Beginning at roughly $\kappa^+ = 0.1$ in the streamwise direction and $\kappa^+ = 0.3$ in the spanwise direction, the energy content devolves into something akin to white noise. This high-wavenumber “white noise” may be an aliasing error produced by linearly interpolating data from an unstructured grid to a uniform line before computing the PSD.

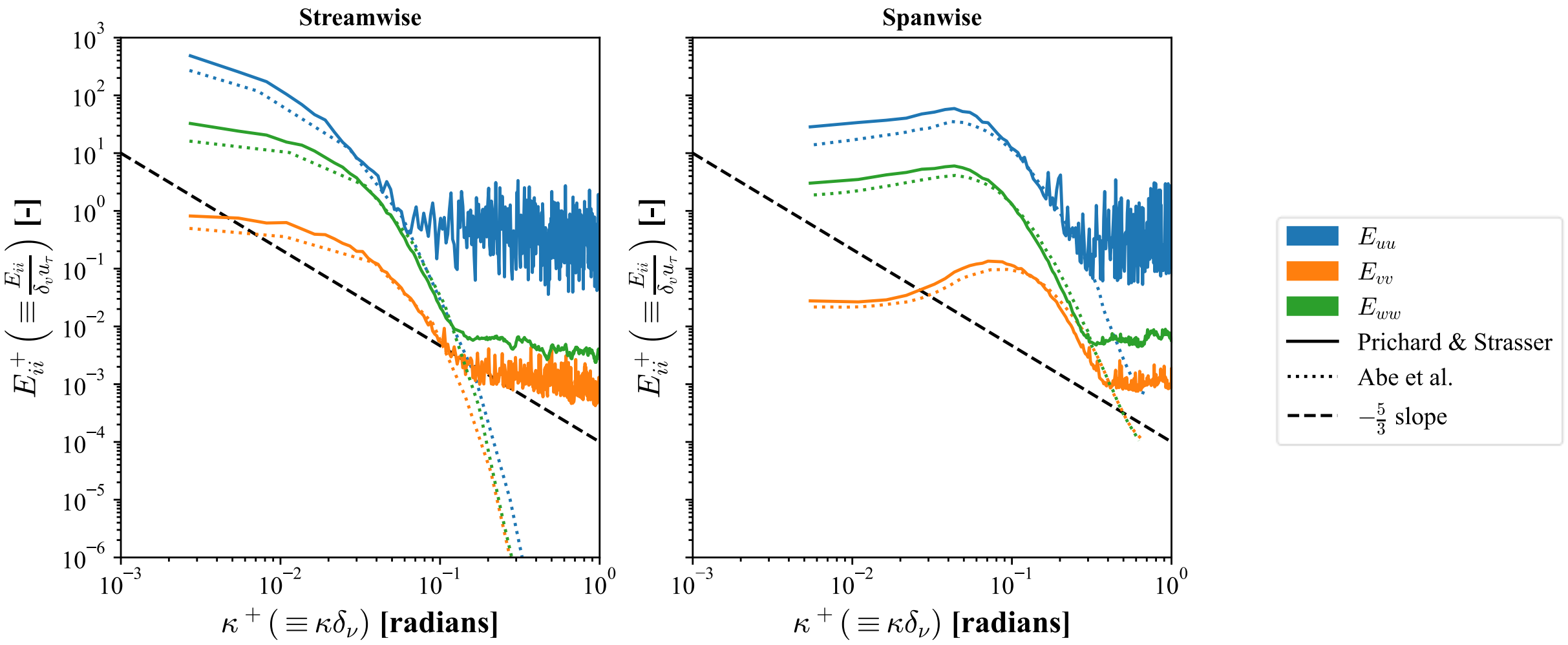

*Figure 12: Turbulent channel energy spectra of grid SP at $y^+ = 5.4$ in comparison with the data from Abe et al..*

Then, to compare our results farther from the wall with those of Vreman & Kuerten, Figure 13 illustrates the spectrum further from the wall at $y^+ = 30$. Compared to Vreman & Kuerten [8], grid S1 with CD underestimates energy at both high and low wavenumbers. In the streamwise direction, there is a small window around $\kappa^+ = 0.1$ where energy is greater than validation data. As with earlier plots for $y^+ = 5.4$, we see pile-up in just the $ww$ component.

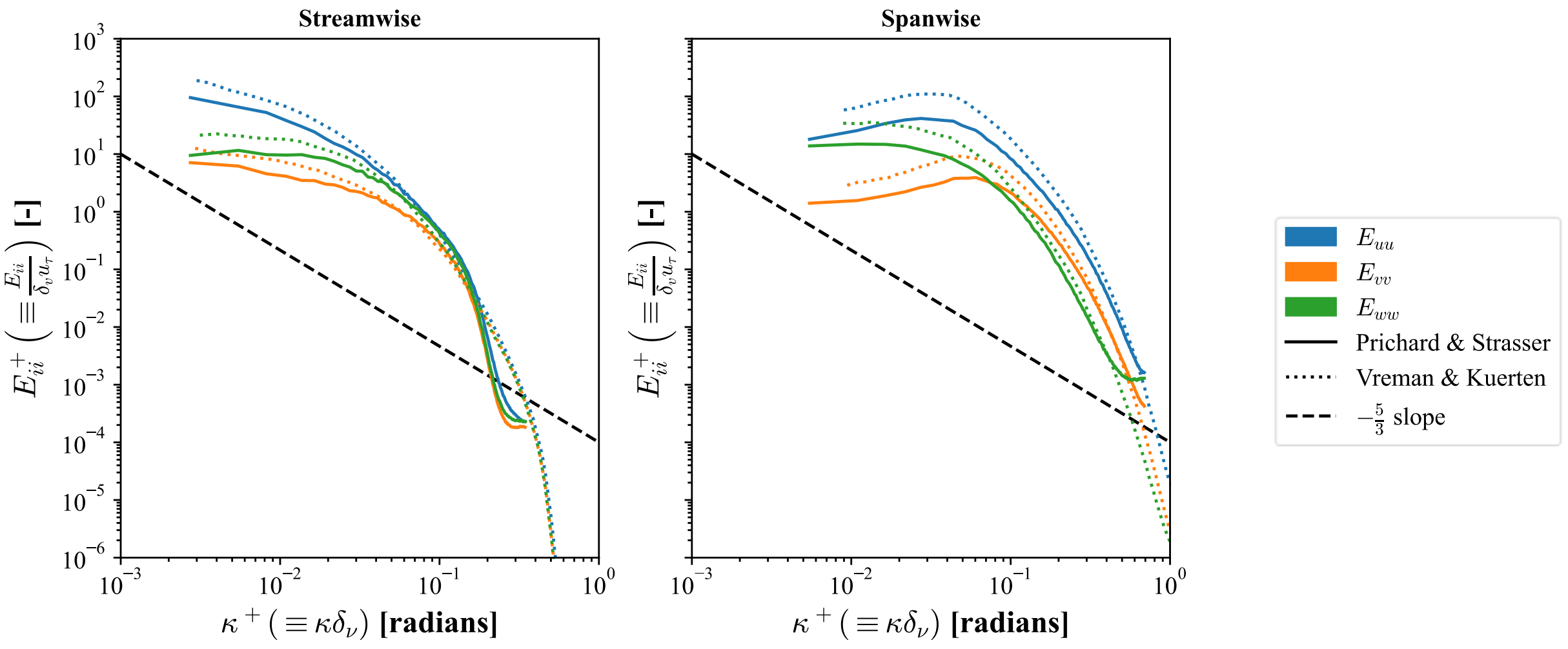

*Figure 13: Turbulent channel energy spectra at $y^{+} = 30$ for grid S1 in comparison with the data from Vreman & Kuerten.*

Figure 14 shows that there is no improvement when moving to the larger domain, which is grid L1. However, grid S2 greatly improves agreement with Vreman & Kuerten as illustrated in Figure 15. Spanwise spectra are a near-perfect match at all but the highest wavenumbers ($\kappa^{+} \geq 0.6$), where pile-up remains apparent. Streamwise spectra also agree well, but energy is mildly overproduced up to $\kappa^{+} = 0.2$. From this point, it dips below the validation curve until pile-up begins at $\kappa^{+} = 0.4$.

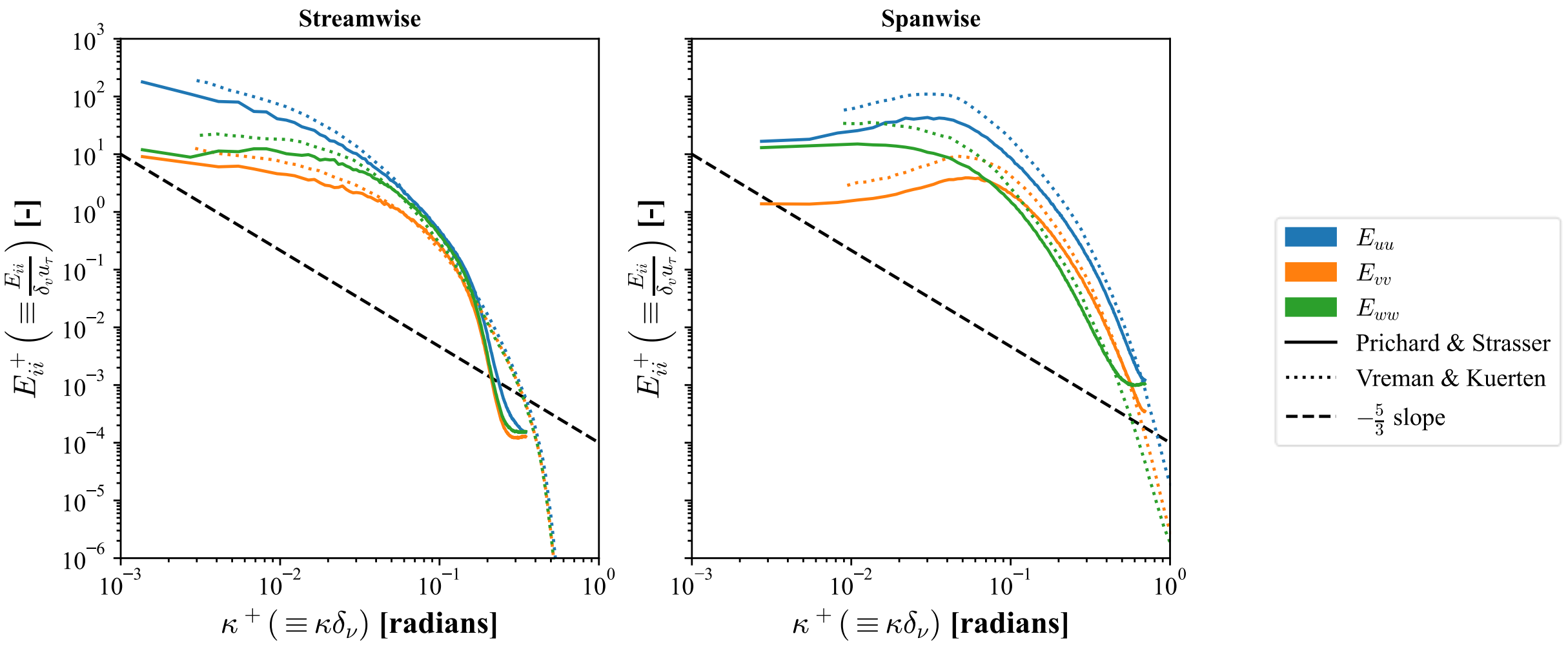

*Figure 14: Turbulent channel energy spectra at $y^+ = 30$ for grid L1 in comparison with the data from Vreman & Kuerten.*

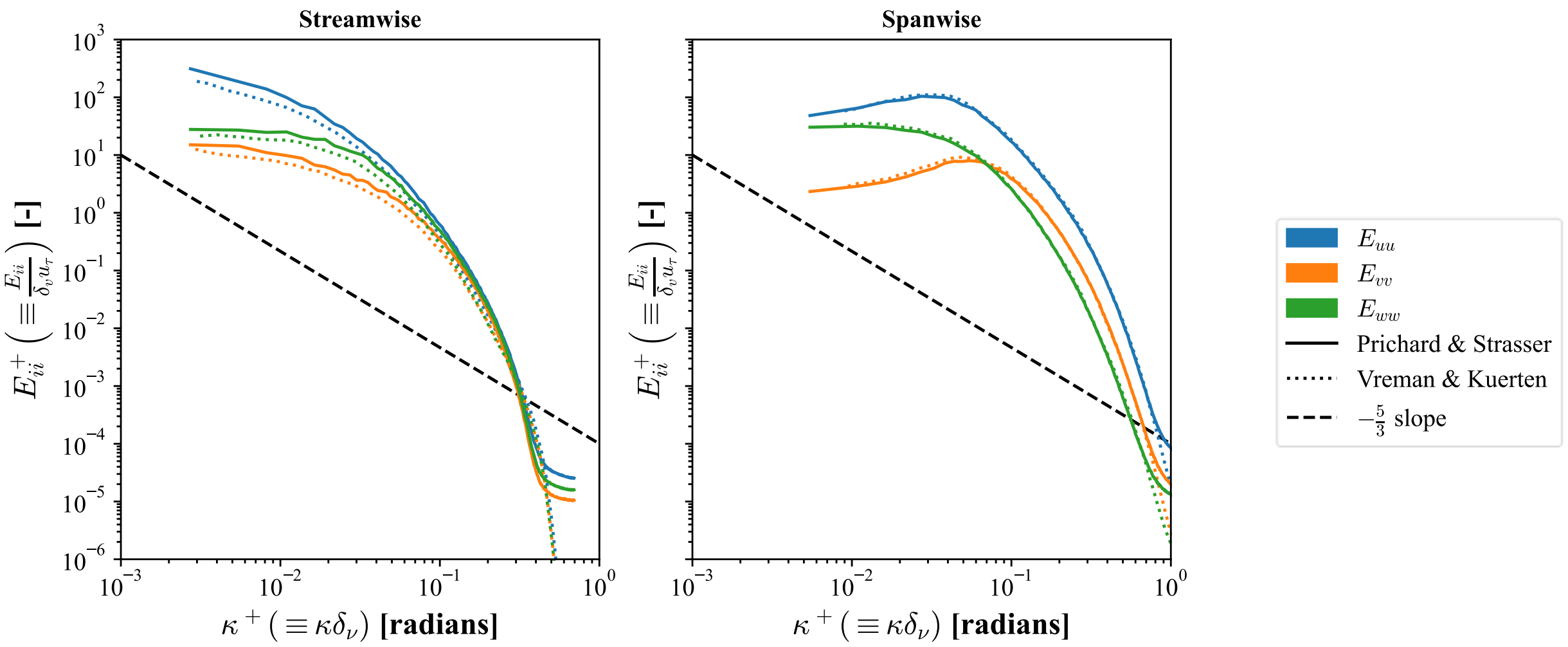

*Figure 15: Turbulent channel energy spectra at $y^+ = 30$ for grid S2 in comparison with the data from Vreman & Kuerten.*

Similar to the prior results nearer the wall, grid SP shows reasonable spectra up to a wavenumber of ~0.1, and the spectra above 0.1 are white noise, in Figure 16. These findings are consistent with those closer to the wall, instructing that finer mesh resolution is the dominant player in matching spectra. Regardless, however, pile-up seems to be unavoidable within the bounds of our evaluations.

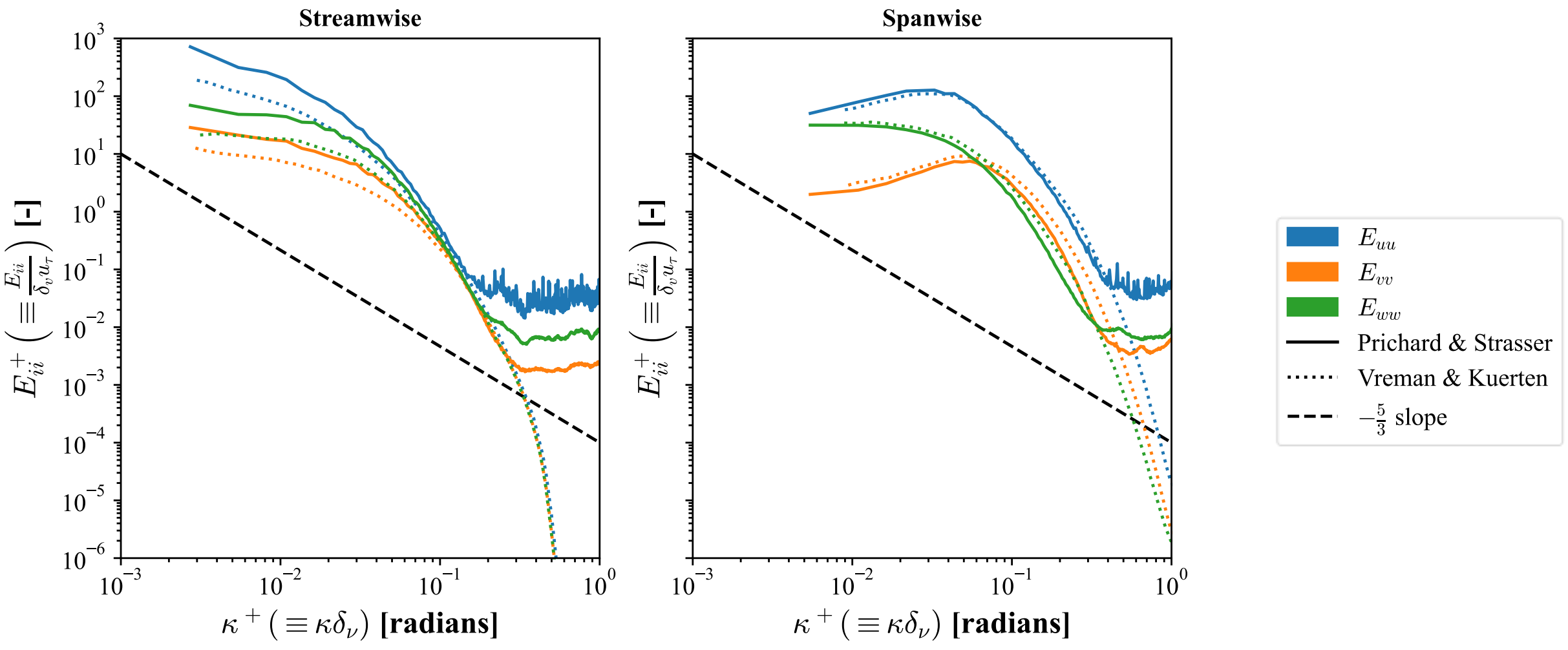

*Figure 16: Turbulent channel energy spectra at $y^+ = 30$ for grid SP in comparison with the data from Vreman & Kuerten.*

Finally, our DNS turbulence kinetic energy (TKE) budgets will be compared with those of Abe et al.. Relatively fewer differences are evident between the different grids' TKE budgets. Representative figures will be provided in this section, but detailed Reynolds stress data are provided in Appendix C. We found statistical variability to be greater in these measures. Despite the prior inaccuracies in the energy spectra, the DNS (solid lines) TKE budgets match Abe et al.'s data (dashed lines) reasonably well. Generally, peaks are somewhat underestimated, with the largest differences seen in production and dissipation. The largest proportional error is seen at the centerline, where production and dissipation are underestimated by around 50%. Large discrepancies are also seen for the velocity-pressure-gradient correlation, but we found this measure to be highly sensitive to statistical variation; it is likely increasing either export frequency or integration time is necessary to produce a precise profile of this value. Grid S1 produces similar results with both BCD (in Figure 17) and pure CD (in Figure 18). Pure CD produced a more accurate dissipation curve, while the other components of the budget were closer to Abe et al.'s data with BCD (Figure 17).

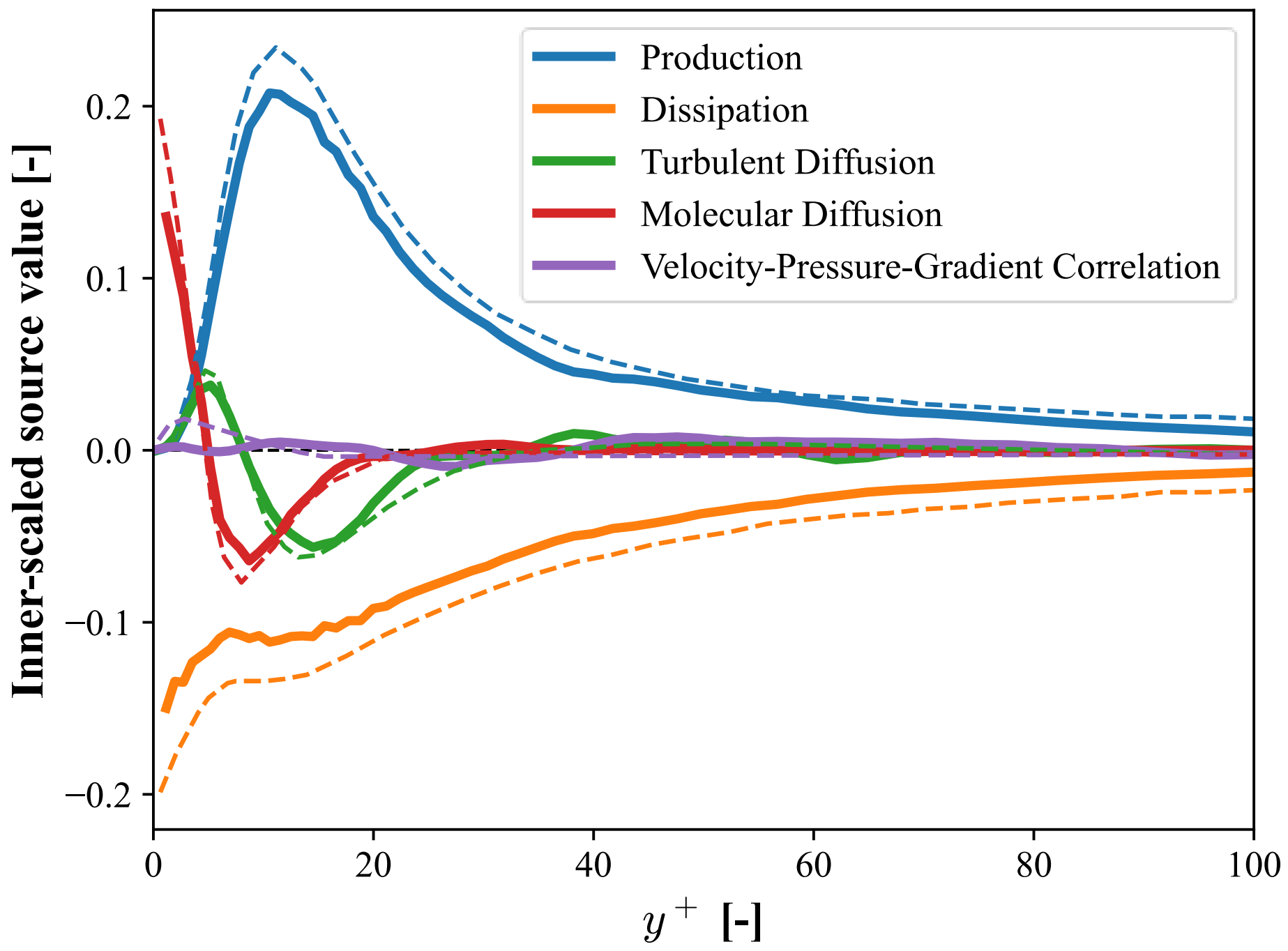

*Figure 17: Turbulent channel turbulence kinetic energy budget for grid S1 with BCD in comparison with the data from Abe et al. (dashed lines).*

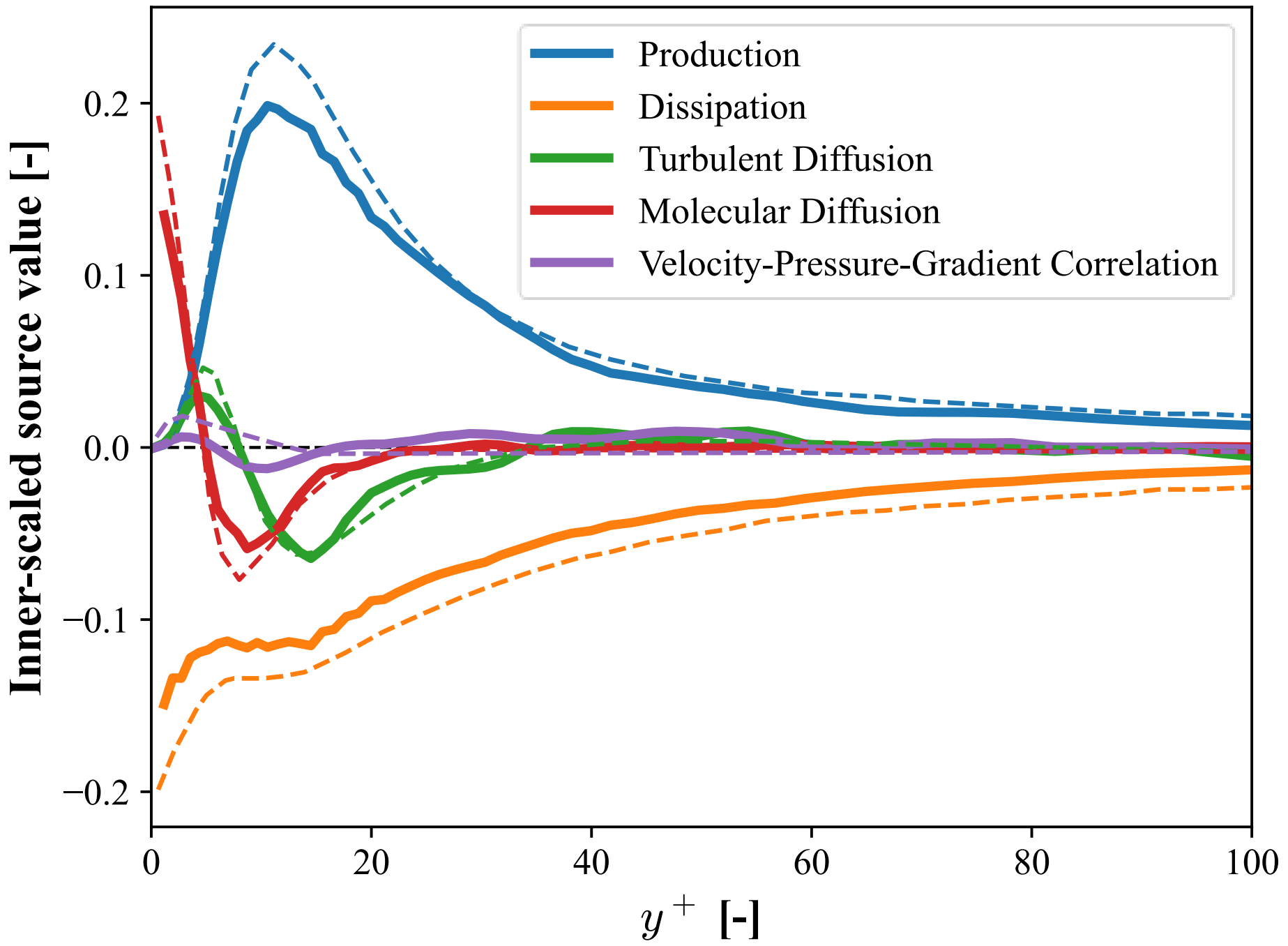

a

*Figure 18: Turbulent channel turbulence kinetic energy budget for grid S1 with pure CD in comparison with the data from Abe et al. (dashed lines).*

In contrast to the other metrics considered thus far, Figure 19 illustrates that increasing the domain size produced a meaningful difference in the TKE budgets. Whereas peak production was underpredicted by ~10% on grid S1, grid L1 produced a production curve matching Abe et al.'s data. This suggests that Abe et al.'s domain was too small, suppressing production of the largest energy-carrying length scales.

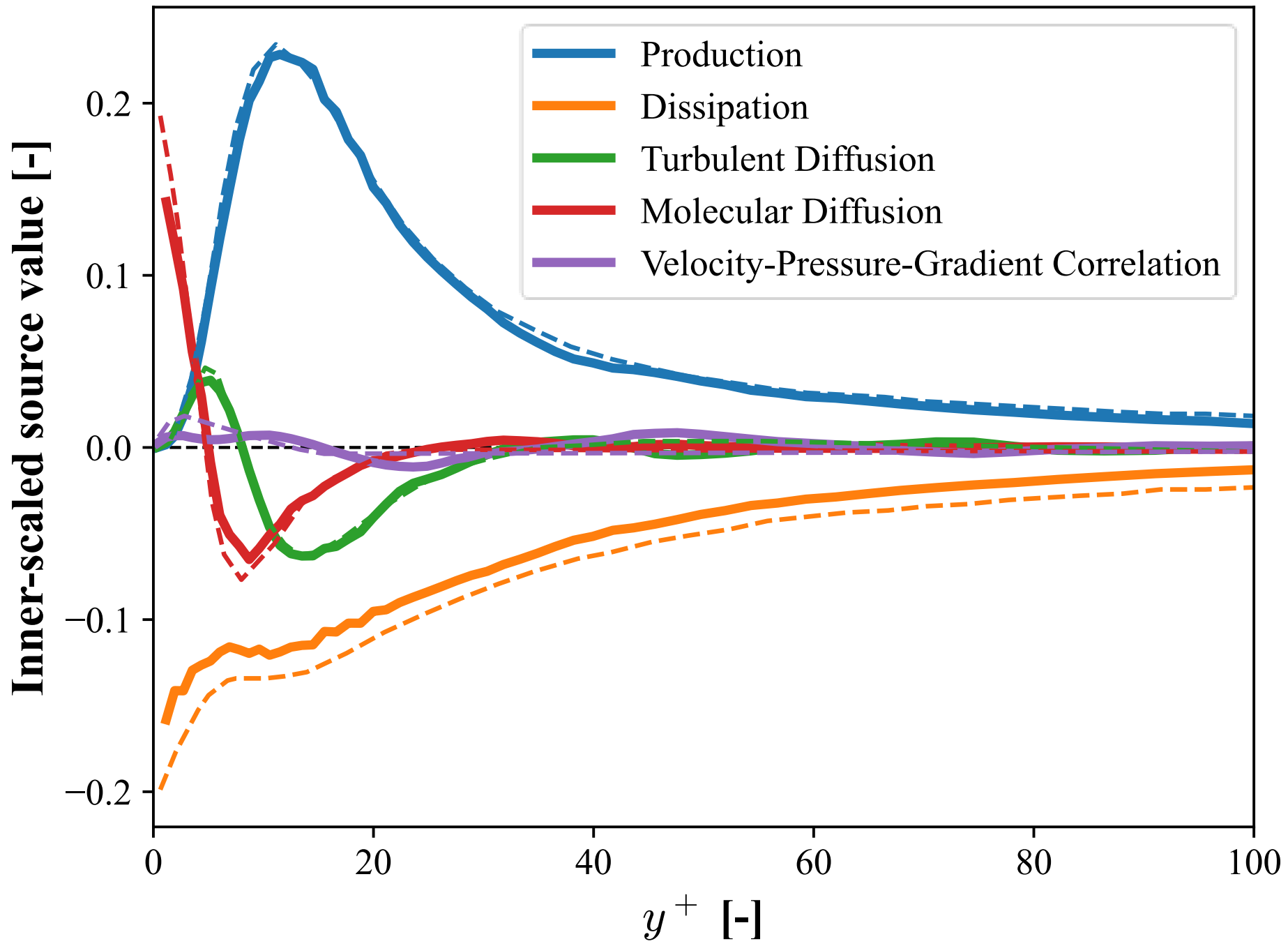

*Figure 19: Turbulent channel turbulence kinetic energy budget for grid L1 in comparison with the data from Abe et al. (dashed lines).*

Except for its production term, grid S2 produced the most accurate TKE budget as depicted in (in Figure 20). Turbulent diffusion and molecular diffusion closely match Abe et al.'s profiles, and dissipation is the closest of all grids tested. Nonetheless, dissipation remains underpredicted throughout. It is possible that, despite its increased cell count compared to Abe et al.'s grid, the physics remain under-resolved. This would also explain the consistent pile-up observed in previously presented spectra. The velocity-pressure-gradient correlation poorly matches Abe et al.'s data in all cases. Again, we found this metric to be very sensitive to statistical variation. CFD data, due to a Fluent limitation, were truncated to 6 significant digits, which may have caused noise to overwhelm the signal here.

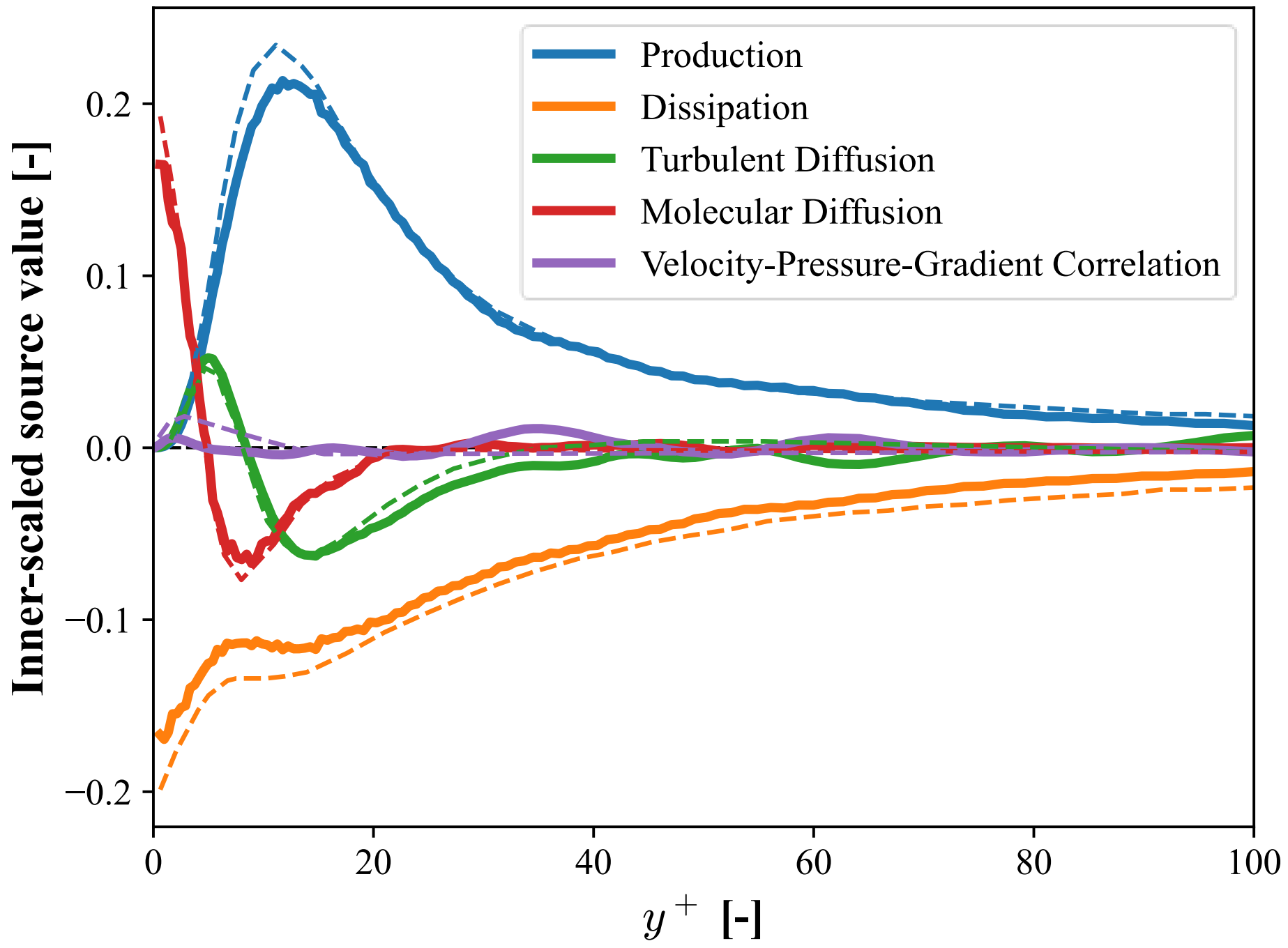

*Figure 20: Turbulent channel turbulence kinetic energy budget for grid S2 in comparison with the data from Abe et al. (dashed lines).*

# Computational Performance

The purpose of this study was not to benchmark Fluent's GPU solver, but the performance advantage of GPU-accelerated processing is worth noting. On grid S1, solver speed was compared between two hardware configurations: a machine equipped with one Nvidia A100 80GB versus a machine with 120 Milan-X cores and 448GB of RAM.

In the GPU case, we used a single CPU core (the CPU does not participate in computations) and the *-gpu_async* command line flag. In the CPU case, we used 118 out of 120 cores; for an explanation on why we did not use all 120, see [29,30]. Both cases completed a 50-time step benchmark with 20 iterations per time step. It took the CPU solver 465s to complete the benchmark. The GPU solver took just 88.8s. Note that scalability of the GPU solver seems to be quite poor and unpredictable, so, at greater scale, it is feasible that a CPU-based system could be the better option.

## Conclusions

Direct numerical simulation of even moderately complex geometry has always been a challenging task, requiring either extensive blocking or the use of an unstructured grid. In some cases, such as a deforming mesh, a structured grid is entirely ruled out, forcing the use of an unstructured grid and corresponding low-order methods. Fluent's GPU-accelerated solver promises to alleviate the increased computational cost accompanying such methods. To date, no studies have rigorously assessed the suitability of an industrial VFM solver (Fluent) and its methods for performing DNS, nor its underlying numerical stencils. This means that any claim Fluent has on resolving DNS-level flow should be approached with caution. We aim to deliver on this particular need and do not claim to replace or supplant any of the well-known CFD or meshing workshops/databases used for validation exercises.

We have shown that Fluent is capable of producing DNS results for a straightforward flow scenario with moderate error. Mean velocity profiles show moderate error near the wall, at up to 8%, while the most refined mesh quickly drops to less than 1% error at $y^+ \sim 50$. Energy spectra are well resolved in lower and moderate wavenumbers, but significant pileup was seen even with a grid of $512 \times 128 \times 512$ cells ($x \times y \times z$). Even though TKE profiles matched validation data, TKE budgets showed error that many would generally be considered unacceptable for DNS validation. Our comparison of momentum schemes found BCD to be the least diffusive option for DNS of an arbitrary grid. We found little difference between BCD with boundedness of 0.25 compared to boundedness of 0; this suggests that the convection boundedness criterion is rarely violated due to the small mesh length scale. Furthermore, our results showed that polyhedral grids – while inferior to hexahedral – produce much less numerical diffusion than tetrahedral grids, which suggests they may be the best option for DNS of complex geometries. Consideration must be given to the size of the domain to ensure the largest turbulent scales are captured. Additionally, the relatively lower-order methods are more stringent in their grid requirements. With BCD, it seems that a grid spacing of $1\eta - 2\eta$, at maximum, is required to produce correct energy spectra, though the coarser grids produced reasonable TKE budgets.

It has also been shown that GPU-based solution methods can provide a key efficiency increase which make DNS feasible. In the case of Fluent, the solver lacked reliability at this stage even while producing significant speed enhancement over its sister CPU-based solver. Performance tips were also provided.

Indications are that $t^+ = 0.08$ for a time step size was overkill relative to $t^+ = 0.32$, unless small differences in high wavenumber spectra matter. In fact, the higher of the two time step size models had stress budgets closer to the experiment. It seems wiser to “spend” that $4x$ difference in run time on flow time-averaging duration or increased cell count. It is likely that any chance of reaching resolved "DNS" would require a multiple of our highest grid resolution (S2), because it still was insufficient.

We conclude that resolved DNS in a commercial FV code is still unattainable without unreasonable computational effort. Moving forward, we plan to extend this validation to more complex physics and ultimately fluid-structure interactions (FSI) dealing with shark denticles [31]. However, without a significant increase in computation power to support the additional mesh elements necessary for sufficient resolution, the authors expect that more complex engineering problems, such as shock-wave boundary layer interaction, separation control, or FSI remain computationally impractical with current industrial FVM solvers.

**ACKNOWLEDGEMENTS**

This work was supported by BWXT Advanced Technologies LLC. Thanks also to Kent Gingerich of F.L.U.I.D. (https://fluidgroup.org/) for his editorial support.

# References

[1] Oswatitsch, K., and Wieghardt, K., 1987, "Ludwig Prandtl and His Kaiser-Wilhelm-Institut," Annual Review of Fluid Mechanics, **19**(Volume 19, 1987), pp. 1–26. https://doi.org/10.1146/annurev.fl.19.010187.000245.

[2] Kim, J., and Leonard, A., 2024, "The Early Days and Rise of Turbulence Simulation," Annu. Rev. Fluid Mech., **56**(1), pp. 21–44. https://doi.org/10.1146/annurev-fluid-120821-025116.

[3] Deardorff, J. W., 1970, "A Numerical Study of Three-Dimensional Turbulent Channel Flow at Large Reynolds Numbers," J. Fluid Mech., **41**(2), pp. 453–480. https://doi.org/10.1017/S0022112070000691.

[4] Moin, P., "LARGE EDDY SIMULATION OF INCOMPRESSIBLE TURBULENT CHANNEL FLOW."

[5] Li, W., and Giometto, M. G., 2025, “Analysis of ANSYS FLUENT for Wall-Modeled Large-Eddy Simulation of Turbulent Channel Flow,” Journal of Fluids Engineering, **147**(1), p. 011503. https://doi.org/10.1115/1.4066485.
[6] Kim, J., Moin, P., and Moser, R., 1987, “Turbulence Statistics in Fully Developed Channel Flow at Low Reynolds Number,” J. Fluid Mech., **177**, pp. 133–166. https://doi.org/10.1017/S0022112087000892.
[7] Abe, H., Kawamura, H., and Matsuo, Y., 2001, “Direct Numerical Simulation of a Fully Developed Turbulent Channel Flow With Respect to the Reynolds Number Dependence,” Journal of Fluids Engineering, **123**(2), pp. 382–393. https://doi.org/10.1115/1.1366680.
[8] Vreman, A. W., and Kuerten, J. G. M., 2014, “Comparison of Direct Numerical Simulation Databases of Turbulent Channel Flow at *Re* τ = 180,” Physics of Fluids, **26**(1), p. 015102. https://doi.org/10.1063/1.4861064.
[9] Menter, F. R., Kolmogorov, D. K., Garbaruk, A. V., and Stabnikov, A. S., 2023, “Direct- and Large Eddy Simulations of Turbulent Flow in CS0 Diffuser on Resolved and Under-Resolved Meshes,” Flow Turbulence Combust, **110**(3), pp. 515–546. https://doi.org/10.1007/s10494-023-00399-1.
[10] Jin, Y., Uth, M. F., and Herwig, H., 2015, “Structure of a Turbulent Flow through Plane Channels with Smooth and Rough Walls: An Analysis Based on High Resolution DNS Results,” Computers & Fluids, **107**, pp. 77–88. https://doi.org/10.1016/j.compfluid.2014.10.012.
[11] De Azevedo, V. W. F., Denner, F., Evrard, F., and Paladino, E. E., 2021, “Performance Evaluation of Standard Second-Order Finite Volume Method for DNS Solution of Turbulent Channel Flow,” J Braz. Soc. Mech. Sci. Eng., **43**(11), p. 513. https://doi.org/10.1007/s40430-021-03234-8.
[12] Komen, E., Shams, A., Camilo, L., and Koren, B., 2014, “Quasi-DNS Capabilities of OpenFOAM for Different Mesh Types,” Computers & Fluids, **96**, pp. 87–104. https://doi.org/10.1016/j.compfluid.2014.02.013.
[13] Zheng, E. Z., Rudman, M., Singh, J., and Kuang, S. B., 2019, “Direct Numerical Simulation of Turbulent Non-Newtonian Flow Using OpenFOAM,” Applied Mathematical Modelling, **72**, pp. 50–67. https://doi.org/10.1016/j.apm.2019.03.003.
[14] Antoniadis, A. F., Drikakis, D., Farmakis, P. S., Fu, L., Kokkinakis, I., Nogueira, X., Silva, P. A. S. F., Skote, M., Titarev, V., and Tsoutsanis, P., 2022, “UCNS3D: An Open-Source High-Order Finite-Volume Unstructured CFD Solver,” Computer Physics Communications, **279**, p. 108453. https://doi.org/10.1016/j.cpc.2022.108453.
[15] Capuano, F., Beratlis, N., Zhang, F., Peet, Y., Squires, K., and Balaras, E., 2023, “Cost vs Accuracy: DNS of Turbulent Flow over a Sphere Using Structured Immersed-Boundary, Unstructured Finite-Volume, and Spectral-Element Methods,” European Journal of Mechanics - B/Fluids, **102**, pp. 91–102. https://doi.org/10.1016/j.euromechflu.2023.07.008.
[16] Marocco, L., and Franco, A., 2017, “Direct Numerical Simulation and RANS Comparison of Turbulent Convective Heat Transfer in a Staggered Ribbed Channel With High Blockage,” Journal of Heat Transfer, **139**(2), p. 021701. https://doi.org/10.1115/1.4034774.

[17] Chekifi, T., Boukraa, M., and Aissani, M., 2021, “DNS Using CLSVOF Method of Single Micro-Bubble Breakup and Dynamics in Flow Focusing,” J Vis, **24**(3), pp. 519–530. https://doi.org/10.1007/s12650-020-00715-1.
[18] F. Zarandi, M. A., Pillai, K. M., and Hasan, A. B. M. R., 2022, “Investigating Liquid-fronts during Spontaneous Imbibition of Liquids in Industrial Wicks. Part II : Validation by DNS,” AIChE Journal, **68**(11), p. e17807. https://doi.org/10.1002/aic.17807.
[19] Moin, P., and Kim, J., 1982, “Numerical Investigation of Turbulent Channel Flow,” J. Fluid Mech., **118**(1), p. 341. https://doi.org/10.1017/S0022112082001116.
[20] Yeung, P. K., and Pope, S. B., 1989, “Lagrangian Statistics from Direct Numerical Simulations of Isotropic Turbulence,” J. Fluid Mech., **207**, pp. 531–586. https://doi.org/10.1017/S0022112089002697.
[21] Smyth, W. D., 1999, “Dissipation-Range Geometry and Scalar Spectra in Sheared Stratified Turbulence,” J. Fluid Mech., **401**, pp. 209–242. https://doi.org/10.1017/S0022112099006734.
[22] Pirozzoli, S., and Orlandi, P., 2021, “Natural Grid Stretching for DNS of Wall-Bounded Flows,” Journal of Computational Physics, **439**, p. 110408. https://doi.org/10.1016/j.jcp.2021.110408.
[23] Mansour, N. N., Kim, J., and Moin, P., 1988, “Reynolds-Stress and Dissipation-Rate Budgets in a Turbulent Channel Flow,” Journal of Fluid Mechanics, **194**, pp. 15–44. https://doi.org/10.1017/S0022112088002885.
[24] 2023, *ANSYS Fluent Theory Guide*, Ansys, Inc.
[25] Rhie, C. M., and Chow, W. L., 1983, “Numerical Study of the Turbulent Flow Past an Airfoil with Trailing Edge Separation,” AIAA Journal, **21**(11), pp. 1525–1532. https://doi.org/10.2514/3.8284.
[26] Syrakos, A., Varchanis, S., Dimakopoulos, Y., Goulas, A., and Tsamopoulos, J., 2017, “A Critical Analysis of Some Popular Methods for the Discretisation of the Gradient Operator in Finite Volume Methods,” Physics of Fluids, **29**(12), p. 127103. https://doi.org/10.1063/1.4997682.
[27] Frigo, M., and Johnson, S. G., 2005, “The Design and Implementation of FFTW3,” Proc. IEEE, **93**(2), pp. 216–231. https://doi.org/10.1109/JPROC.2004.840301.
[28] Kacinski, R., Strasser, W., Leonard, S., Prichard, R., and Truxel, B., 2023, “Validation of a Human Upper Airway Computational Fluid Dynamics Model for Turbulent Mixing,” Journal of Fluids Engineering, **145**(12), p. 121203. https://doi.org/10.1115/1.4063061.
[29] Prichard, R., and Strasser, W., 2024, “When Fewer Cores Is Faster: A Parametric Study of Undersubscription in High-Performance Computing,” Cluster Comput. https://doi.org/10.1007/s10586-024-04353-2.
[30] Prichard, R., and Strasser, W., 2024, “An Intuitive Multi-Objective, Multi-Variable High-Performance Computing Optimisation Methodology,” International Journal of Computational Fluid Dynamics, **38**(8–9), pp. 610–616. https://doi.org/10.1080/10618562.2025.2513607.
[31] Prichard, R., Strasser, W., and Kacinski, R., 2025, “A Fundamental Viscous Flow Study to Understand a Separation Control Mechanism,” Journal of Fluids Engineering, **147**.
[32] Jasak, H., Weller, H. G., and Gosman, A. D., 1999, “High Resolution NVD Differencing Scheme for Arbitrarily Unstructured Meshes,” Int. J. Numer. Meth. Fluids, **31**(2), pp.

431–449. https://doi.org/10.1002/(SICI)1097-0363(19990930)31:2%253C431::AID-FLD884%253E3.0.CO;2-T.

[33] Martha, C. S., Blaisdell, G. A., and Lyrintzis, A. S., 2013, "Large Eddy Simulations of 2-D and 3-D Spatially Developing Mixing Layers," Aerospace Science and Technology, **31**(1), pp. 59–72. https://doi.org/10.1016/j.ast.2013.09.007.

[34] Adedoyin, A. A., Walters, D. K., and Bhushan, S., 2015, "EVALUATION OF TURBULENCE MODEL AND NUMERICAL SCHEME COMBINATIONS FOR PRACTICAL FINITE-VOLUME LARGE EDDY SIMULATIONS," Mississippi State University.

[35] Leonard, B. P., 1979, "A Stable and Accurate Convective Modelling Procedure Based on Quadratic Upstream Interpolation," Computer Methods in Applied Mechanics and Engineering, **19**(1), pp. 59–98. https://doi.org/10.1016/0045-7825(79)90034-3.

[36] Padway, E., and Nishikawa, H., 2022, "Resolving Confusion Over Third Order Accuracy of U-MUSCL," AIAA Journal, **60**(3), pp. 1415–1439. https://doi.org/10.2514/1.j060773.

[37] Chu, S., Kurganov, A., Mohammadian, S., and Zheng, Z., 2023, "Fifth-Order A-WENO Path-Conservative Central-Upwind Scheme for Behavioral Non-Equilibrium Traffic Models," CiCP, **33**(3), pp. 692–732. https://doi.org/10.4208/cicp.OA-2022-0263.

[38] Lu, C. L., Li, Q. S., Huang, S. H., Chen, F. B., and Fu, X. Y., 2012, "Large Eddy Simulation of Wind Effects on a Long-Span Complex Roof Structure," Journal of Wind Engineering and Industrial Aerodynamics, **100**(1), pp. 1–18. https://doi.org/10.1016/j.jweia.2011.10.006.

[39] Adedoyin, A. A., Walters, D. K., and Bhushan, S., 2015, "Investigation of Turbulence Model and Numerical Scheme Combinations for Practical Finite-Volume Large Eddy Simulations," Engineering Applications of Computational Fluid Mechanics, **9**(1), pp. 324–342. https://doi.org/10.1080/19942060.2015.1028151.

[40] Sharma, V., and Dutta, S., 2023, "Experimental and Numerical Investigation of Bio-Inspired Riblet for Drag Reduction," Journal of Fluids Engineering, **145**(2), p. 021207. https://doi.org/10.1115/1.4056185.

[41] Sterken, L., Sebben, S., and Löfdahl, L., 2016, "Numerical Implementation of Detached-Eddy Simulation on a Passenger Vehicle and Some Experimental Correlation," Journal of Fluids Engineering, **138**(9), p. 091105. https://doi.org/10.1115/1.4033296.

[42] Walters, D. K., Bhushan, S., Alam, M. F., and Thompson, D. S., 2013, "Investigation of a Dynamic Hybrid RANS/LES Modelling Methodology for Finite-Volume CFD Simulations," Flow Turbulence Combust, **91**(3), pp. 643–667. https://doi.org/10.1007/s10494-013-9481-9.

# Appendix A: Fluent's Numerical Schemes

Fluent makes available six schemes for momentum discretization: Upwind Differencing (UD), Linear Upwind Differencing (LUD), Central Differencing (CD), Bounded Central Differencing (BCD), Quadratic Upstream Interpolation for Convective Kinematics (QUICK), and Monotone Upstream-centered Scheme for Conservation Laws (MUSCL). Most of these schemes are known outside of Fluent, but Fluent's implementation details merit discussion. In this section, we will use the convention described in Figure 21 except where stated otherwise. Furthermore, "scalar" will be used to describe a strict scalar or a vector component.

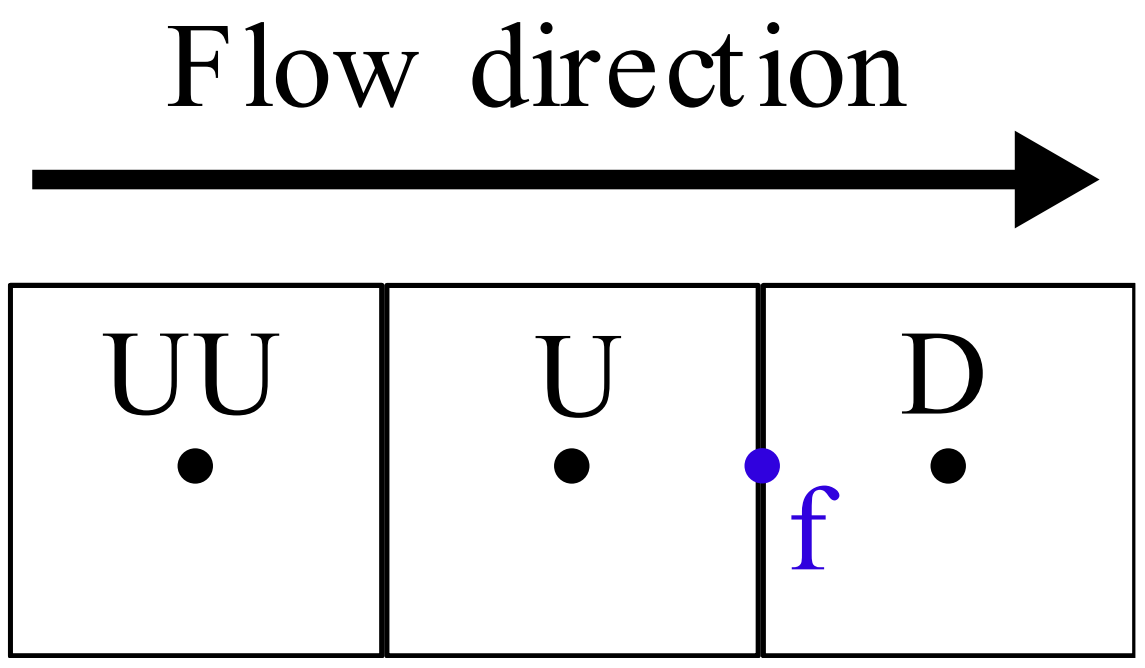

*Figure 21: Labeling convention used in discussion of spatial discretization schemes. When computing a scalar value on face $f$, cells $U$ and $D$ are those immediately upstream and downstream, respectively. For a structured grid, cell $UU$ can be identified as the far upstream cell.*

### *Upwind Differencing (UD)*

UD ("first-order upwind") simply sets a scalar's face value, $\phi_f$, equal to the computed value at the upwind cell center, $\phi_u$, as shown in Equation 28. This is widely known to be highly diffusive, making it a poor choice for DNS. Nonetheless, it will be included in the comparison as a reference point.

$$\phi_{f,UD} = \phi_U \tag{28}$$

### *Linear Upwind Differencing (LUD)*

LUD ("second-order upwind" in Fluent nomenclature) adds a gradient term to UD shown in Equation 29,

$$\phi_{f,LUD} = \phi_U + \psi\,\overrightarrow{\nabla\phi}_U \cdot \left(\vec{x}_f - \vec{x}_U\right) \tag{29}$$

where $\psi$ is a limiter function (by default, one derived from Barth and Jespersen), and $\vec{x}$ represents a position vector ($\vec{x}_f$ being the position of the face, and $\vec{x}_U$ being the position of cell U's center). Despite its name in Fluent, Fluent's LUD implementation is not necessarily second-order. Depending on the gradient reconstruction scheme used and the nature of the grid, the gradient term can be as low as zeroth-order. Syrakos et al. have shown that this error can diminish the order of convergence of the finite volume method as a whole [26]. More discussion on gradient reconstruction and $\psi(r)$ will follow.

### *Central Differencing (CD)*

Fluent's CD scheme uses a deferred correction approach, where face values are calculated with an implicit UD term (as defined in Equation 28) and an explicit correction term provided by Equation 30 (and again Equation 28).

$$\phi_{f,CD}^{dc} = \left(\phi_{f,UD}\right)_{\text{implicit}} + \left(\phi_{f,CD} - \phi_{f,UD}\right)_{\text{explicit}} \tag{30}$$

Where $\phi_{f,CD}$ is the face value computed with a traditional, non-deferred correction approach:

$$\phi_{f,CD} = \frac{1}{2}(\phi_U + \phi_D) + \frac{1}{2}\left[\overrightarrow{\nabla\phi}_U \cdot (\vec{x}_f - \vec{x}_U) + \overrightarrow{\nabla\phi}_D \cdot (\vec{x}_f - \vec{x}_D)\right] \tag{31}$$

This approach is intended to suppress dispersion. If the solution converges, this should reduce to pure central differencing, as the explicit and implicit UD approximations will be identical.

### *Bounded Central Differencing (BCD)*

Ansys does not divulge many details about their BCD scheme, but the Fluent Theory Guide describes it to be a blend of "pure central differencing, a blended scheme of a central differencing and an upwind scheme, and the first-order upwind scheme." A user-tunable "boundedness" constant, which ranges from 0 to 1, controls how BCD switches between these schemes. Ansys states that setting boundedness to 0 reduces BCD to pure central differencing, while setting boundedness to 1 strictly enforces the Convection Boundedness Criterion (CBC). Briefly, the CBC puts restrictions on the normalized cell value of a scalar, $\tilde{\phi}_C$ depicted in Equation 32:

$$\tilde{\phi}_C = \frac{\phi_C - \phi_U}{\phi_D - \phi_U} \tag{32}$$

The CBC requires that $0 \leq \tilde{\phi}_C \leq 1$. Equivalently, this restriction poses that $\phi_C$ must fall in the range $[\phi_U, \phi_D]$. Note that, in this context, $U$ and $D$ do not represent the pair of cells adjacent to a given face; instead, they represent the cells upwind and downwind of cell $C$ as illustrated in Figure 22.

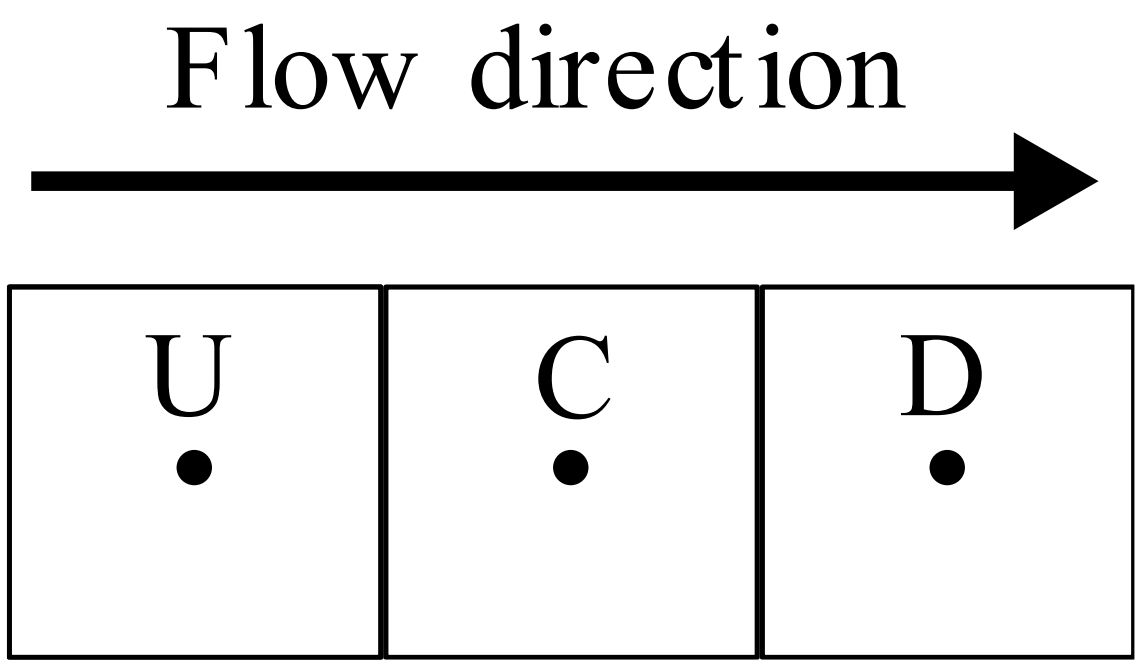

*Figure 22: A depiction of the convention used in the CBC.*

Selecting cells $U$ and $D$ is untenable on an unstructured grid, as flow can be entering and exiting a cell simultaneously through many faces. Therefore, Fluent must be using a modified definition. It is known that the BCD scheme is based on Jasak et al.'s "Gamma" scheme [32]. Jasak et al. present the following formulation for estimating a normalized variable on an arbitrary grid:

$$\tilde{\phi}_C \approx 1 - \frac{\phi_D - \phi_C}{2\overrightarrow{\nabla\phi}_C \cdot \hat{d}} \tag{33}$$

Where $\hat{d}$ is the unit vector pointing from the cell $C$'s centroid to that of its neighbor, $D$. The Gamma scheme enforces the CBC by switching between UD and CD based on the normalized variable at a face's upstream cell, $\tilde{\phi}_U$ (note that here we revert to the convention described in Figure 21) , shown in Equation 34:

$$\tilde{\phi}_U \approx 1 - \frac{\phi_D - \phi_U}{2\overrightarrow{\nabla\phi}_U \cdot \hat{d}} \tag{34}$$

If $\tilde{\phi}_U \leq 0$ or $\tilde{\phi}_U \geq 1$, UD is used, and the Gamma scheme blends between UD and CD from $0 < \tilde{\phi}_C < \beta_m$, where $\beta_m$ is a tunable constant ranging from 0 to 1/2. Pure CD is then used on the range $\beta_m \leq \tilde{\phi}_C < 1$. Blending is linear based on a blending factor, $\gamma$, where $\gamma = 0$ produces pure UD and $\gamma = 1$ produces pure CD. Jasak et al. vary $\gamma$ linearly (Equation 35) with $\tilde{\phi}_C$:

$$\gamma = \frac{\tilde{\phi}_C}{\beta_m} \tag{35}$$

In summation, the Gamma scheme produces the following relation for $\phi_f$ in Equations 36 and 37

$$\phi_f = \begin{cases} \phi_U & \tilde{\phi}_U \leq 0,\ \ \tilde{\phi}_U \geq 1 \\ [1-\gamma(1-f_x)]\phi_U + \gamma(1-f_x)\phi_D & 0 < \tilde{\phi} < \beta_m \\ f_x\phi_U + (1-f_x)\phi_D & \beta_m \leq \tilde{\phi}_U < 1 \end{cases} \tag{36}$$

where

$$f_x = \frac{|\vec{x}_U - \vec{x}_f|}{|\vec{x}_U - \vec{x}_D|} \tag{37}$$

Fluent’s BCD almost certainly has some differences compared to Gamma. An internal Fluent setting, “bcd-tanh-blending”, suggests that Fluent uses a hyperbolic tangent blending function instead of the Gamma scheme’s linear blending. In literature, some claim that BCD incorporates LUD [33,34], but we find no evidence to support or contradict this point. Nothing in the Gamma scheme from the literature matches the described behavior of BCD’s boundedness parameter, suggesting a further modified implementation in Fluent.

### *QUICK*

Quadratic Upwind Interpolation for Convective Kinematics (QUICK), well known in literature since its introduction by Leonard [35], calculates face values as a weighted combination of second-order central and second-order upwind approximations. Fluent's QUICK implementation takes on the generalized form in Equation 38:

$$\begin{aligned} \phi_{f,QUICK} &= \theta\phi_{f,\mathrm{central}} + (1-\theta)\phi_{f,\mathrm{upwind}} \\ &= \theta\frac{\Delta x_D \phi_U + \Delta x_U \phi_D}{\Delta x_D + \Delta x_U} + (1-\theta)\left(\frac{\Delta x_{UU} + 2\Delta x_U}{\Delta x_{UU} + \Delta x_U}\phi_U - \frac{\Delta x_U}{\Delta x_{UU} + \Delta x_U}\phi_{UU}\right) \end{aligned} \tag{38}$$

A constant value of $\theta = 1/8$ produces a traditional form of QUICK, but Fluent varies $\theta$ based on solution values to improve boundedness. Note that unlike Fluent's CD and LUD schemes, its implementation of QUICK does not incorporate reconstructed gradients. QUICK is only applicable to quadrilateral/hexahedral cells, with Fluent falling back to LUD as described in equation 29 on faces belonging to other cell types. For this reason, QUICK will only be considered on fully hexahedral grids. Fluent similarly falls back on LUD on partition boundaries, so all discretization studies were performed using a single core.

### *MUSCL*

Finally, Fluent's Monotone Upwind-centered Scheme for Conservation Laws (MUSCL) implementation is straightforward, consisting of a weighted combination of CD and LUD as in Equation 39

$$\phi_{f,MUSCL} = \frac{1}{3}\phi_{f,CD} + \frac{2}{3}\phi_{f,LUD} \tag{39}$$

In contrast to QUICK, these CD and LUD terms rely on reconstructed gradients. The CD and LUD terms are exactly as described in Equations 31 and 29 with one exception: no gradient limiter is applied to the LUD portion. In theory, this should reduce false diffusion compared to LUD. Ansys claims that this implementation of MUSCL is third-order accurate, but due to the propagation of error from the gradient reconstruction scheme, this claim is taken to be generally false. Nonetheless, it is possible on uniform Cartesian grids that this particular combination eliminates gradient reconstruction's second-order error contribution. It should be noted that the MUSCL formulation used in Fluent does not refer to the classical flux reconstruction proposed by van Leer for use in Riemann-based solution methods on structured grids, but rather to a method for reconstruction of face variable states on general, unstructured grids. On uniform Cartesian grids, the Fluent MUSCL scheme reconstructs face variable states to third order in smooth regions of the flowfield, for which the slope limiter of the LUD portion is equal to one. On more general grids the MUSCL reconstruction is second-order accurate [36].

An important note here is that FVM are not inherently limited to $2^{nd}$ order. High-order FVM are well established in the literature. We would like to specifically call attention to [37] which remains a standard reference for high-order FVM approaches and often achieves the same accuracy as some second-order schemes at lower computational cost. However, beyond $2^{nd}$ order, storage requirements can become prohibitively expense.

# Appendix B: Detailed Comparisons of Fluent's Schemes on Multiple Grids

## Diffusion Study Setup

The literature lacks a comparison of these schemes as implemented in Fluent, so we performed this evaluation by comparing all six schemes in an inviscid shear layer, whose domain and boundary conditions are depicted in Figure 23. Flow is diagonally downward in the $(1, -1)$ direction with a velocity magnitude of 2 m/s on the top boundary and 1 m/s on the left boundary. Flow exits via outlets with a pressure of 0 Pa. These comparisons were performed in a steady, inviscid solver.

Six 2D grids are compared, described in Table 6 and depicted in part in Figure 23. To avoid any differences that may exist between Fluent's 2D and 3D implementations, these 2D grids were extruded one cell in the transverse direction and the comparison performed in Fluent's 3D solver. Two grids were created for each cell type – one with 25 cells along each edge, and one with 100 cells along each edge. In the case of the quadrilateral grids, this produced cell counts of around 650 and 10,000, respectively, while the corresponding triangular grids had twice the cell count. The polygonal grids were formed from the corresponding triangular grids via Fluent's built-in conversion tool. For simplicity of language, we will refer to these prismatic 3D grids by the extruded 2D faces' topology. In all cases, gradient reconstruction was performed using the least-squares cell-based method, which will be described later.

*Table 6: Grids on which Fluent's discretization schemes' effects were studied. Two grid sizes, 25x25 and 100x100, were compared for each cell type.*

| **Mesh** | **Cells (25 x 25)** | **Cells (100 x 100)** |
|---|---|---|
| Triangular (wedge) | 1,250 | 20,000 |
| Hexagonal (polyhedral prism) | 678 | 10,203 |
| Quadrilateral (hexahedral) | 625 | 10,000 |

Data were gathered along two lines, which will be denoted "normal" and "aligned". Each line consists of 140 evenly spaced points, with data at each point interpolated from nearby cell values. Ideally, the profile along the "normal" line would be a square step function increasing from 1 m/s to 2 m/s at its midpoint. Its spatial derivative, which will also be shown, should be the Dirac delta function centered on $x = 0.5$. Finally, the profile along the "aligned" line should be flat, with a constant value of 1.5.

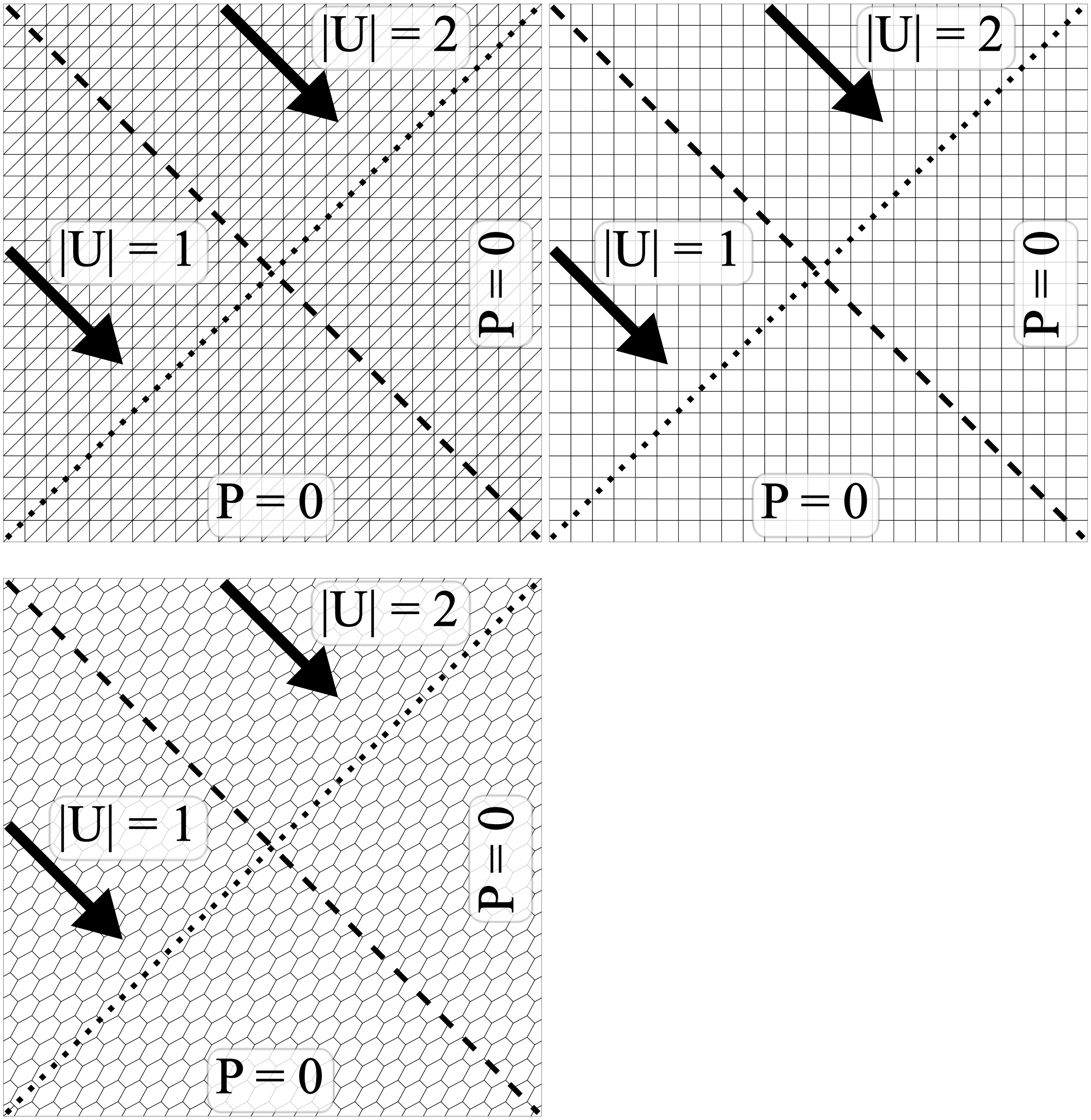

*Figure 23: 25x25 grids used to compare numerical dispersion and false diffusion. Flow is from top left to bottom right with a magnitude of 1 m/s on the bottom half and 2 m/s on the top half, with zero pressure boundary conditions on the outlets. Each 3D mesh is formed by extruding these one-cell thick into the page. The dotted line denotes the "normal" sampling line (whose direction moves from top left to bottom right), and the dashed line denotes the*

*"aligned" sampling line (whose direction moves from bottom left to top right). The 100x100 grids have the same relative cell distribution and are not repeated here.*

## Diffusion Study Results: Triangular Grid

It is well known that triangular/tetrahedral grids produce substantial false diffusion due to the certainty that at least one face of each cell will be misaligned with the flow direction. We certainly see this to be the case in Figure 24, where the triangular grids exhibit substantial smearing of the normal velocity profile. Increasing grid resolution from 25x25 to 100x100 sharpens the profile but leaves significant diffusion; this increase also worsens the undershoot seen with UD and CD on the low-velocity side of the shear layer. On these grids, LUD and MUSCL perform nearly identically, producing the least false diffusion without creating overshoots/undershoots. While LUD and MUSCL produce the most accurate centerline velocity profile, all methods produce a centerline velocity generally greater than the expected value.

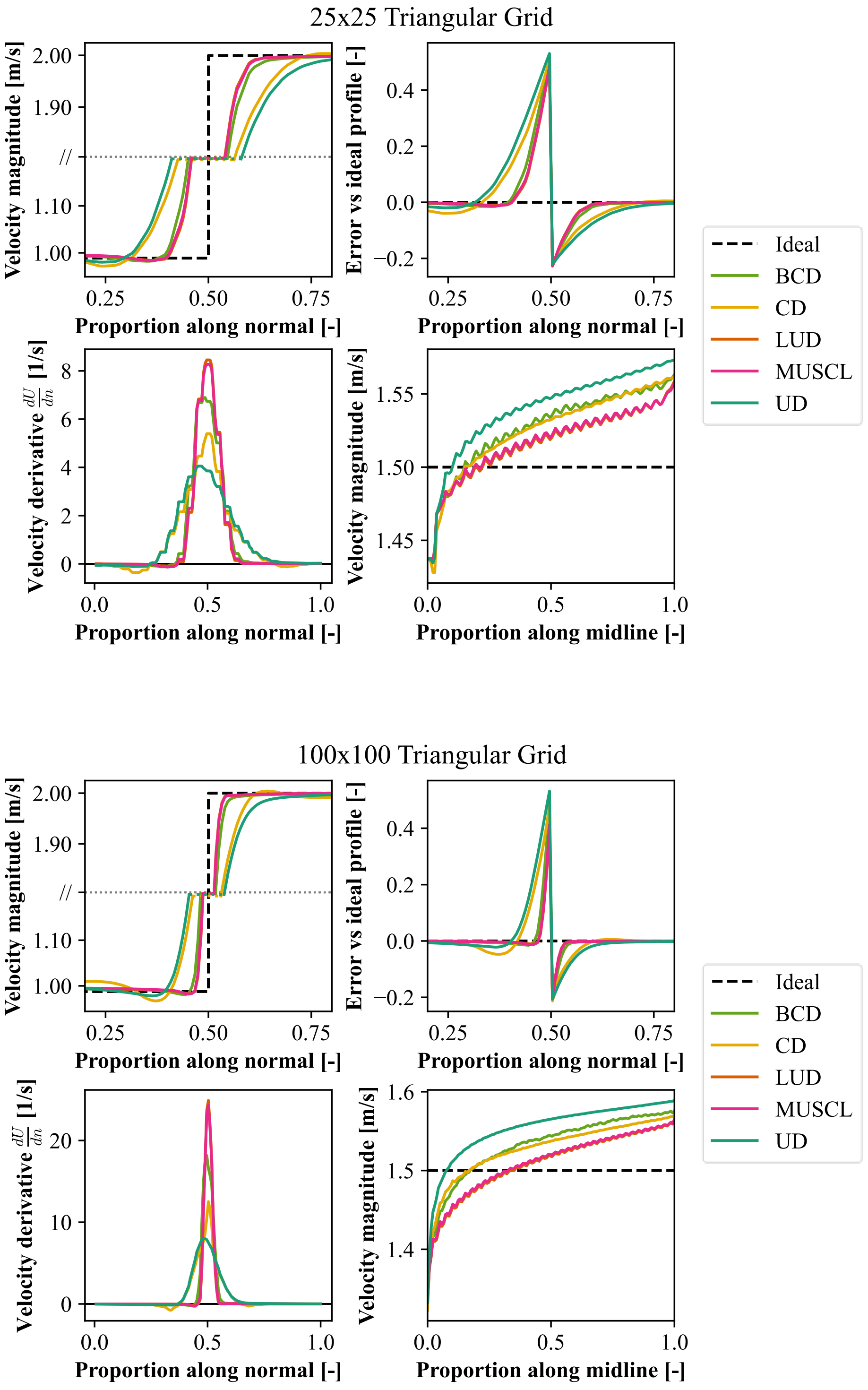

*Figure 24: Comparison of Fluent's momentum discretization schemes on coarse (top) and refined (bottom) grids consisting of triangular (wedge) cells. Significant false diffusion is*

*unavoidable on a triangular grid, but LUD, MUSCL, and BCD minimize diffusion without producing dispersive error. In contrast, CD and BCD produce more diffusion and more dispersion.*

### Diffusion Study Results: Polygonal Grid

Even with half the cell count, the polygonal meshes (Figure 25) are a great improvement over their corresponding triangular grids. In particular, the false diffusion of BCD and CD drops dramatically, positioning them as the two least diffusive methods. While CD exhibits mild dispersion on the low-velocity side of the shear layer, BCD shows no dispersion whatsoever and only slightly more diffusion. However, CD and BCD both underpredict the centerline velocity.

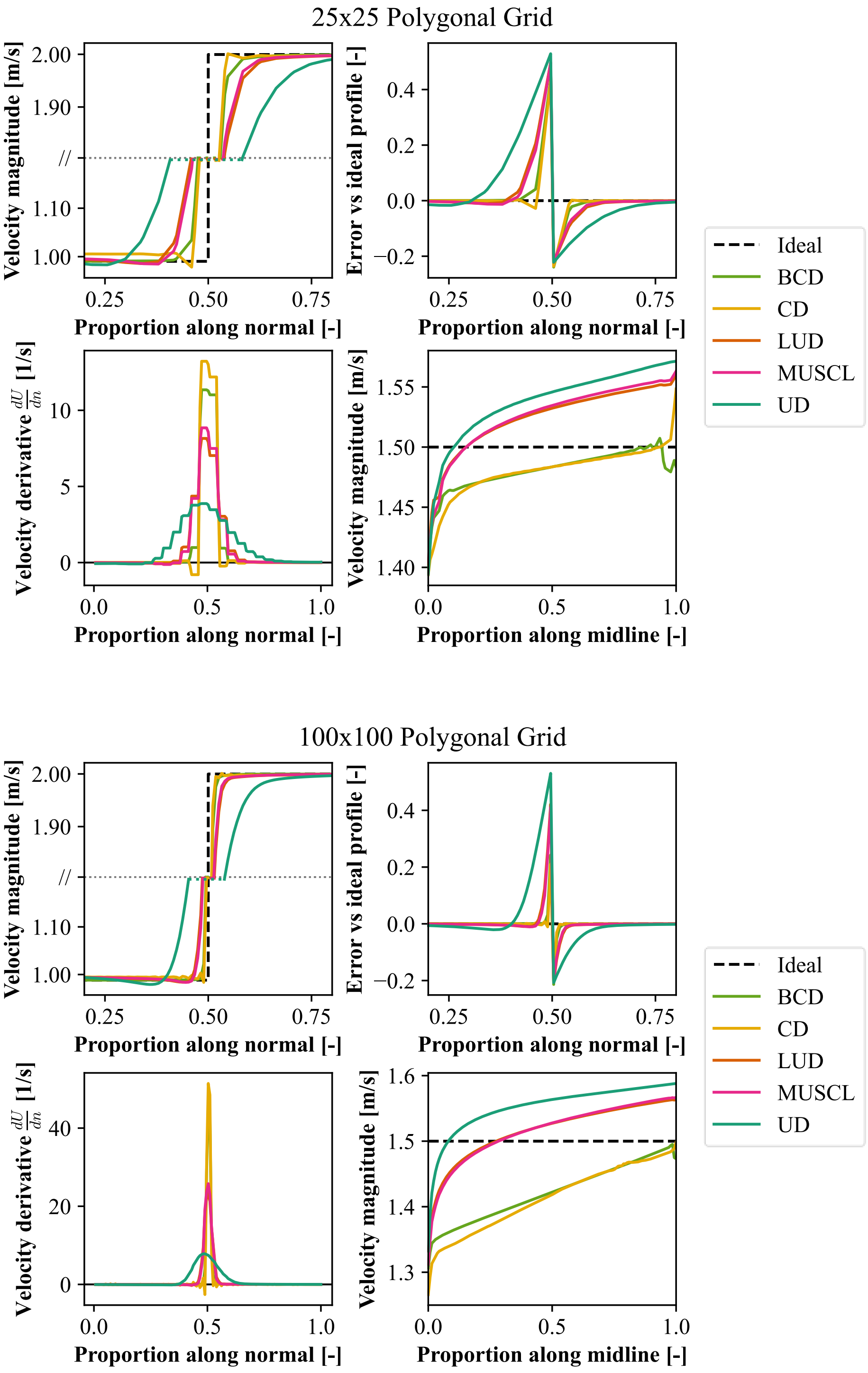

*Figure 25: Comparison of Fluent's momentum discretization schemes on coarse (top) and refined (bottom) grids consisting of polygonal cells. False diffusion is greatly reduced*

*compared to the triangular grids. In particular, CD and BCD are far less diffusive, producing the least diffusion overall. CD exhibits mild dispersion.*

## Diffusion Study Results: Comparison of Triangular/Polygonal Grids

BCD and LUD, two of the least diffusive methods on triangular and polyhedral grids, are compared on the two grid types in Figure 26 and Figure 27. With BCD, the 25x25 polygonal grid produces a normal profile nearly as sharp as the 100x100 triangular grid, and the 100x100 polygonal grid surpasses it. In both polygonal grids, dispersion is nonexistent. When using LUD, the best-case scenario for the triangular grid, the triangular grids have only a slight edge over their polygonal counterparts.

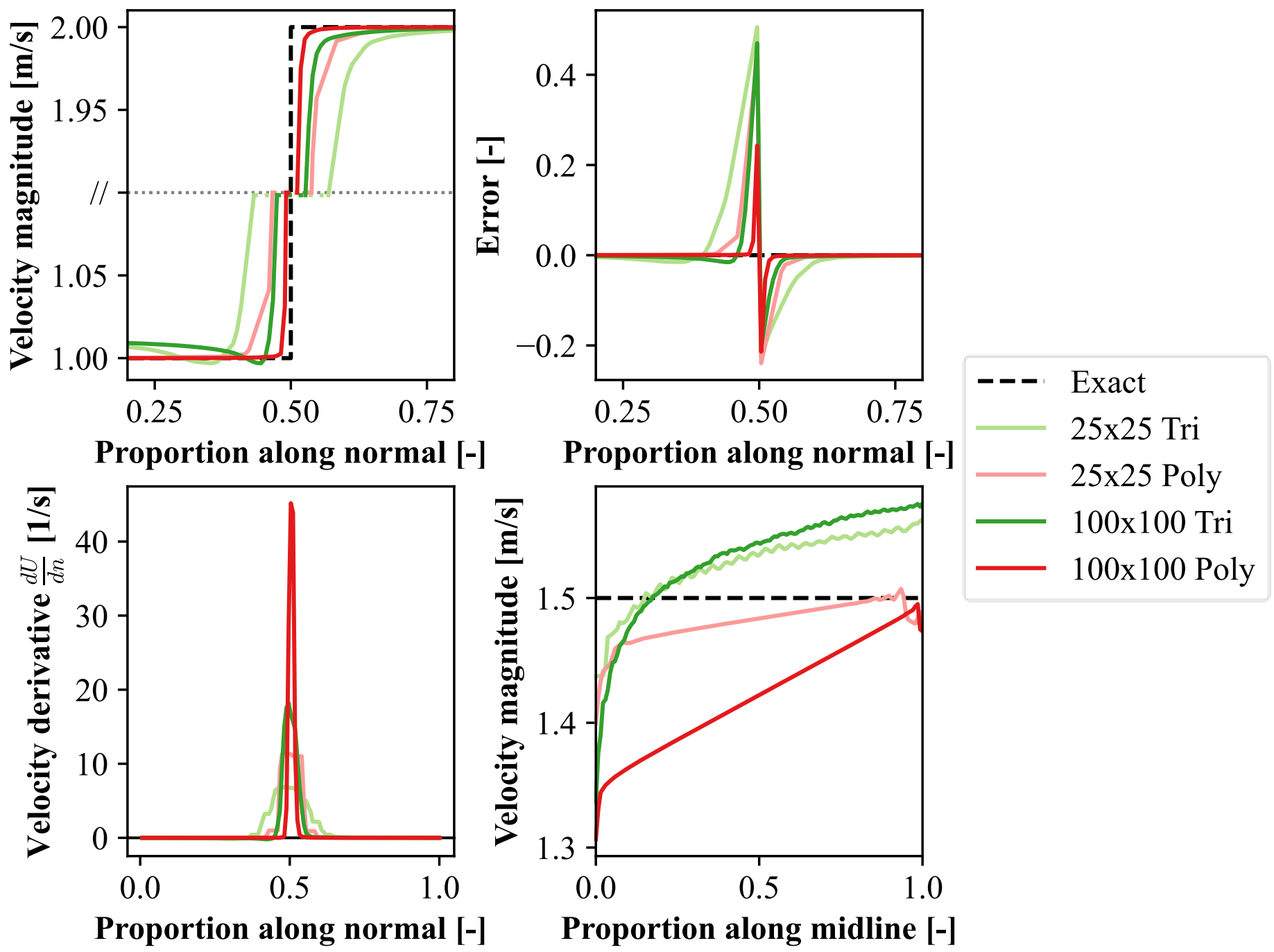

*Figure 26: Comparison of BCD on triangular and polygonal grids. Despite having ~1/30 the cell count, the 25x25 polygonal grid nearly matches the performance of the 100x100 triangular grid.*

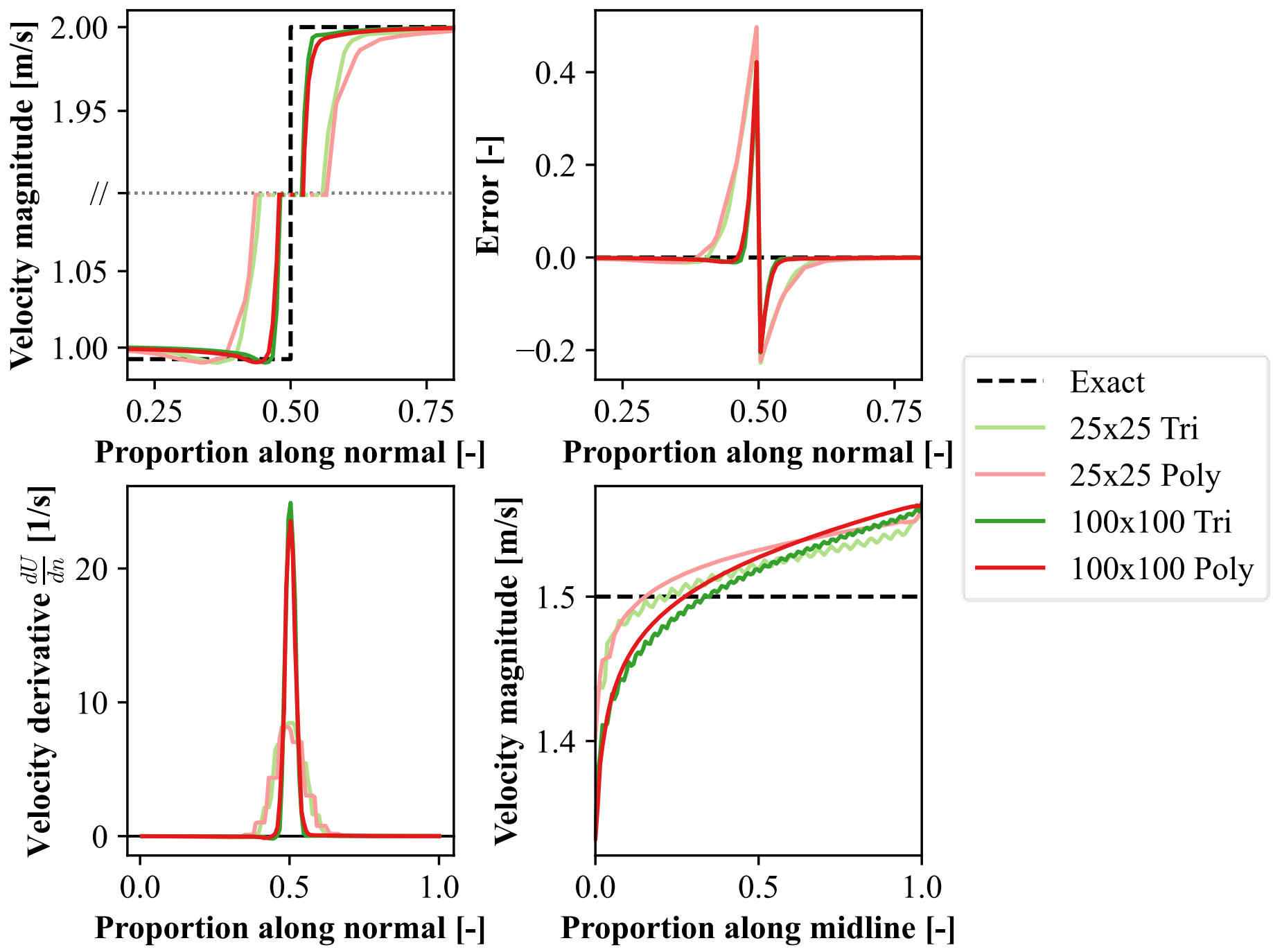

*Figure 27: Comparison of LUD on triangular and polygonal grids. LUD had the least diffusion on the triangular grids while distantly trailing CD and BCD on polygonal grids, lending a major advantage to the triangular grids in this comparison. Nonetheless, the polygonal grids closely trail their corresponding triangular grids, despite the triangular grids having twice the cell count.*

## Diffusion Study Results: Quadrilateral Grid

Relatively minor changes in dispersiveness/diffusiveness occur when moving through various types of polygons. LUD and MUSCL become somewhat less diffusive, CD and BCD become somewhat more diffusive, and UD is effectively unchanged. See Figure 28. With quadrilaterals, QUICK is available, which shows itself to be marginally less diffusive and more dispersive than LUD and MUSCL. Yet again, CD and BCD come out on top, with the least diffusion and least dispersion of Fluent's methods. Like with the polygonal mesh, they underpredict the centerline velocity; LUD/MUSCL/QUICK remain the most accurate in that regard.

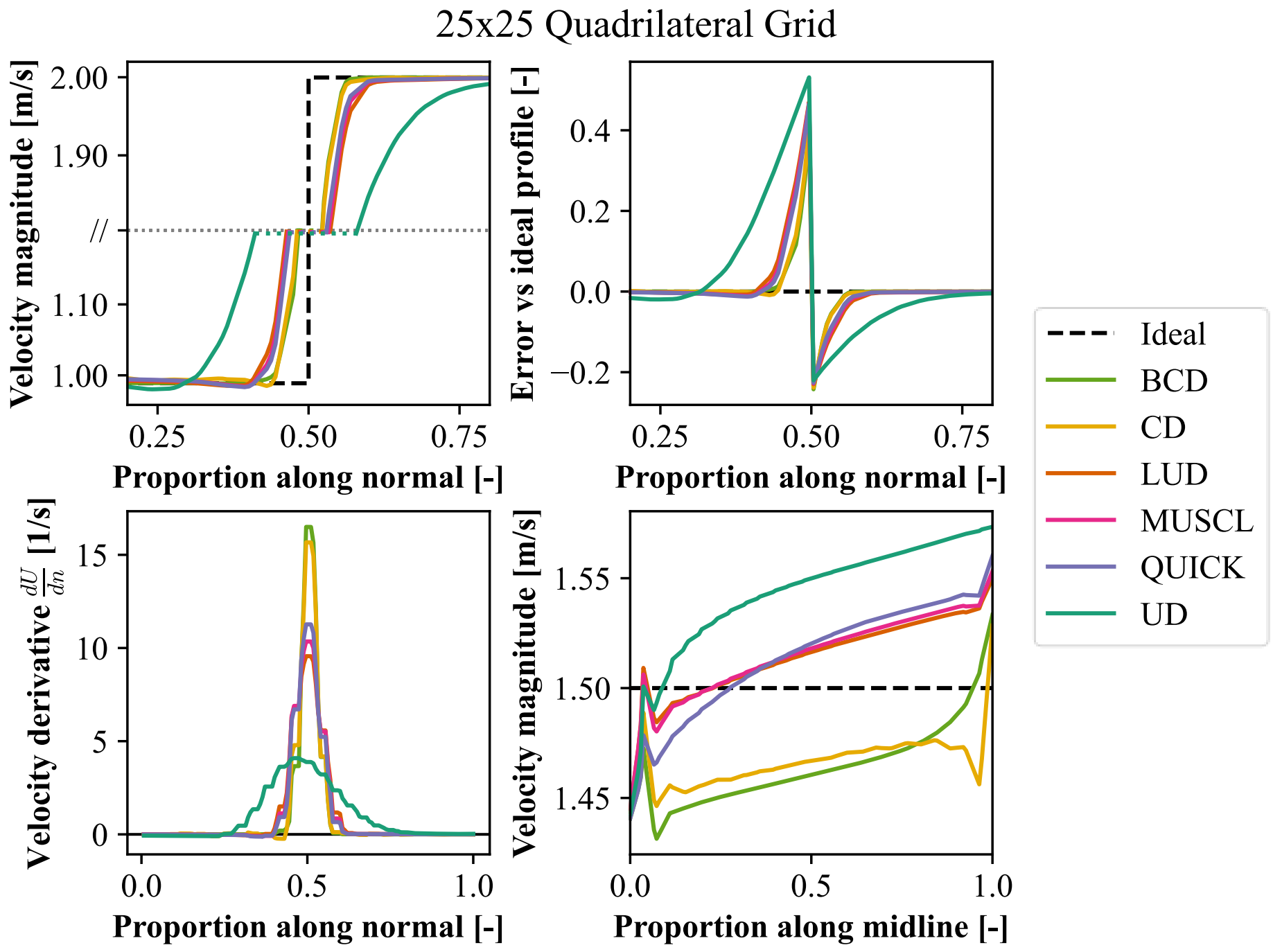

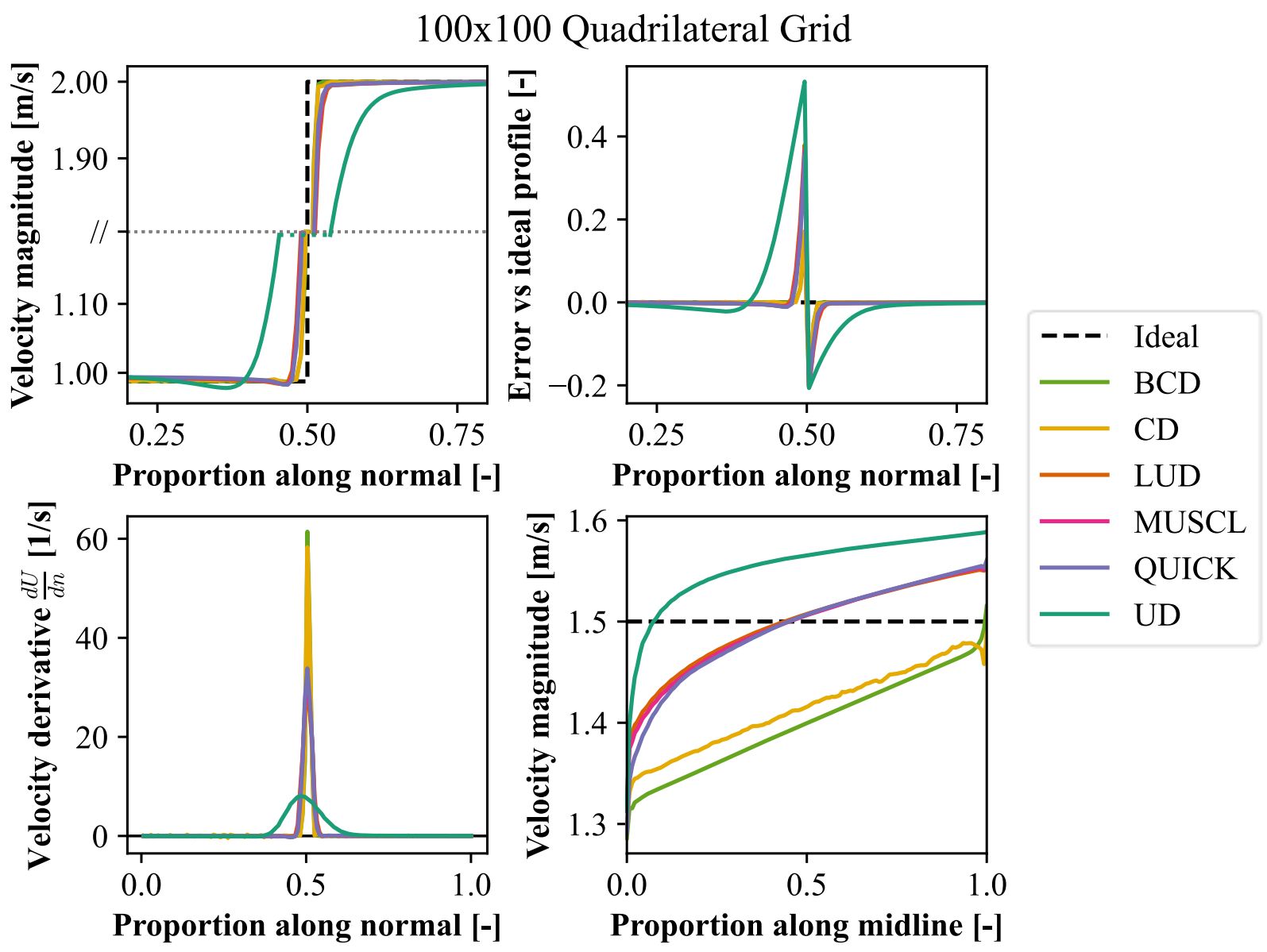

*Figure 28: Comparison of Fluent's momentum discretization schemes on coarse (top) and refined (bottom) grids consisting of quadrilateral cells. Despite flow being maximally misaligned with the grid, false diffusion is comparable to the polygonal grid, with CD and BCD resolving a greater peak velocity gradient while exhibiting less dispersive error.*

This flow scenario represents the worst case for a structured quadrilateral grid and for QUICK, with the flow maximally misaligned with the grid. Despite this fact, numerical diffusion and dispersion were comparable when comparing A) QUICK to Fluent's other "third-order" method (MUSCL) on the quadrilateral grid and B) each method on the quadrilateral grid to itself on the polygonal grid. In other words, QUICK and/or a quadrilateral grid will, at worst, perform similarly to other available options. At best (with flow aligned with the grid), it is well known that they outperform other options.

## Examining Central Differencing and Bounded Central Differencing

Despite QUICK's low diffusion, it requires a hexahedral grid. Our goal is to demonstrate Fluent's ability to perform DNS on an arbitrary grid, so QUICK was removed from consideration. This leaves CD and BCD as the least diffusive options. Figure 29 and Figure 30 compare their performance on all tested grids. Generally, BCD seems to produce better results, with both less diffusion and less dispersion. However, a comparison of these results will follow on a grid-by-grid basis.

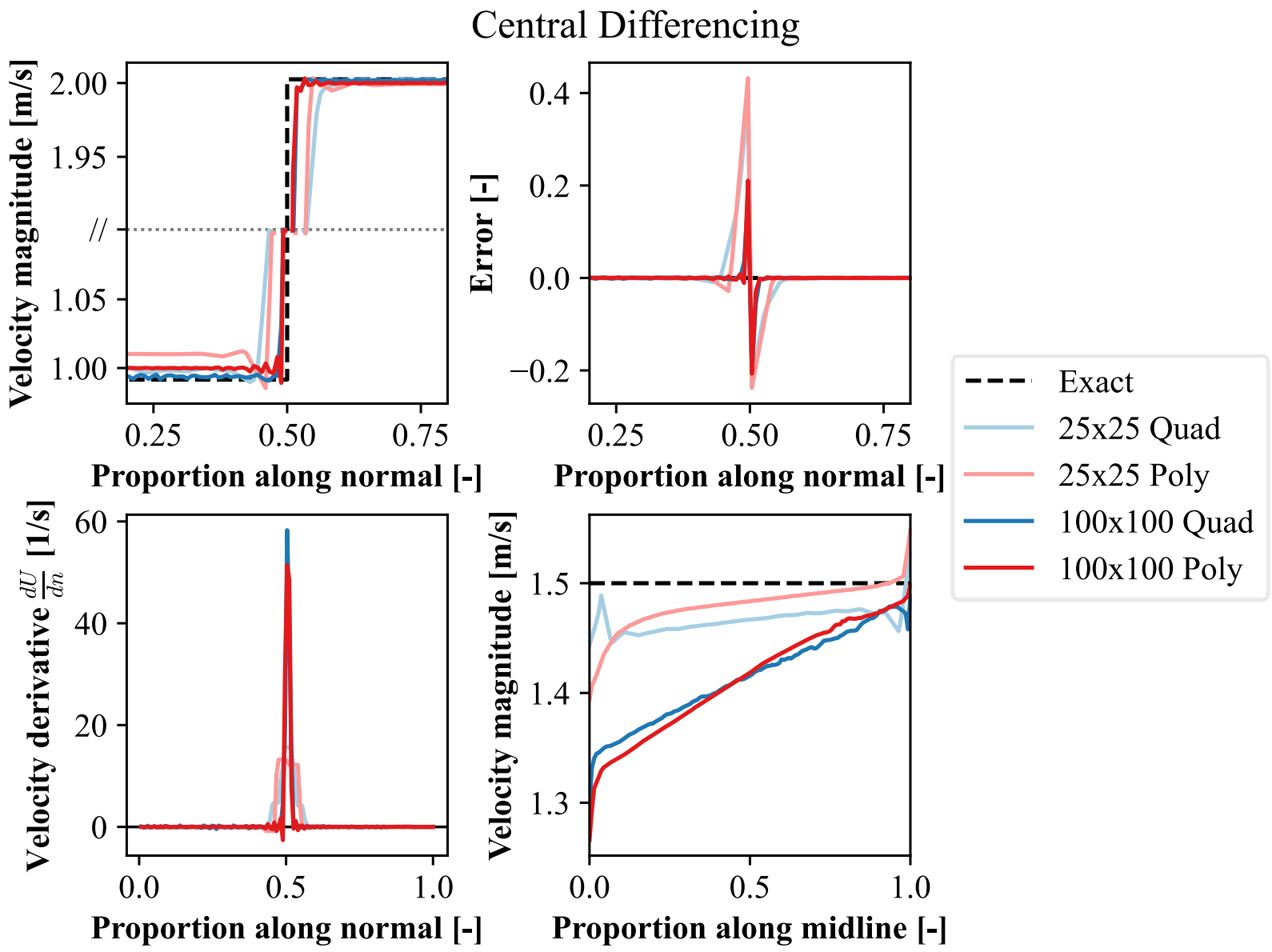

*Figure 29: Comparison of CD on quadrilateral and polygonal grids. The triangular grids exhibit the greatest magnitude of diffusive and dispersive error, with the coarsest polygonal grid producing less diffusive error and less dispersive error than the finest triangular grid (32x the cell count). The 100x100 polygonal grid is comparable to the 100x100 quadrilateral grid, though the latter resolves a higher peak velocity gradient and produces less dispersive error.*

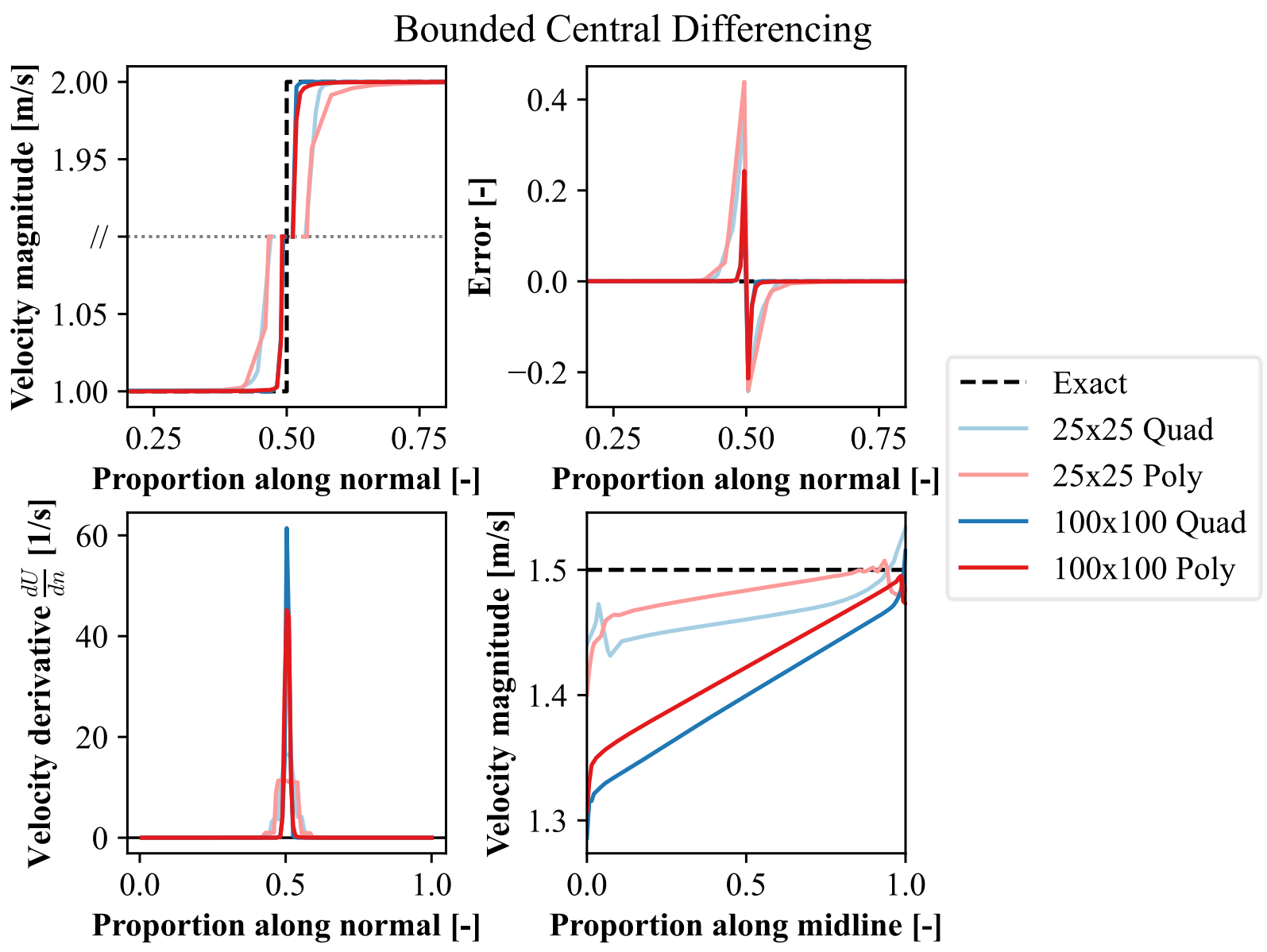

*Figure 30: Comparison of BCD with default boundedness (1.0) on all tested grids. Results are similar to those with CD, but dispersion is nearly entirely suppressed. Sharper gradients are resolved except in the polygonal grids, where the maximum resolved gradient is somewhat lessened.*

BCD is tunable via a “boundedness” parameter. Ansys states that setting boundedness to 0 reduces BCD to pure central differencing, while a boundedness of 1 (the default value, shown in prior plots) strictly applies the CBC. Despite the importance of this parameter and BCD’s widespread use in scale-resolving simulations [33,38–42], no fundamental study of its influence on diffusion/dispersion has been found in the literature. Consequently, we will compare five values of boundedness: 0, 0.25, 0.5, 0.75, and 1.0. For brevity, these will be referred to as “BCD-#”, where ‘#’ is the boundedness value. Since we have established that polygonal grids are less diffusive than triangular grids when using

BCD, we will omit discussion of the boundedness criterion on triangular grids. The interested reader will find those plots in Appendix C.

On the coarse quadrilateral grid, shown in Figure 31, CD, BCD-0.75, and BCD-1.0 behave similarly (with diffusiveness increasing in that order). However, at BCD-0.5, some dispersion is seen, which becomes significant in BCD-0.25 and BCD-0. There is not much pattern to the accuracy of the centerline velocity prediction, but it appears that lower boundedness generally produces a more accurate result there. This trend aligns with the prior observation that upwind-biased schemes more accurately predicted the centerline velocity. Moving to the finer quadrilateral grid (Figure 32) we see similar behavior: while all schemes resolve the gradient quite well, dispersion is evident in BCD with boundedness of 0.5 and lower. All schemes again perform similarly at predicting the centerline velocity, though error is greater than with the coarser mesh.

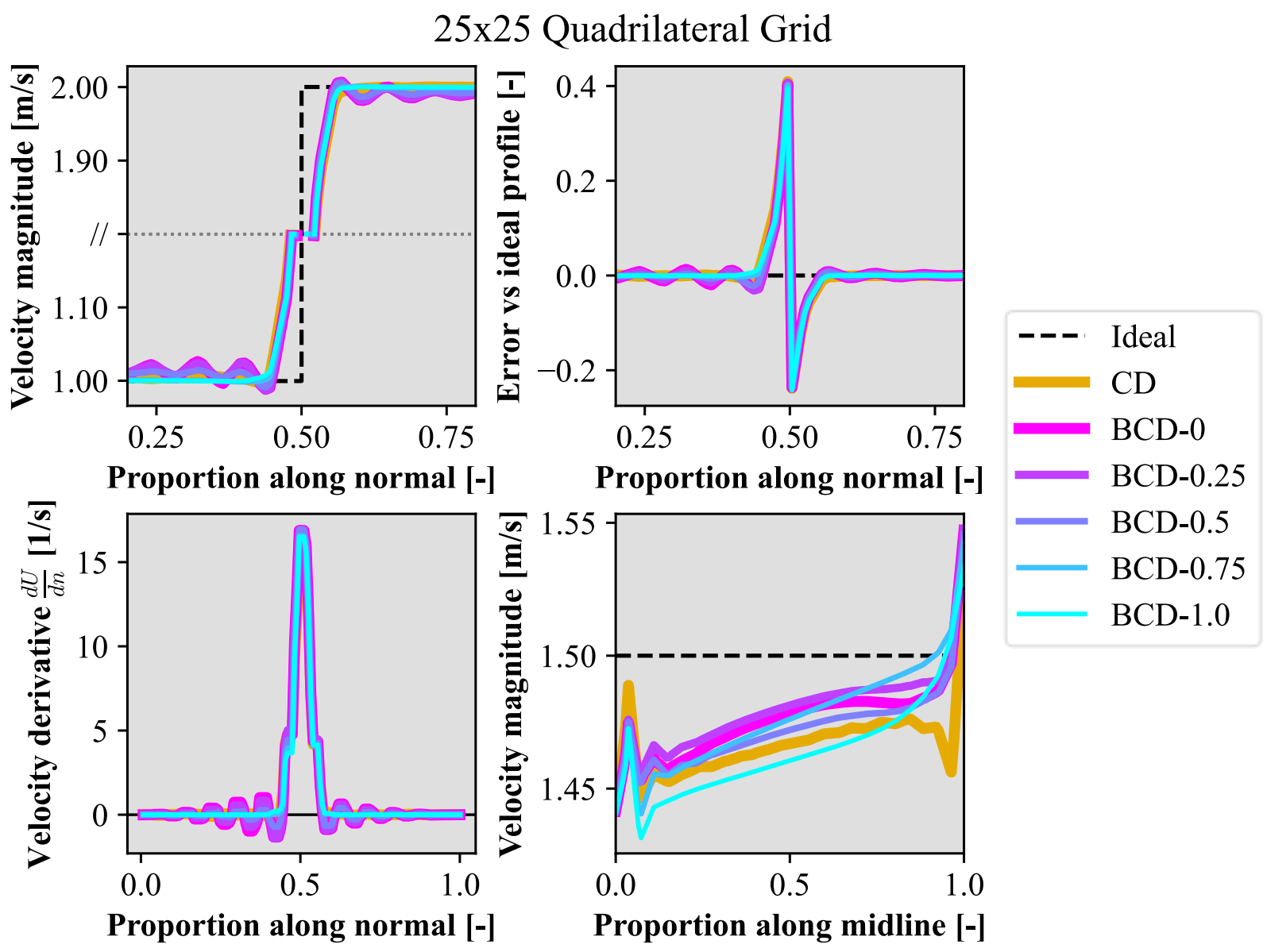

*Figure 31: Comparison of CD and BCD with boundedness from 0-1 on a 25x25 quadrilateral grid.*

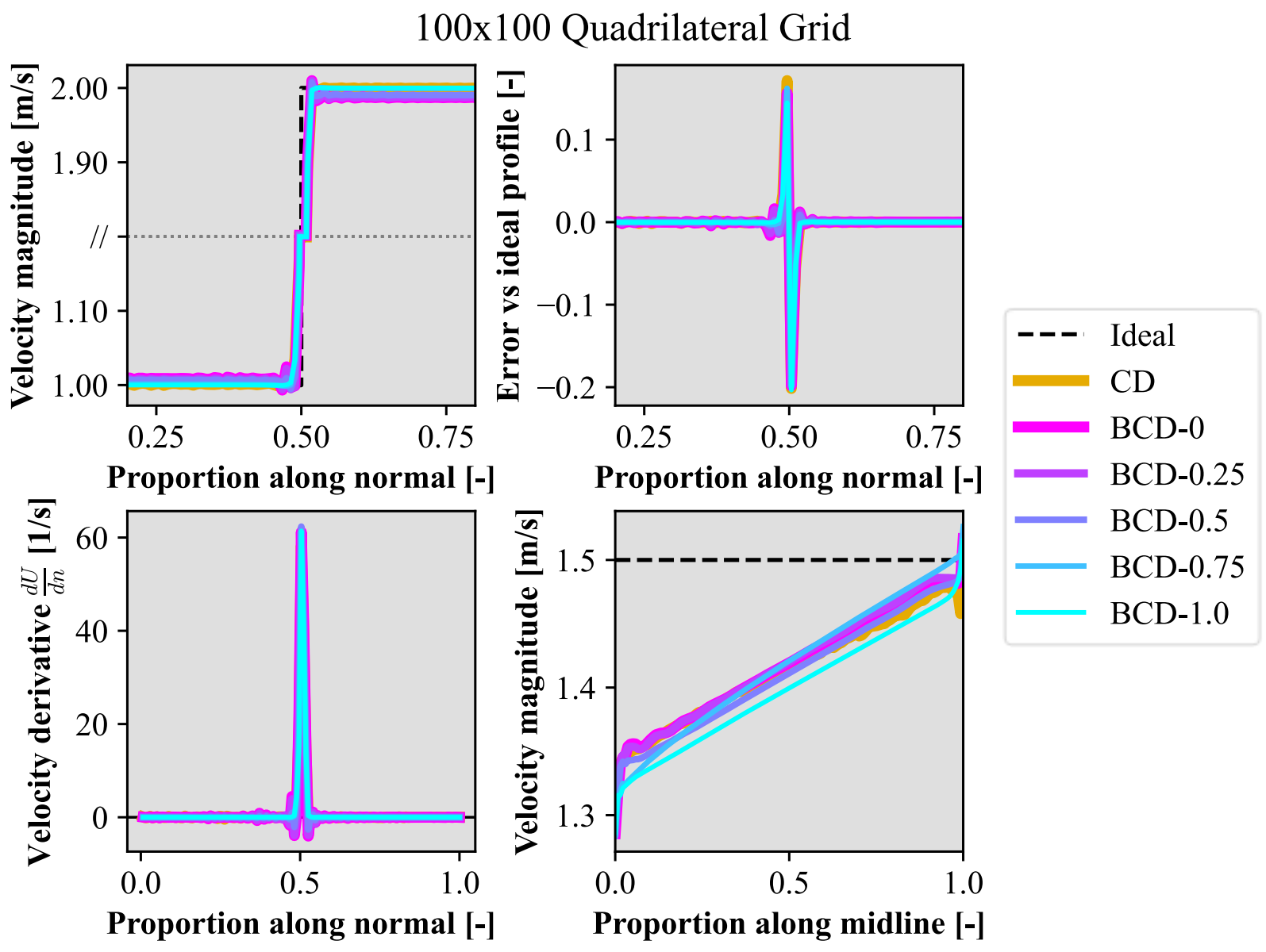

*Figure 32: Comparison of CD and BCD with boundedness from 0-1 on a 100x100 quadrilateral grid.*

The polyhedral grids, shown in Figure 33 and Figure 34, display the same qualitative behavior as the quadrilateral grids. CD and BCD-1.0 produce similar results, with dispersive error growing as boundedness falls. However, with the polyhedral grids, lower boundedness produces significantly greater centerline gradients.

Since Ansys claims that BCD-0 reduces to pure central differencing [24], BCD-0 and CD should produce equivalent results. Generally, this is shown by the data, though minor differences can be seen. For example, in Figure 33 it can be seen that CD produces a slightly higher peak gradient.

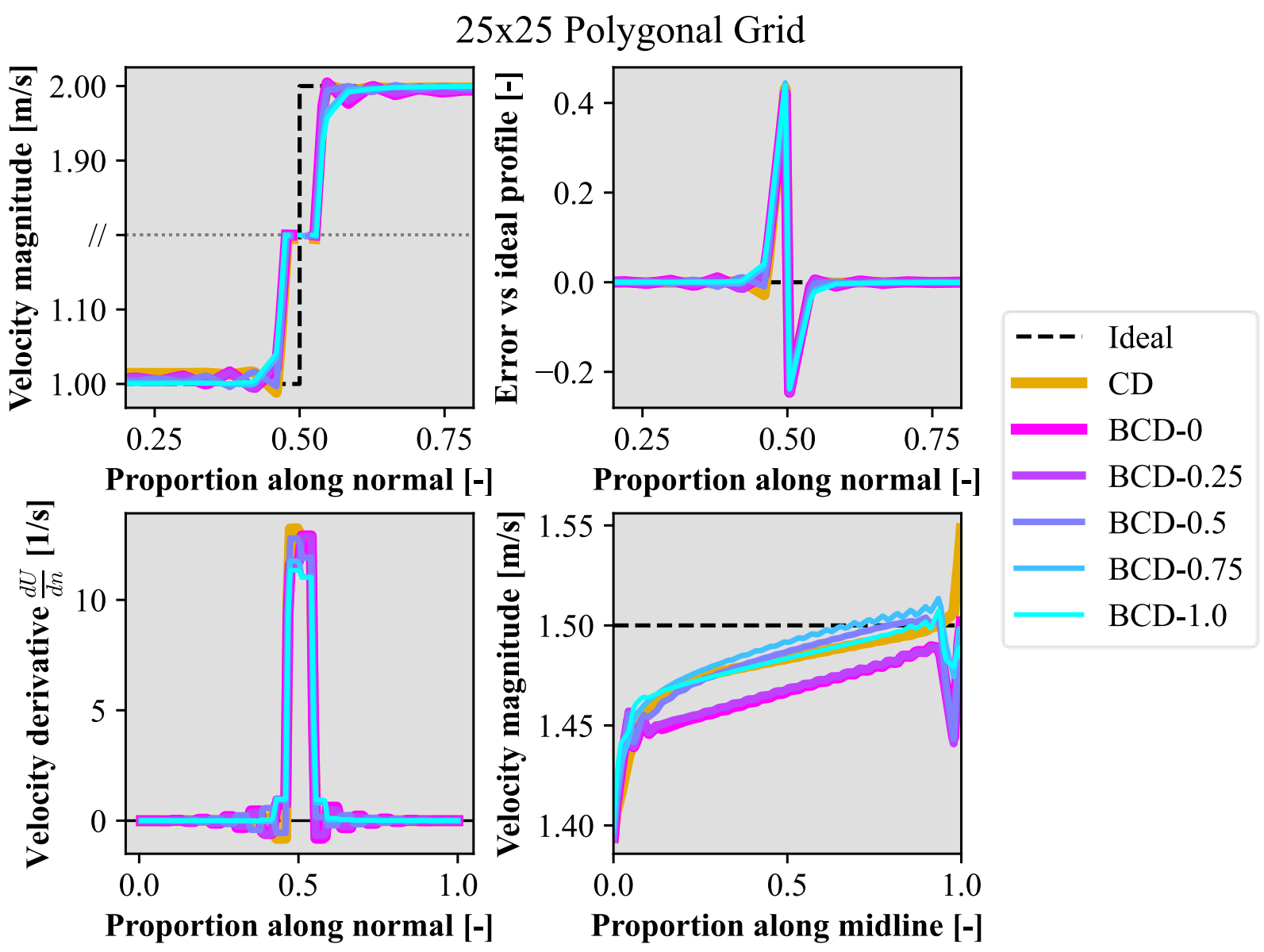

*Figure 33: Comparison of CD and BCD with boundedness from 0-1 on a 25x25 polygonal grid.*

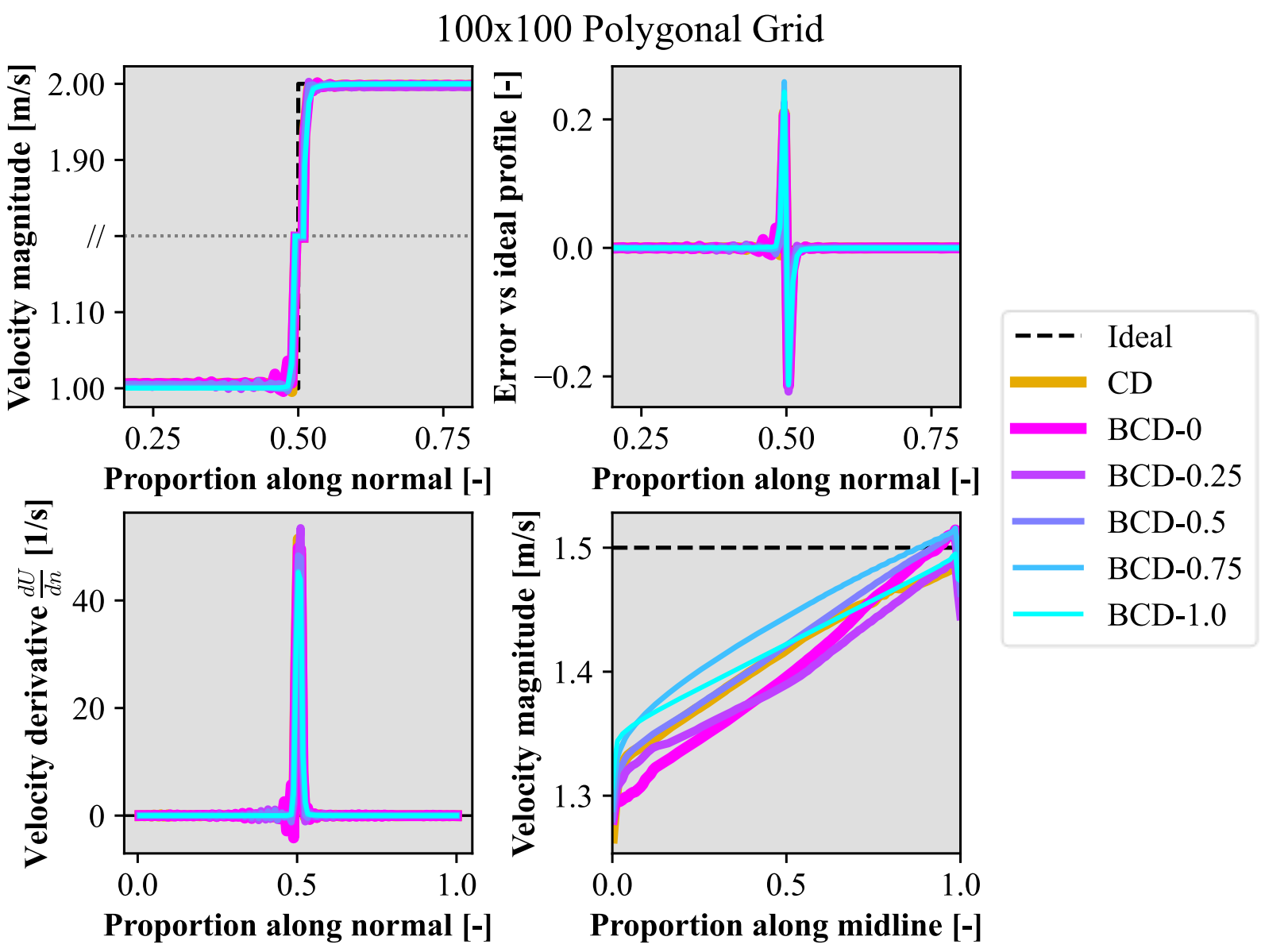

*Figure 34: Comparison of CD and BCD with boundedness from 0-1 on a 100x100 polygonal grid.*

# Appendix C: Additional Plots

Figures 35-37 illustrate time step independence of the results presented in this work. Figure 35 shows results with a time step of 0.08 viscous timescales. Figure 36 doubles this value, and Figure 37 doubles it again. While this paper used a time step of $0.08t^+$ for conservatism, the energy spectra shown here suggest that a time step of even larger than 0.32 viscous timescales may be adequate.

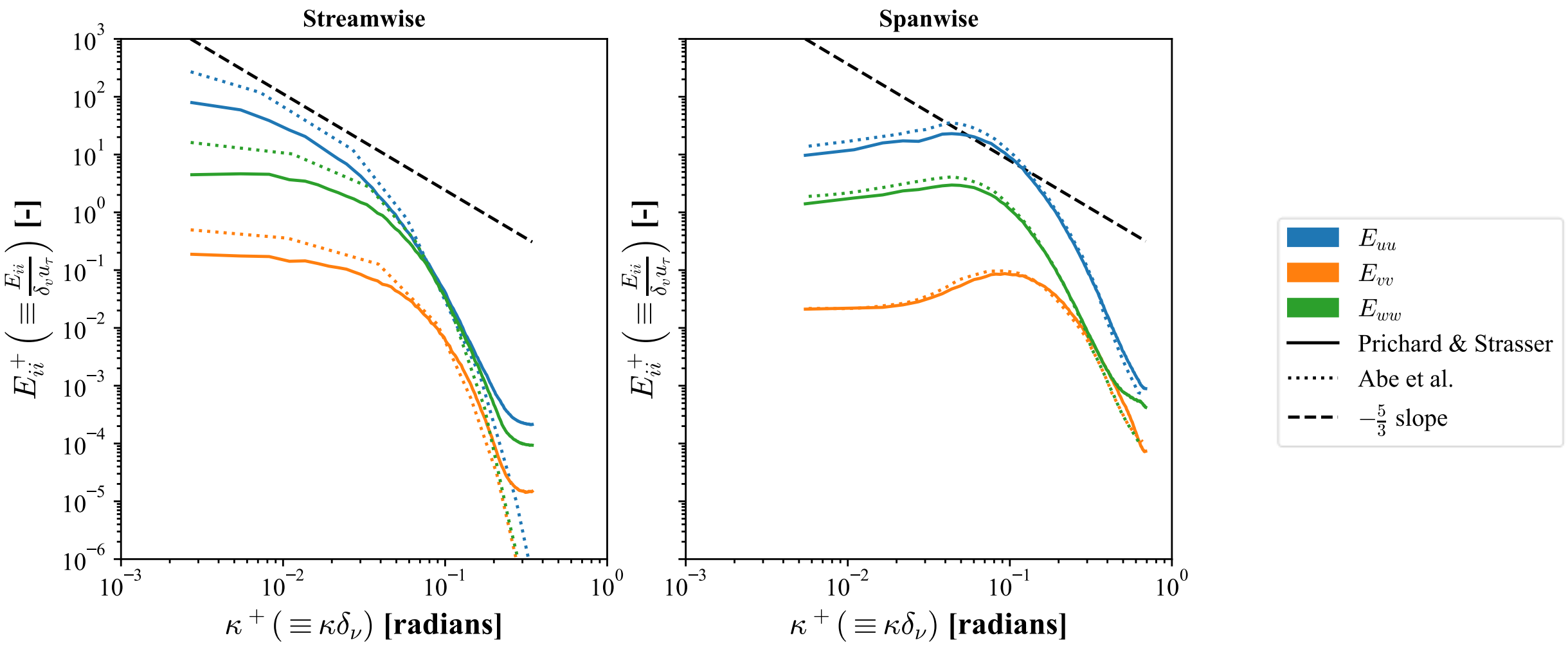

*Figure 35: Spectral densities at $y^+ = 5.4$ for grid S1 with a time step size of $0.08t^+$, which was used in all discussed results.*

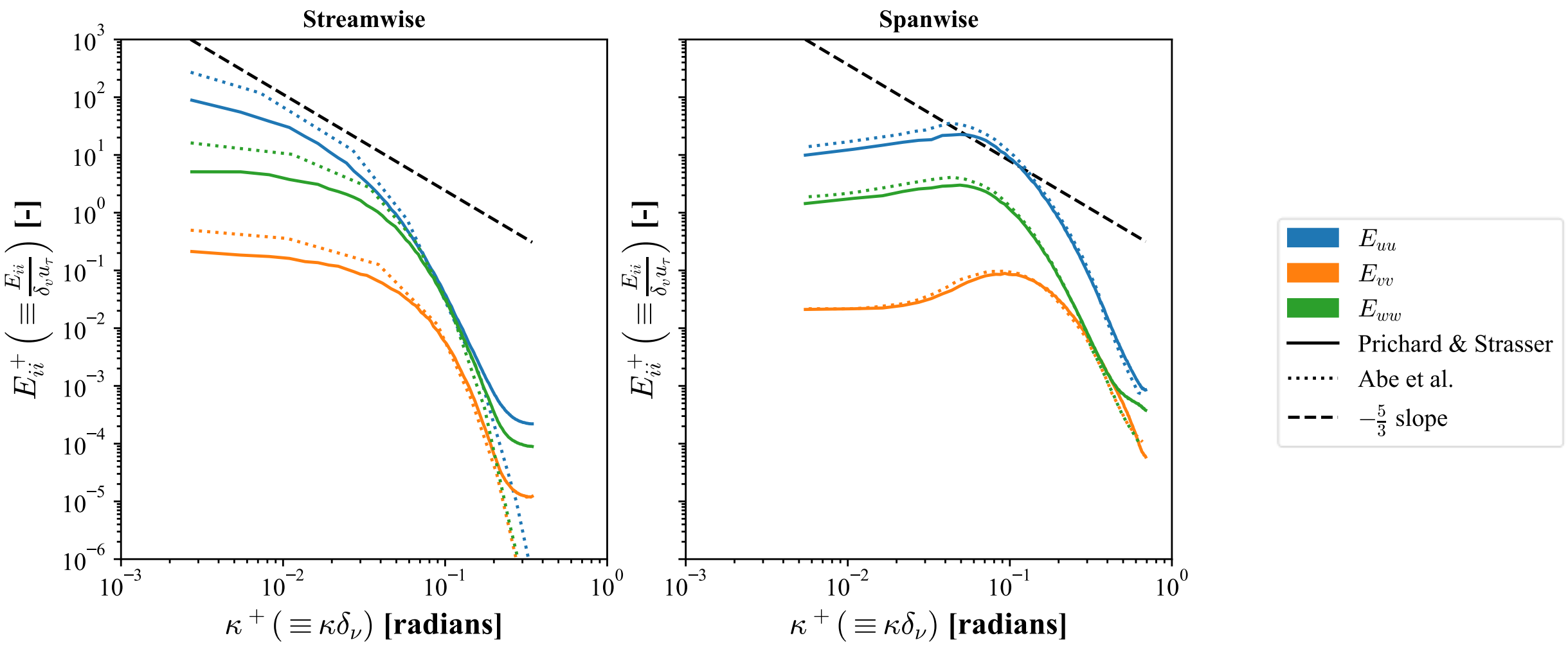

*Figure 36: Spectral densities at $y^+ = 5.4$ for grid S1 with a time step of $0.16t^+$ , double the value used in this paper. There is no significant change from the prior figure, which depicts a time step half this size, so the combination of time step size, iteration count, and pressure-velocity coupling scheme was deemed sufficient for convergence.*

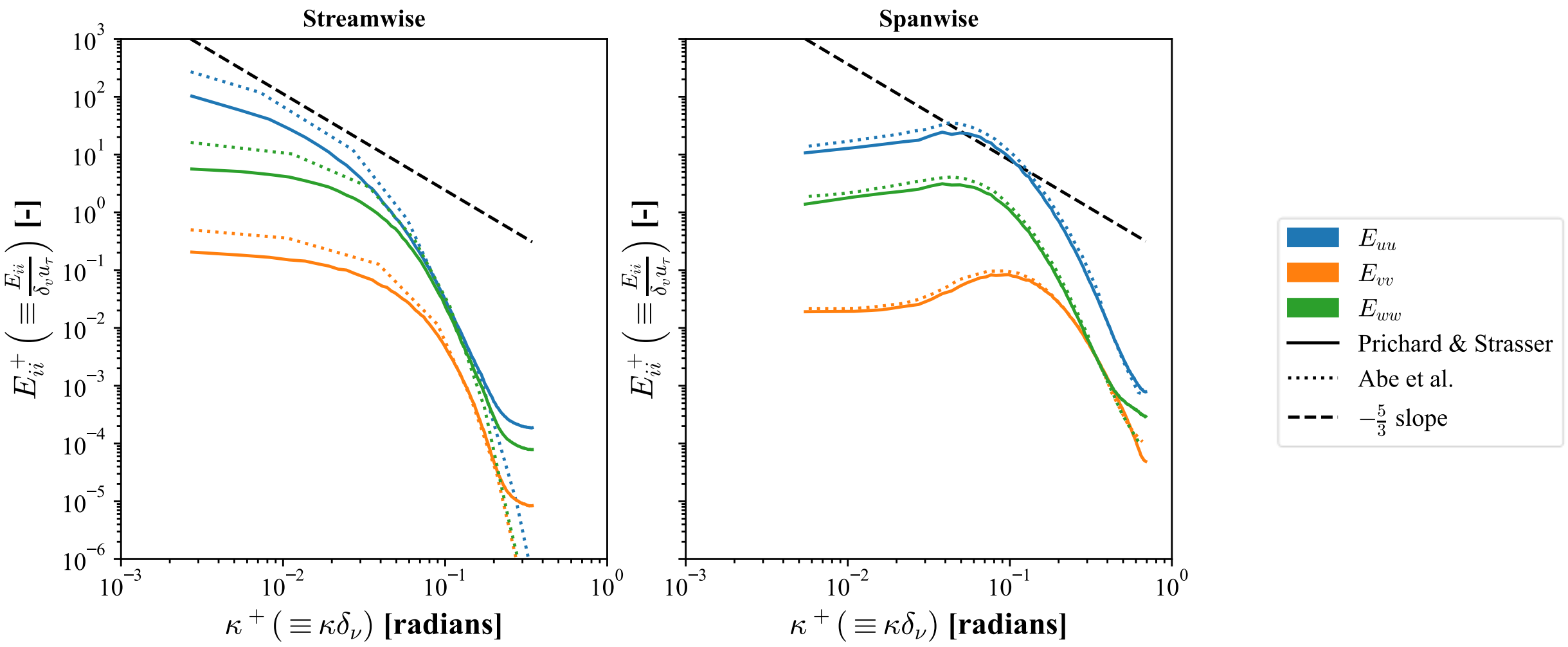

*Figure 37: Spectral densities at $y^+ = 5.4$ for grid S1 with a time step of $0.32t^+$, four times the value used in this paper. Changes remain minimal, consisting primarily of a small reduction in high-wavenumber energy.*

While triangular/tetrahedral grids have fallen out of favor with the maturation of polyhedral grids, Figure 38 shows the effect of boundedness on a triangular grid. In contrast to the two other grid types, it is noteworthy that greater boundedness reduced diffusion and dispersion. Without public access to definitive knowledge of Fluent's implementation, it is unclear why this might be.

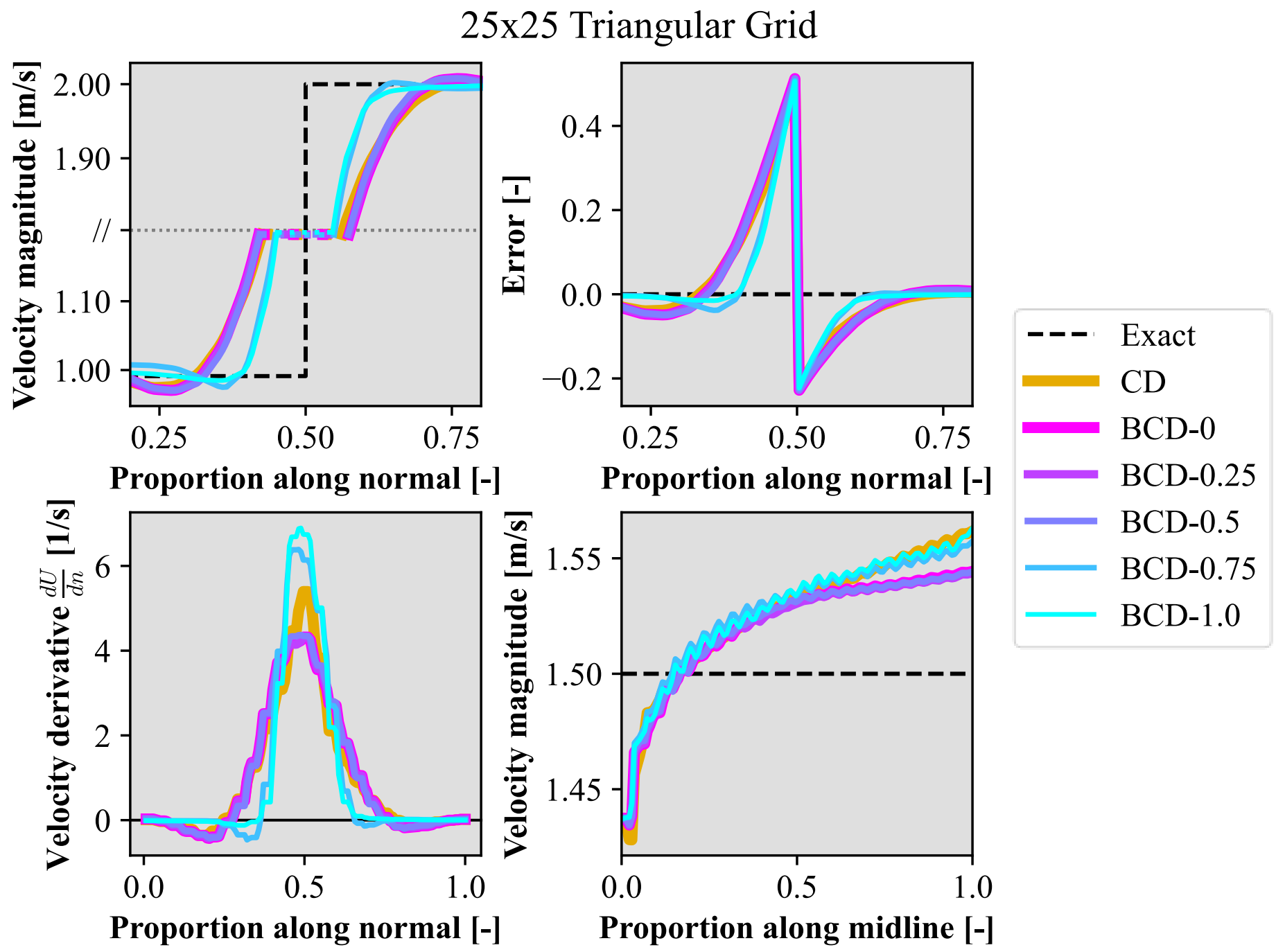

25x25 Triangular Grid
Velocity magnitude [m/s]
2.00
1.90
//
1.10
1.00
0.25
0.50
0.75
Proportion along normal [-]
Error [-]
0.4
0.2
0.0
−0.2
0.25
0.50
0.75
Proportion along normal [-]
Velocity derivative $\frac{dU}{dn}$ [1/s]
6
4
2
0
0.0
0.5
1.0
Proportion along normal [-]
Velocity magnitude [m/s]
1.55
1.50
1.45
0.0
0.5
1.0
Proportion along midline [-]
Exact
CD
BCD-0
BCD-0.25
BCD-0.5
BCD-0.75
BCD-1.0

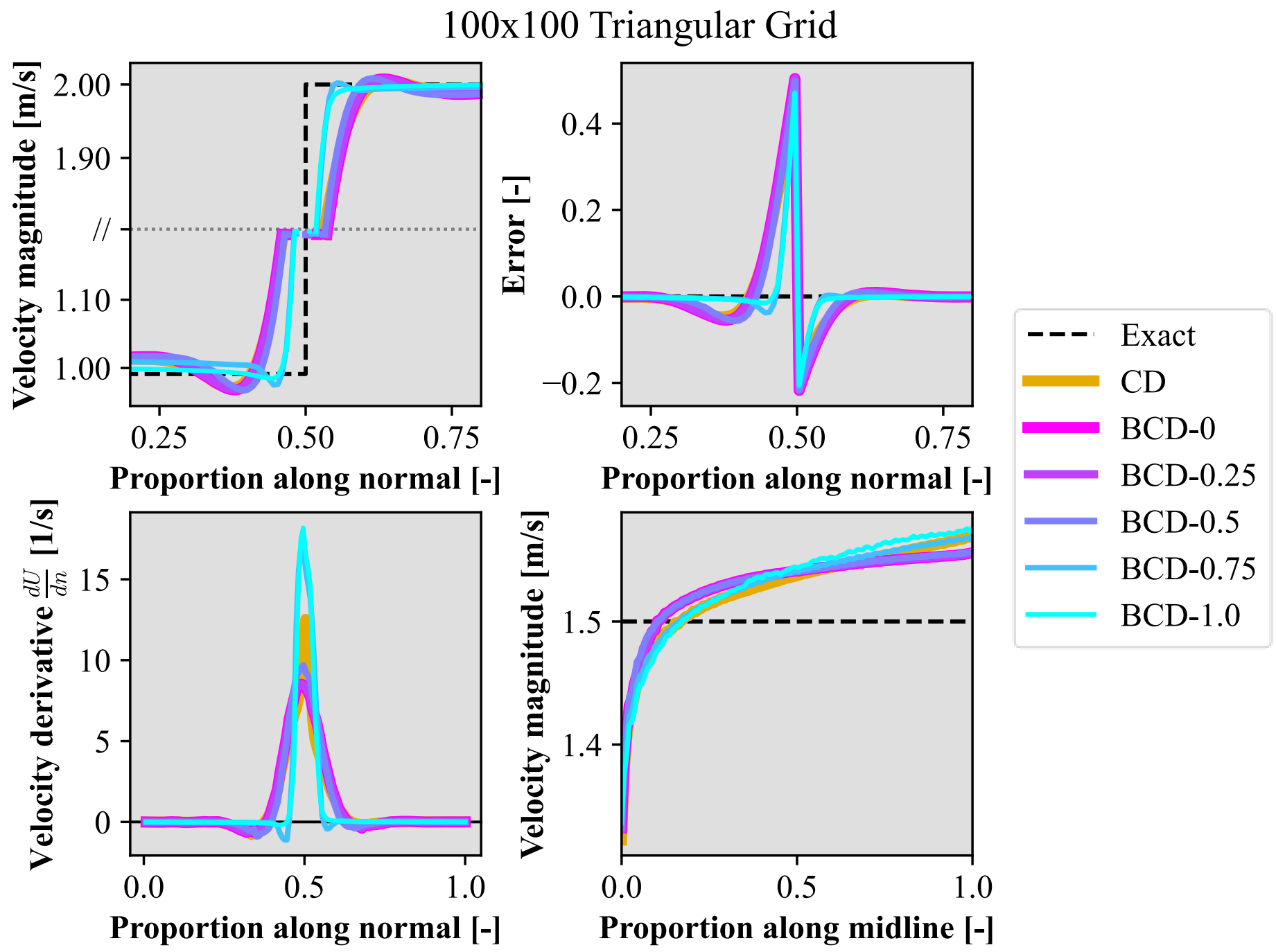

*Figure 38: Comparison of BCD boundedness on grids composed of triangular cells. While Fluent's documentation states that greater boundedness is more diffusive, here we see the least dispersion and least diffusion with the two greatest boundedness settings.*

Figures 39-43 present the Reynolds Stress Tensor (RST) profiles from the wall to the centerline for all structured grids and for various times and discretizations for which we have no experimental data. As expected in channel flow, TKE peaks at $y+$ of 15-20, while $< uv >$ peaks at roughly $y+$ = 30. Its redistributive effect on the normal stresses is seen, as $< vv >$ and $< ww >$ peak slightly farther from the wall. Differences between grids are minimal, suggesting grid requirements for TKE profiles are much less stringent than those for an accurate energy spectrum. Additionally, time step size and BCD versus CD does not appear to make much difference. The largest variations are seen in $< uw >$ minima and especially in $< uw >$. Of course, $< uw >$ should be zero at sufficient time-averaging time, and quadrupling the time step size tends to exacerbate this deviation.

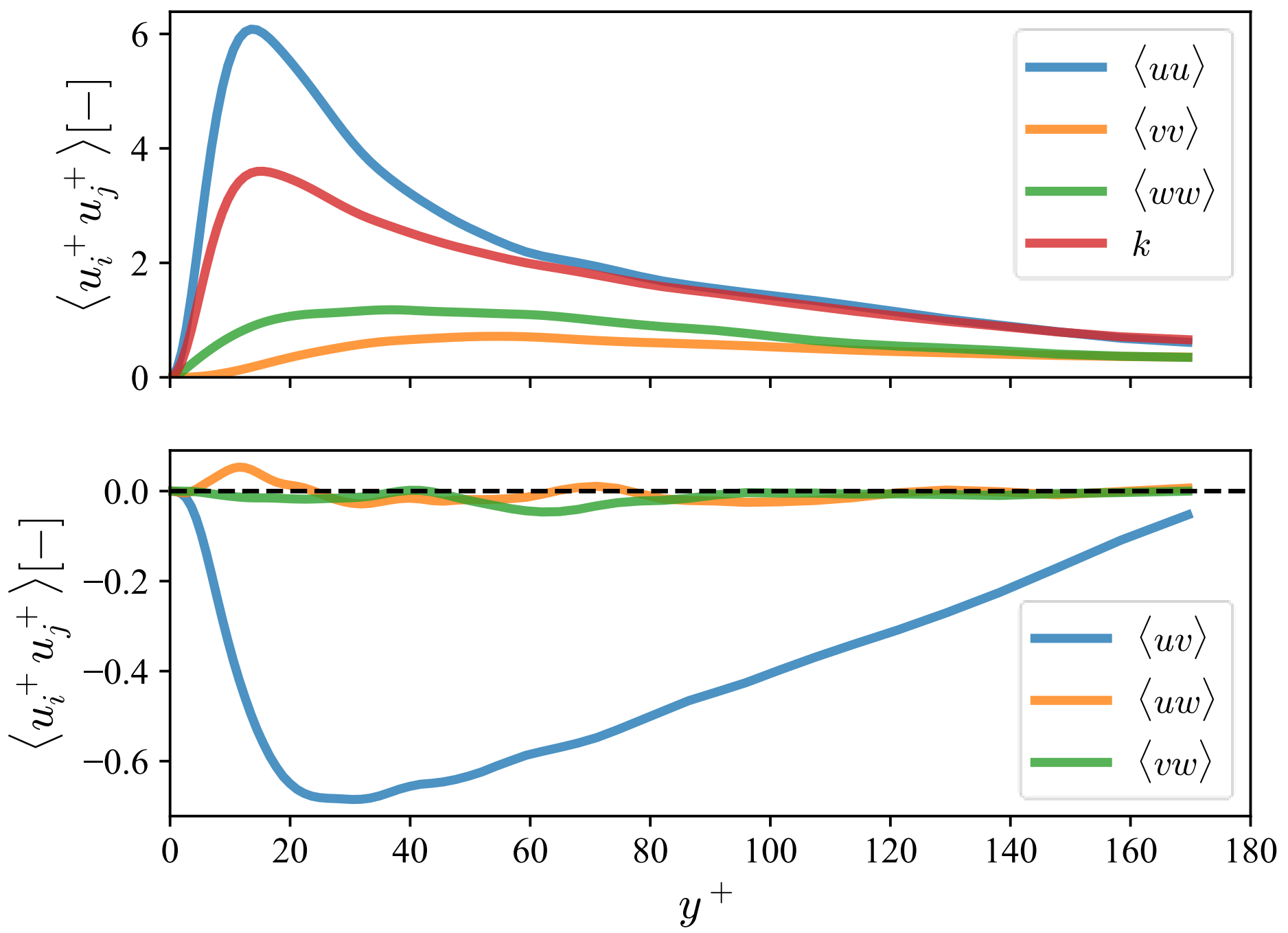

*Figure 39: Grid S1 – RST and TKE profiles using BCD with* $t^+ = 0.08$.

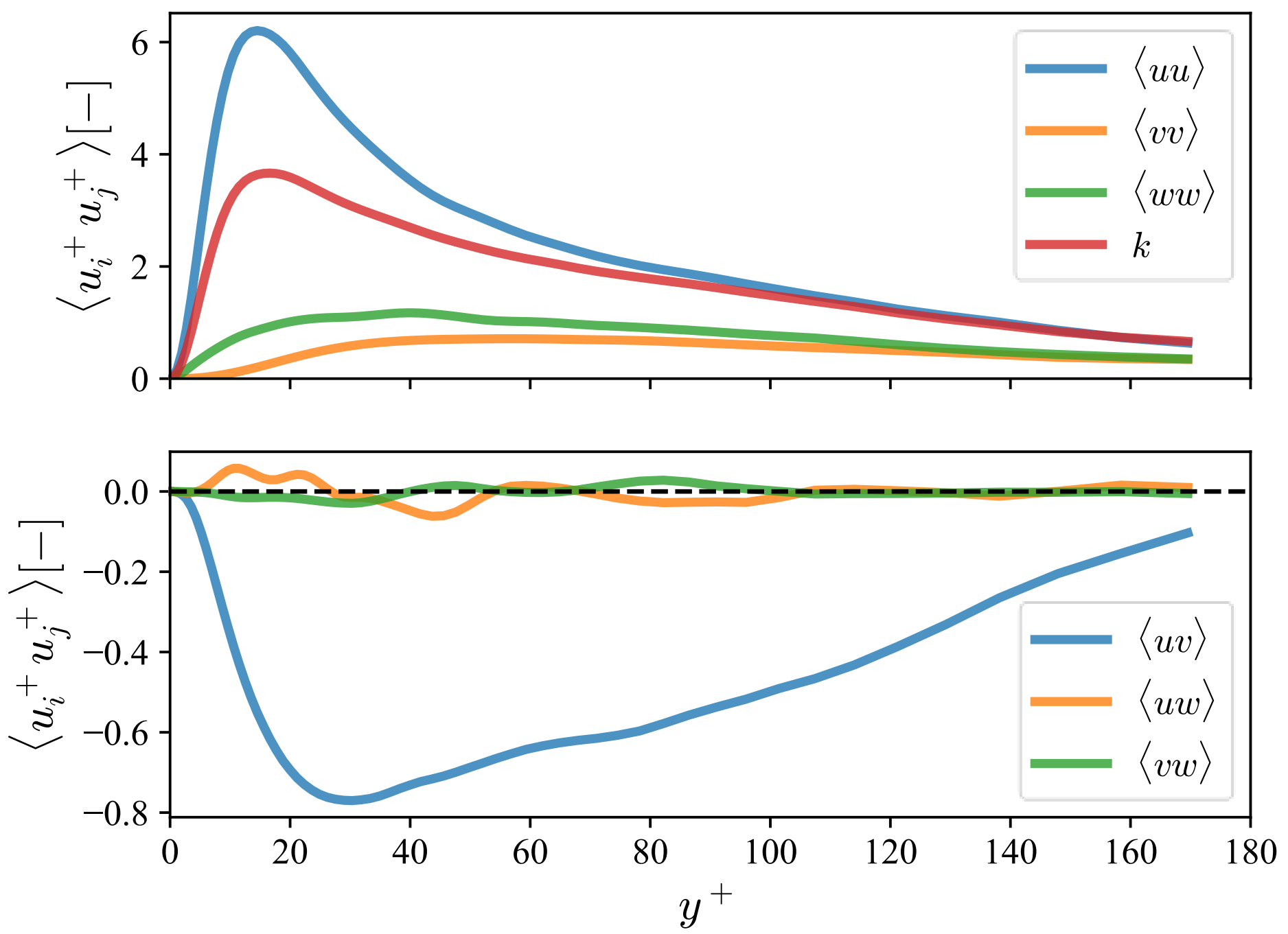

*Figure 40: Grid S1 – RST and TKE profiles using pure CD with* $t^+ = 0.08$.

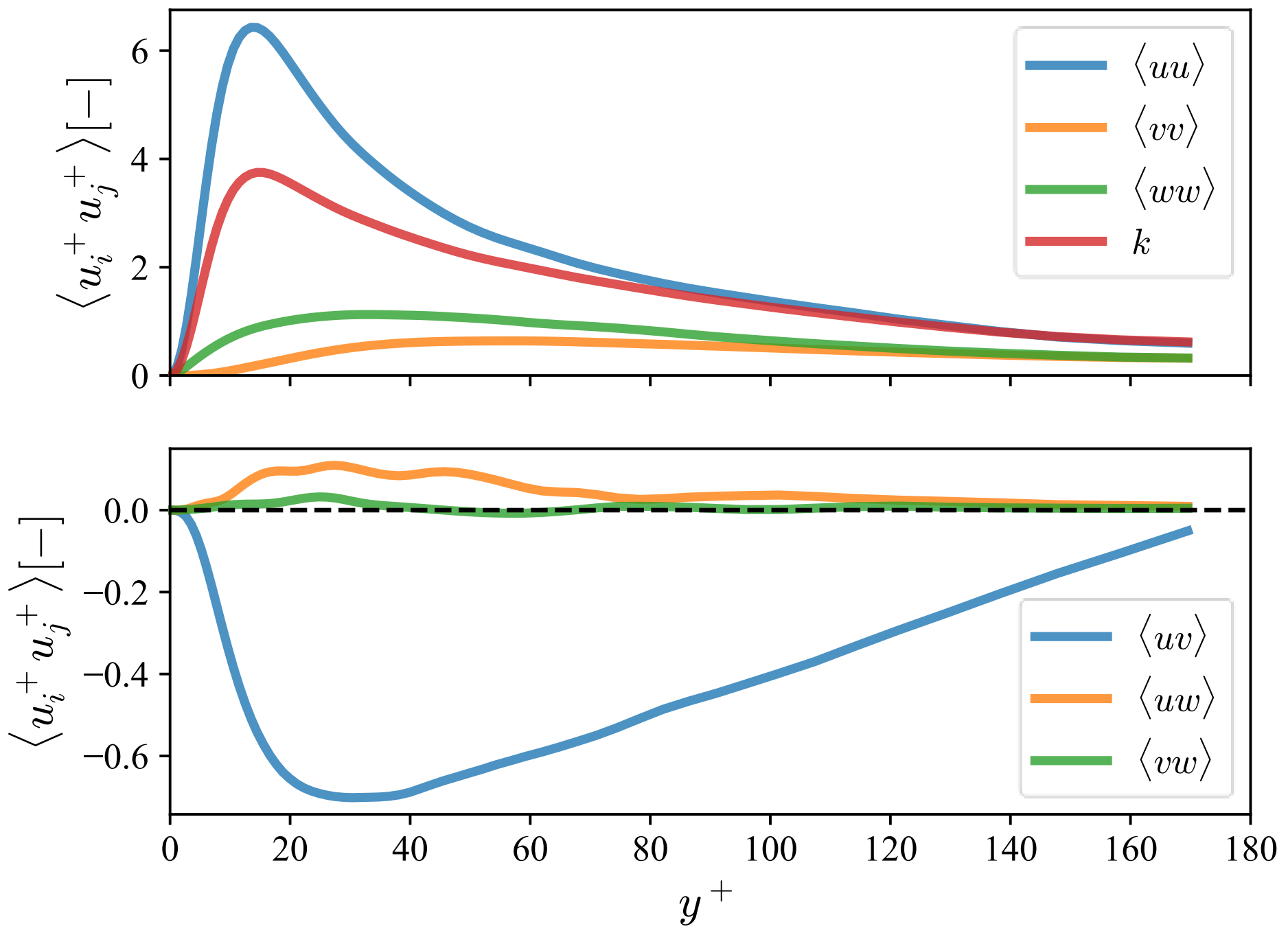

*Figure 41: Grid S1 – Grid S1 – RST and TKE profiles using BCD with* $t^+ = 0.32$.

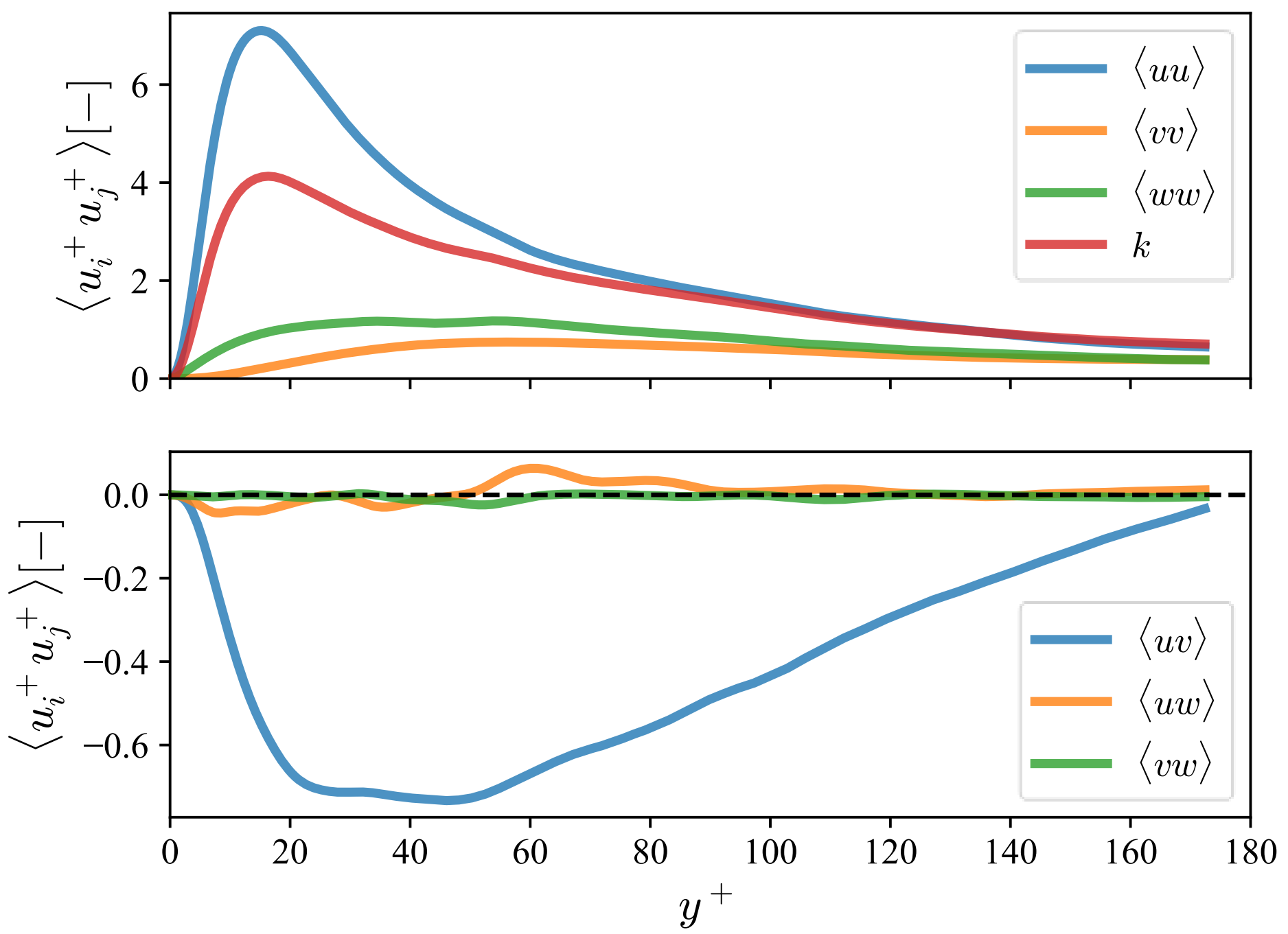

*Figure 42: Grid S2 – RST and TKE profiles using BCD with* $t^+ = 0.08$.

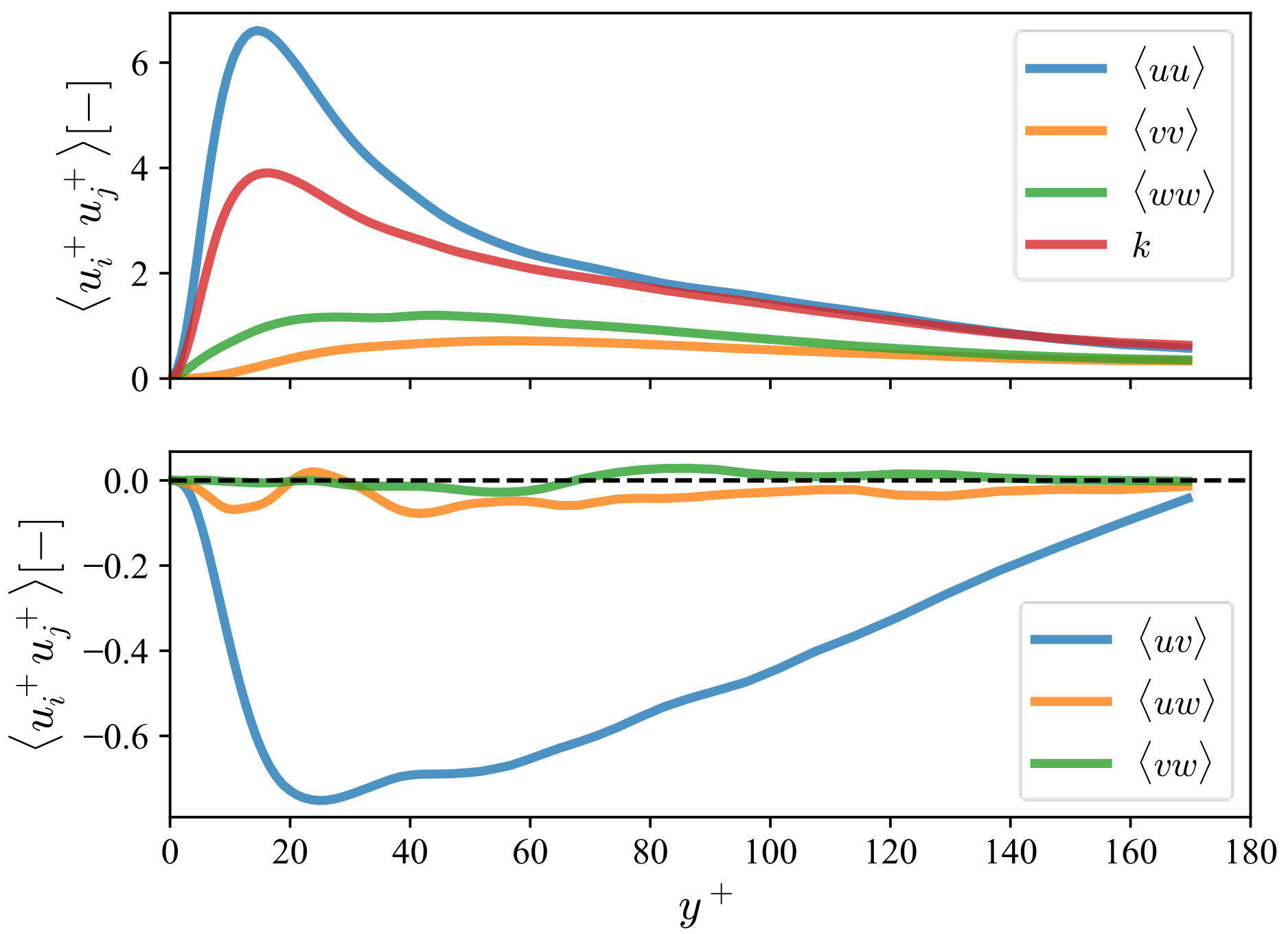

*Figure 43: Grid L1 – RST and TKE profiles using BCD with* $t^+ = 0.08$.

Finally, Figures 44 and 45 compare mesh length scale to computed Kolmogorov length scale profiles for grid S1 with CD and grid L1. No meaningful differences are evident when compared to Figure 33.

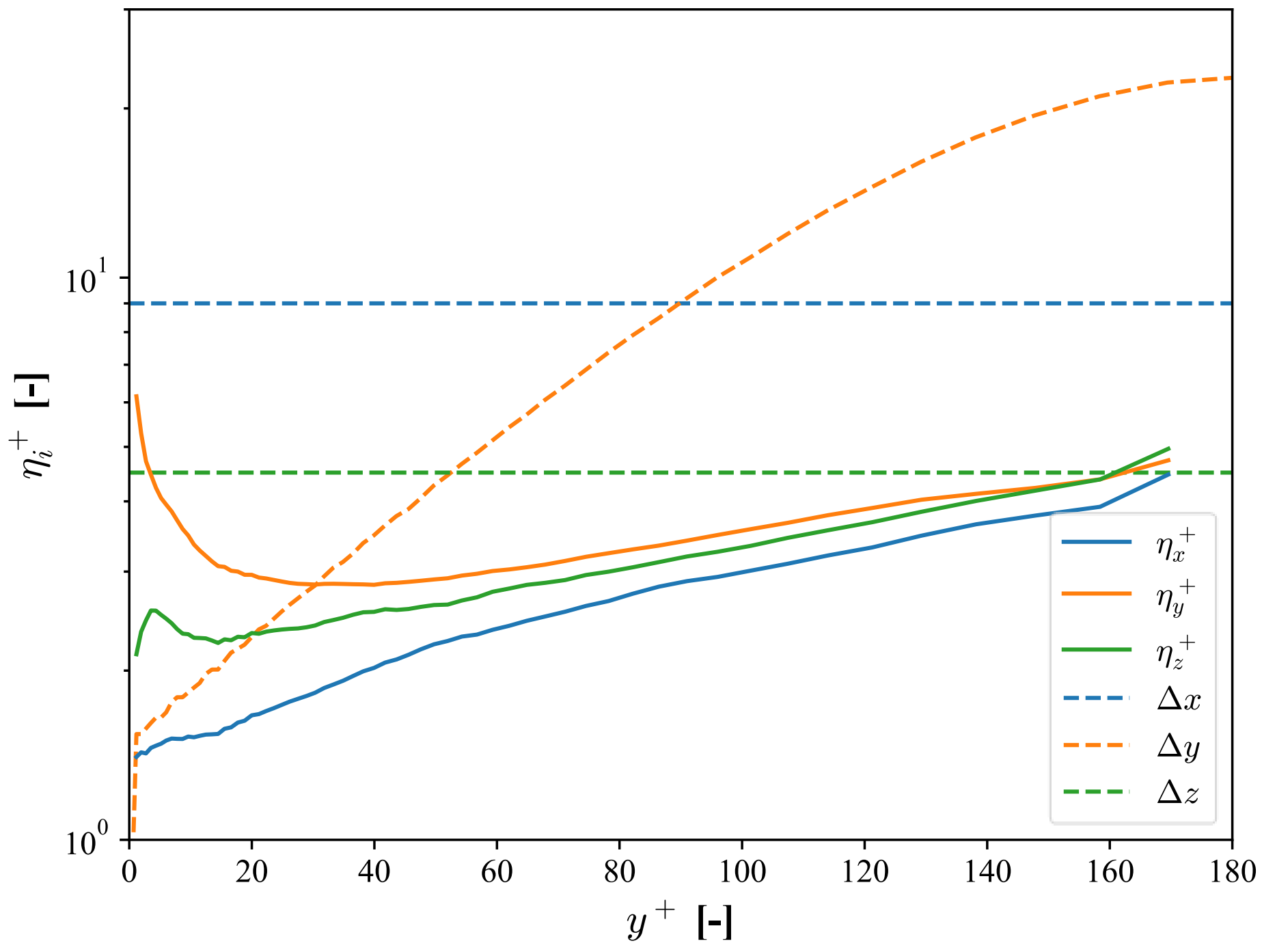

*Figure 44: Kolmogorov length scales calculated by grid S1 with pure CD.*

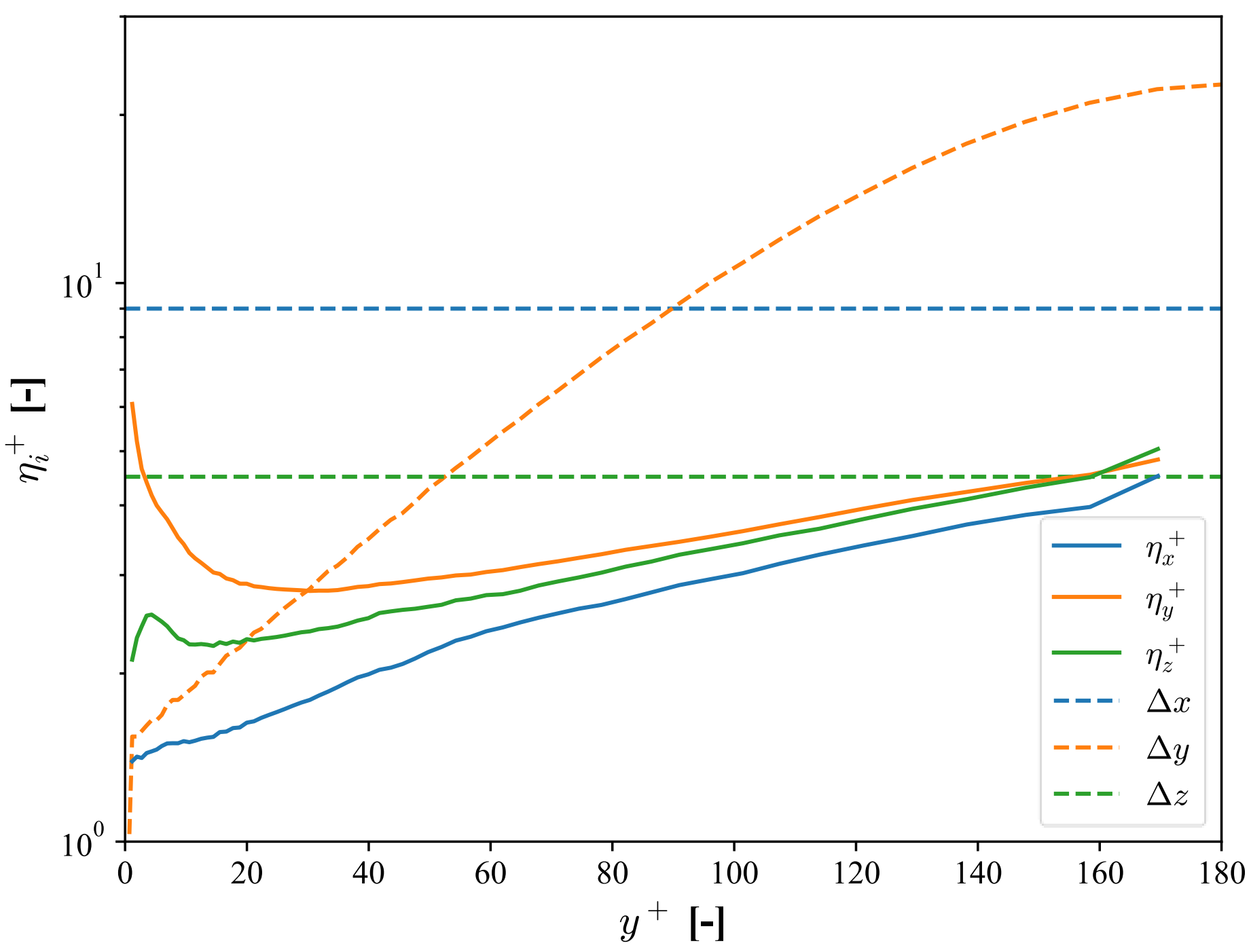

*Figure 45: Kolmogorov length scales calculated by grid L1.*

# Appendix D: Turbulent Flow Visualization (grid S1 BCD_SOI_DP_T0.08)

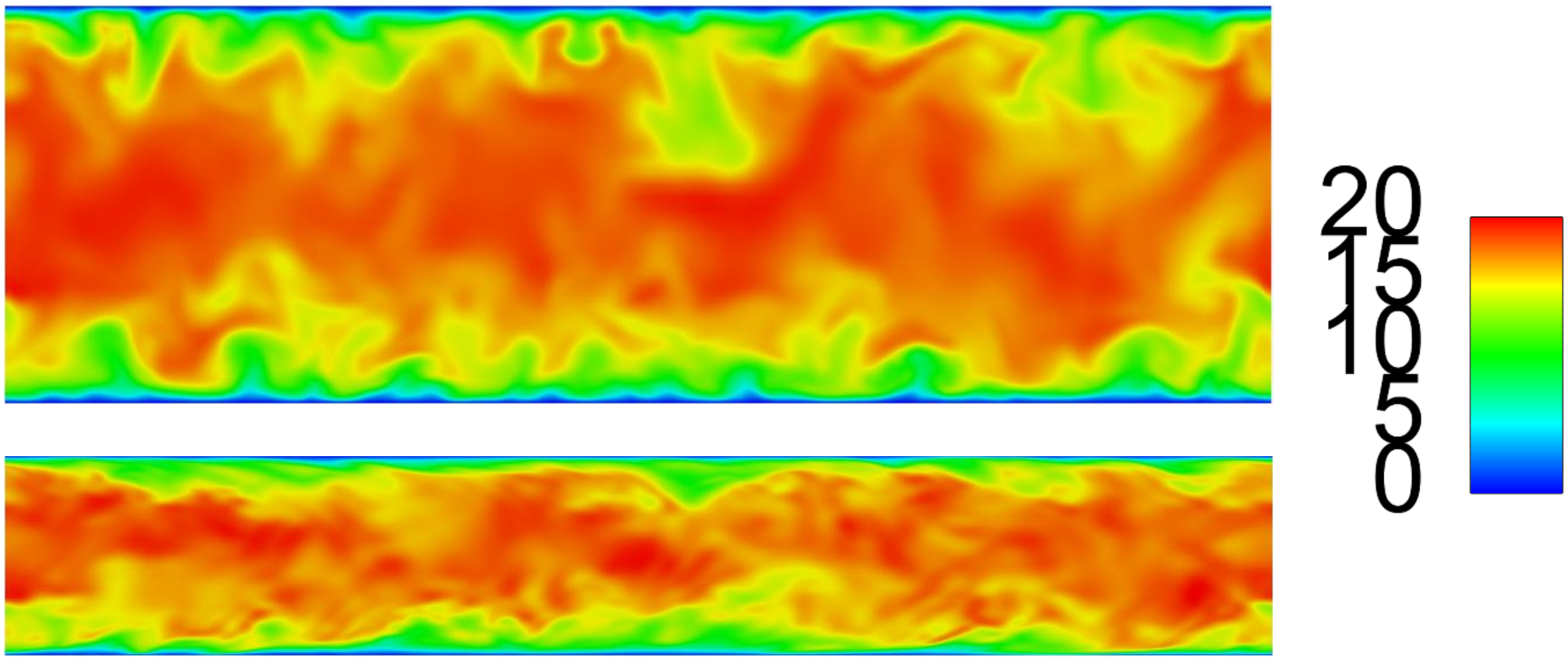

Figure 46 displays contours of dimensionless velocity magnitude at a single snapshot in time. As expected for a turbulent flow, a broad range of chaotic length scales are evident. Figure 47 and Figure 48 show isosurfaces of the second invariant of the velocity gradient tensor, also known as the Q criterion. These isosurfaces reveal many types of near-wall turbulent structures, including streaks and hairpin vortices. While instantaneous solution values, such as these graphics, are not of great validation value, they show that a true turbulent solution was attained.

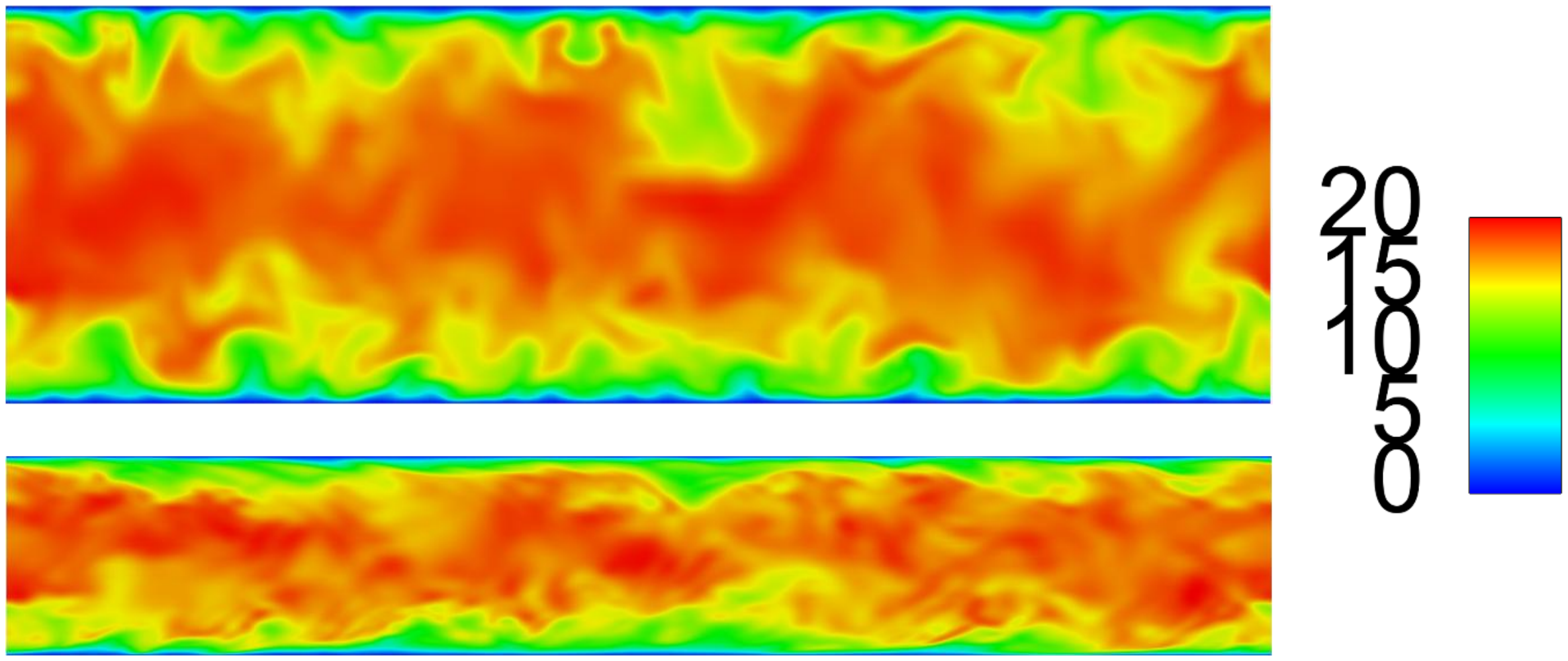

*Figure 46: Contours of instantaneous dimensionless velocity magnitude on the domain front plane (top) and side plane (bottom).*

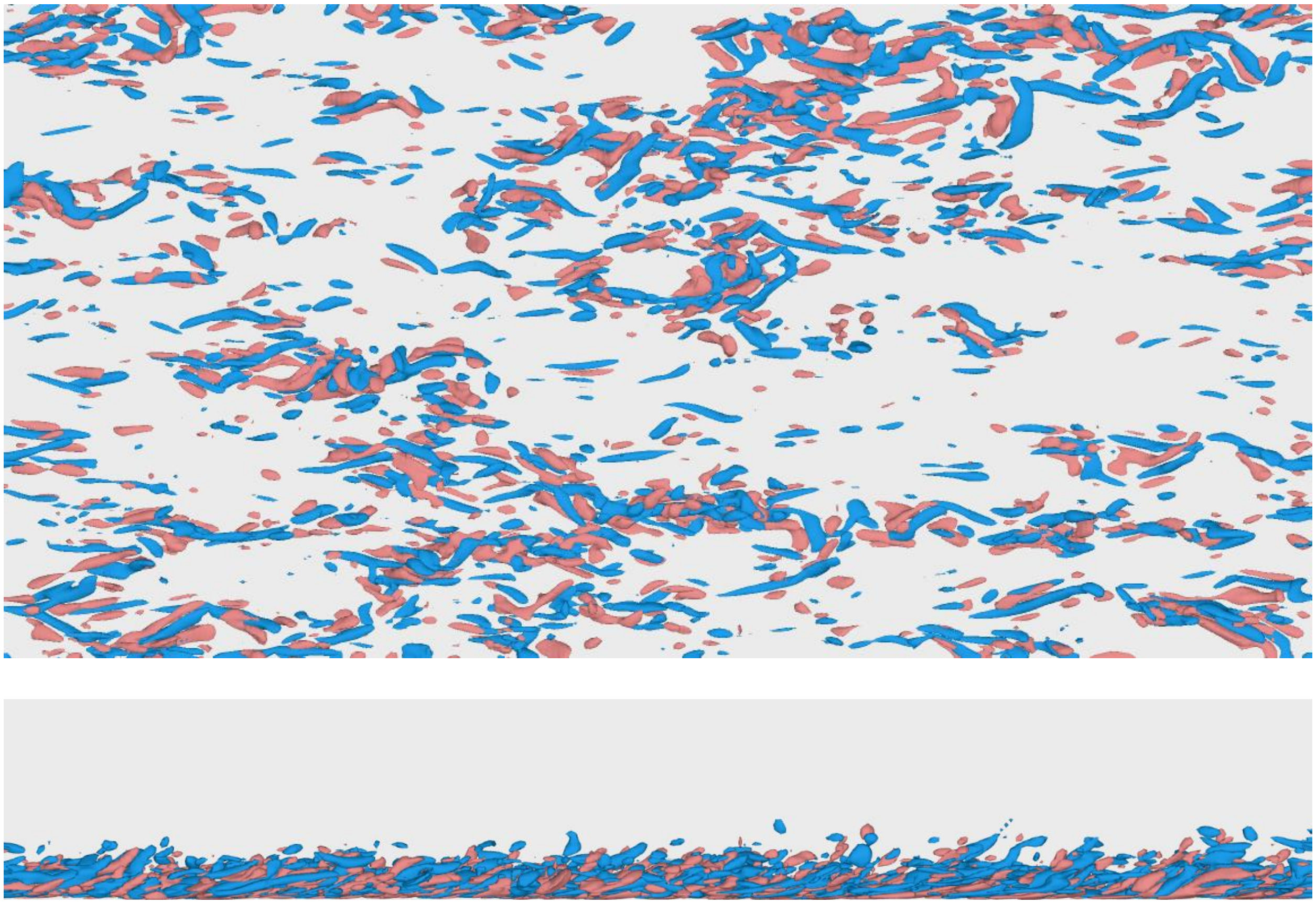

*Figure 47: Isosurfaces of Q criterion. Blue surfaces are at Q = 1500, and red surfaces are at Q = -1500. Top: Top view from centerline of channel. Bottom: View from side of channel. Flow proceeds from left to right in both graphics.*

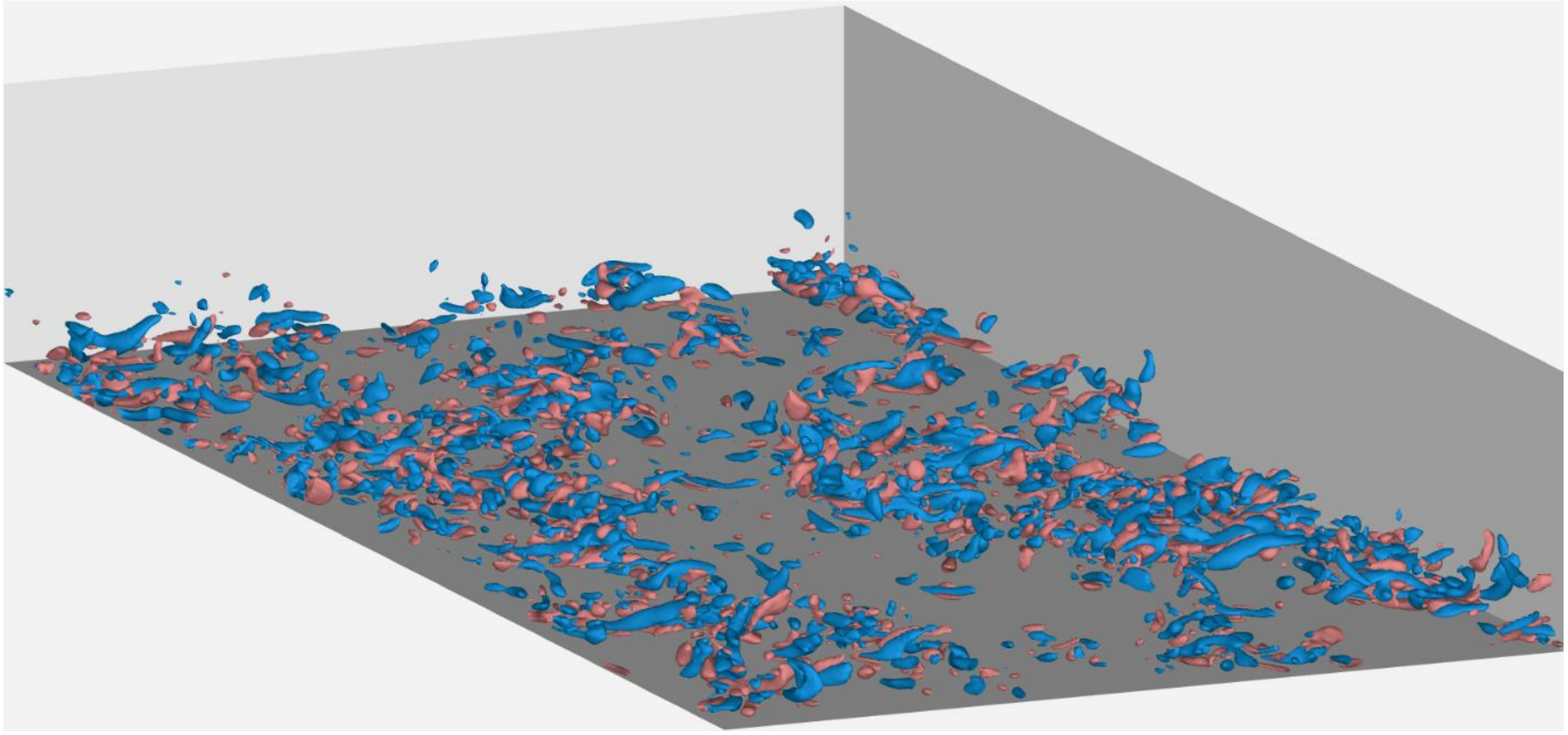

*Figure 48: Isosurfaces of Q criterion from an oblique view. Blue surfaces are at Q = 1500, and red surfaces are at Q = -1500. Flow is from top left to bottom right.*

# Appendix E: Practical Data Exfiltration Considerations

Surprisingly, one of the most challenging parts of using Fluent for DNS was extracting solution data from Fluent for analysis. Because Fluent's native GPU solver does not support Fluent's "user-defined functions" (arbitrary user-written C/C++ code that is hooked to the solver), and because some values of interest required temporal correlations, it was not possible to perform the necessary statistical analyses natively in the solver. One might expect that it would be trivial to write results to a file for later analysis, but this was not the case.

First is the issue of getting data out of the solver in any form whatsoever. While it is true that Fluent contains "auto-export" functionality to periodically write solution data to file, the intense preponderance of bugs in the GPU solver (despite its “fully released”, non-beta status) made this no small task.

The primary challenge we faced was exported data failing to update after each time step. Though the solution data could be seen updating in Fluent’s interface as the solver ran, each consecutive export contained a carbon copy of the same solution field. After much experimentation, we found that stopping and resuming the solver caused the exported solution field to update – once. Consequently, the only solution we found was to indefinitely restart the solver. We accomplished this with a Scheme loop:

```
(do () ((file-exists? "break-execution")) (ti-menu-load-string
"/solve/dual-time-iterate 30 20"))
```

This Scheme command repeatedly runs 30 time steps with 20 iterations each indefinitely, or until it sees a file named “break-execution” in the current working directory. While this was the best solution we found, we discovered that it occasionally causes Fluent to crash with an error stating that Scheme’s heap had filled up. This is despite the fact that the loop does not seem to contain any heap allocations. Fortunately, Fluent provides the `-h` command line argument to increase heap space.

While less pervasive than the prior problem, we also observed the solution field freezing indefinitely. This was fixed by switching from the “batch” (command-line interface) solver to the graphical interface, albeit at a *4x reduction* in solution speed.

After creating a way to get data out of Fluent, another roadblock was encountered. Though Fluent provides the user with many data export formats, most of the text-based (as opposed to binary) export formats write data to file with a single core. The amount of inter-process communication required for this to occur tremendously slowed down the solver. There is an export format tailored to the GPU solver, but its documentation is incomplete, so it was unusable. We found that Fluent's EnSight export format has a parallelized version that recovers solver speed, but we soon encountered the final major obstacle:

Regardless of solver precision, Fluent's exports dramatically truncate numerical precision of the results. In most cases, data was truncated to 6 decimal significant figures. For reference, a single-precision floating point number represents 7-8 digits of decimal precision, while a double-precision float represents 15-17 digits of decimal precision. Since we were running in double precision, this meant we were losing roughly 10 digits of precision in writing the data to file. While 6 digits may be sufficient for many problems, it was completely unusable for power spectra. No export format allows the user to specify precision. Until the data precision export limitation was breached, all energy spectra were plagued with white-noise-like pile-up at high wavenumbers.

Despite its impact on solver speed, we found that the "ASCII" export format was the only option for sensitive measures such as energy spectra, with its 10 digits of precision. However, for less sensitive measures (e.g. the root-mean-squared (RMS) fluctuating component of the velocity field) we were able to use the parallel EnSight export to improve solver speed.

# Appendix F: Least-Squares Gradient Reconstruction

The LS scheme computes the cell gradient of a scalar $\phi$ via a linear regression, which relates a cell's scalar value $\phi_c$ to the value $\phi_i$ at each neighboring cell $i$. In vector form, the relation with each neighboring cell $i$ is given by:

$$\nabla\phi_c \cdot \vec{r}_{c,c_i} = \phi_i - \phi_c \tag{40}$$

where $\vec{r}_{c,c_i}$ is the position vector from the centroid of the present cell to the centroid of neighbor $i$. When all neighboring cells are considered ($i \in [1, n]$), the relationship can be written as the following linear system:

$$\boldsymbol{A}\vec{x} = \vec{b} \tag{41}$$

with:

$$\boldsymbol{A} = \begin{bmatrix} \Delta x_{c,c_1} & \Delta y_{c,c_1} & \Delta z_{c,c_1} \\ \vdots & \vdots & \vdots \\ \Delta x_{c,c_n} & \Delta y_{c,c_n} & \Delta z_{c,c_n} \end{bmatrix}, \qquad \vec{x} = (\nabla\phi)_c = \begin{bmatrix} \frac{\partial\phi}{\partial x} \\ \frac{\partial\phi}{\partial y} \\ \frac{\partial\phi}{\partial z} \end{bmatrix}_c, \qquad \vec{b} = \begin{bmatrix} \phi_{c_1} - \phi_c \\ \vdots \\ \phi_{c_n} - \phi_c \end{bmatrix} \tag{42}$$

The solution vector, $\vec{x}$, represents the reconstructed gradient at the cell center, which is known to be first-order accurate on an arbitrary grid and second-order accurate on a structured grid [26]. Generally, this system is over-constrained, so it must first be orthogonalized before solution. Fluent accomplishes this with the Gram-Schmidt process.